\documentclass[lettersize,journal]{IEEEtran}
 
    \usepackage{jhoagg}
    \usepackage{mathtools}
    \usepackage{amsfonts}
    \usepackage{amssymb,latexsym}
    \usepackage[psamsfonts]{eucal}
    \usepackage{amsthm}
    \usepackage{graphicx}
    \usepackage[inline]{enumitem}
    \usepackage{indentfirst}
    \usepackage{setspace}
    \usepackage{microtype}
    \usepackage{threeparttable}
    \usepackage{float}
    \usepackage{cleveref}
    \usepackage{afterpage}
    \usepackage{placeins}
    \usepackage{cancel}
    \usepackage{leftidx}
    \usepackage{lipsum}
    \usepackage{multicol}
    \usepackage{breqn}
    \usepackage[export]{adjustbox}
    \usepackage{comment}
    \usepackage[ruled, linesnumbered, lined]{algorithm2e}
    \usepackage[dvipsnames]{xcolor}
    \usepackage{xcolor}
    \usepackage[most]{tcolorbox}
    \usepackage{soul}
    \usepackage[noadjust]{cite}
    \usepackage{standalone}
    \usepackage{textgreek}

    \usepackage{cleveref}

    \usepackage{tikz}
    \usetikzlibrary{shapes,arrows,positioning,fit}

    \usepackage{pgf}
    \usepackage{pgfplots}
    \usepgflibrary{plothandlers}
    \usepgfplotslibrary{external} 
    \pgfkeys{/pgf/number format/.cd,1000 sep={}}  
    \usetikzlibrary{calc,shapes.misc,plotmarks}
    \pgfplotsset{compat=1.13}
    
    \let\originalleft\left
    \let\originalright\right
    \renewcommand{\left}{\mathopen{}\mathclose\bgroup\originalleft}
    \renewcommand{\right}{\aftergroup\egroup\originalright}

    \newcounter{thm} 
    \theoremstyle{definition}
    \newtheorem{theorem}[thm]{\indent Theorem}
    
    \newtheorem{assumption}{\indent Assumption}
    
    \theoremstyle{definition}
    \newtheorem{proposition}{\indent Proposition}    
    
    \newtheorem{lemma}{\indent Lemma}
    
    \newtheorem{remark}{\indent Remark}
    
    \newtheorem{corollary}{\indent Corollary}
    
    \newtheorem{definition}{\indent Definition}
    
    \newtheorem{example}{\indent Example}
    
    \newtheorem{fact}{\indent Fact}
    
    \newtheorem{conjecture}{\indent Conjecture}
    
    \newtheorem{experiment}{\indent Experiment}

    \allowdisplaybreaks

    \newlist{enumA}{enumerate}{1}
    \setlist[enumA,1]{label=(A\arabic*),leftmargin=1cm}

    \newlist{enumC}{enumerate}{1}
    \setlist[enumC,1]{label=(C\arabic*),leftmargin=1cm}

    		\newcommand\xqed[1]{%
      \leavevmode\unskip\penalty9999 \hbox{}\nobreak\hfill
      \quad\hbox{#1}}
    \newcommand\exampletriangle{\xqed{$\triangle$}}

    \usepackage{cite}
    \usepackage{accents}
    
    \newlength\figureheight 
    \newlength\figurewidth

    \newcommand{\norm}[1]{\left\lVert#1\right\rVert}

    \DeclareMathOperator{\sgn}{sgn}

    \allowdisplaybreaks
    
    \graphicspath{ {Figures/} }

    \DeclareMathAlphabet{\mathcal}{OMS}{cmsy}{m}{n} 

    \crefname{equation}{}{}
    
    \usepackage[ruled, linesnumbered, lined]{algorithm2e}
    \SetKwInput{KwInit}{Initialize}
    \SetKwFunction{SafeSet}{SafeSet}
    \SetKwFunction{GetSafe}{GetSafe}
    \SetKwFunction{GetUnsafe}{GetUnsafe}
    
    \SetCommentSty{mycommfont}
    \newcommand{\ubar}[1]{\underaccent{\bar}{#1}}
    \newlist{enumalph}{enumerate}{1}
    \setlist[enumalph]{label=\textit{(\alph*)}}

\begin{document}
\title{Electromagnetic Formation Flying Using Alternating Magnetic Field Forces and Control Barrier Functions for State and Input Constraints}
\author{Sumit S. Kamat, T. Michael Seigler, and Jesse B. Hoagg
\thanks{S. S. Kamat, T. M. Seigler, and J. B. Hoagg are with the Department of Mechanical and Aerospace Engineering, University of Kentucky, Lexington, KY, USA. (e-mail: sumits.kamat@uky.edu, tmseigler@uky.edu, jesse.hoagg@uky.edu).}
}
                                                                                                                                         
\maketitle

\begin{abstract}
This article presents a feedback control algorithm for electromagnetic formation flying with constraints on the satellites' states and control inputs. 
The algorithm combines several key techniques. 
First, we use alternating magnetic field forces to decouple the electromagnetic forces between each pair of satellites in the formation. 
Each satellite's electromagnetic actuation system is driven by a sum of amplitude-modulated sinusoids, where amplitudes are controlled in order to prescribe the time-averaged force between each pair of satellites.
Next, the desired time-averaged force is computed from a optimal control that satisfies state constraints (i.e., no collisions and an upper limit on intersatellite speeds) and input constraints (i.e., not exceeding satellite's apparent power capability). 
The optimal time-averaged force is computed using a single relaxed control barrier function that is obtained by composing multiple control barrier functions that are designed to enforce each state and input constraint. 
Finally, we demonstrate the satellite formation control method in numerical simulations. 
\end{abstract}


\section{Introduction}\label{sec:introduction}

Electromagnetic formation flying (EMFF) is accomplished using satellites equipped with electromagnetic coils.
Each satellite's electromagnetic coils generates a magnetic field, which interacts with the magnetic fields of the other satellites to create magnetic field forces.  
These magnetic field forces can be used to control the relative positions of satellites \cite{Kong2004,KWON2010,Porter2014}.
Control of 2 satellites with EMFF actuation is addressed in \cite{Elias2007} and demonstrated experimentally in \cite{Kwon2011}. 
Other work on EMFF control of 2 satellites include \cite{Song2022,Lei2021}.

EMFF for more than 2 satellites is challenging because the intersatellite forces are nonlinear functions of the magnetic moments generated by all satellites in the formation as well as the relative positions of all satellites. 
In other words, there is complex coupling between the electromagnetic fields generated by all satellites in the formation.

EMFF for more than 2 satellites is addressed in \cite{Ahsun2006,Schweighart2010,Cai2023}.
However, these approaches require centralization of all measurement information and solving nonconvex constrained optimization problems that do not scale well to a large number of satellites. 
A decentralized EMFF method is presented in \cite{Abbasi2022}. 
In this method, each satellite has access to measurements of its position and velocity relative to only its local neighbor satellites. 
Notably, \cite{Abbasi2022} addresses the complex intersatellite force coupling by using alternating magnetic field forces (AMFF). 
The key idea of AMFF is that a pair of alternating (e.g., sinusoidal) magnetic moments results in a nonzero average interaction force between the pair of satellites if and only if those alternating magnetic moments have the same frequency \cite{Youngquist2013,Nurge2016}.
Thus, \cite{Abbasi2022} uses a sum of frequency-multiplexed sinusoidal magnetic moments to achieve desired intersatellite forces between every pair of satellites.  
The AMFF approach combined with traditional reaction-wheel actuation is used in \cite{Song2023,Song2023CCC} to develop a centralized algorithm for relative position and attitude control. 
Relative position and attitude control with AMFF is addressed in \cite{Takahasi2022} using a centralized algorithm and in \cite{Abbasi2020} using a decentralized approach. 
However, this previous work on EMFF with AMFF (i.e., \cite{Abbasi2022,Youngquist2013,Nurge2016,Song2023,Song2023CCC,Takahasi2022,Abbasi2020}) does not address state constraints (e.g., no satellite collisions, intersatellite speed limit) or input constraints (e.g., power limitations on the electromagnetic actuation system).

Control barrier functions (CBFs) can be used to determine controls that satisfy state constraints (e.g., \cite{wieland2007,ames2016,xiao2023,tan2021}). 
Specifically, CBFs are used to develop constraints that guarantee forward invariance of a set in which that the state constraints are satisfied.
CBFs are often implemented as constraints in real-time optimization control methods (e.g., quadratic programs) in order to guarantee state-constraint satisfaction while also minimizing a performance cost \cite{ames2016}. 
Although CBFs are used to address state constraints rather than input constraints, \cite{rabiee2024b,rabiee2024a} presents a CBF-based approach that simultaneously addresses state constraints and input constraints. 
This approach uses control dynamics, which allows the input constraints to be transformed into controller-state constraints, and a log-sum-exponential soft minimum to compose state-constraint CBFs and input-constraint CBFs into a single relaxed CBF.

This article presents a feedback control approach for EMFF using AMFF and a composite CBF to generate intersatellite forces that achieve formation and satisfy state and input constraints. 
We use the piecewise-sinusoidal magnetic moment approach in \cite{Abbasi2022} to prescribe the desired intersatellite forces between every pair of satellites; this approach is reviewed in \Cref{sec:Problem_Formulation}.
Then, \Cref{Section:Allocation} presents a new construction for the piecewise-sinusoidal amplitudes that achieve a prescribed intersatellite force. 
\Cref{sec:Formation control with state and input constraints} presents the main contribution of this article, namely, a feedback control for EMFF. 
We use linear quadratic regulator (LQR) to compute desired intersatellite forces, which achieve formation but may not respect state and input constraints. 
Then, we use the method from \cite{rabiee2024b,rabiee2024a} to construct a composite CBF that can be used to respect all state and input constraints. 
The desired intersatellite forces (from LQR) and the composite CBF are used together to construct a safe and optimal control.  
\Cref{sec:Formation Flying Simulation Results} demonstrates the EMFF approach in simulation.
Some preliminary results from this article appeared in the conference article \cite{Kamat2025}.
However, the current article goes beyond the preliminary work by developing an optimal desired formation control, addressing gravitational forces (e.g., formation in orbit), analyzing closed-loop performance, and demonstrating the effectiveness of the approach in different simulation scenarios.

\section{Notation}

Physical vectors are denoted with bold symbols, for example,~$\mathbf{r}$. 
The magnitude of $\mathbf{r}$ is denoted by $| \mathbf{r} |$.
A frame is a collection of mutually orthogonal physical unit vectors.
If $\mathcal{F}$ is a frame, then $[\mathbf{r} ]_{\mathcal{F}}$ is $\mathbf{r}$ resolved in $\mathcal{F}$.

Let $\| \cdot \|$ be the 2-norm.
Let $1_m \in \BBR^m$ denote the vector of ones, and let $I_m \in \BBR^{m \times m}$ be the $m \times m$ identity matrix. 
Kronecker product is denoted by $\otimes$.
If $a \in \mathbb{R}^3$, where $a= \begin{bsmallmatrix} a_1 &a_2 &a_3 \end{bsmallmatrix}^{\mathrm{T}}$, then
\begin{equation*}
    [ a ]_{\times} \triangleq \begin{bsmallmatrix}
        0 &-a_3 &a_2\\
        a_3 &0 &-a_1\\
        -a_2 &a_1 &0
    \end{bsmallmatrix}.
\end{equation*}

Let $\mathbb{N}$ denote the set of nonnegative integers.
We let $\mathcal I \triangleq \{1, ..., n\}$, where $n \in \mathbb{N}$ is the number of satellites in the formation, and $\mathcal P \triangleq \{(i,j) \in \mathcal I \times \mathcal I : i \neq j\}$, which is the set of ordered pairs.
Unless otherwise stated, statements in this paper that involve the subscript $i$ are for all $i \in \mathcal I$, and statements that involve the subscript $ij$ are for all $(i,j) \in \mathcal{P}$.

Let $\xi:\BBR^n \to \BBR$ be continuously differentiable. 
Then, $\xi^\prime :\BBR^n \to \BBR^{1 \times n}$ is defined as $\xi^\prime(x) \triangleq \pderiv{\xi(x)}{x}$. 
The boundary of the set $\SA \subseteq \BBR^n$ is denoted by $\mbox{bd }\SA$.

\begin{figure}[t]
    \centering
    \includegraphics[width=0.45\textwidth]{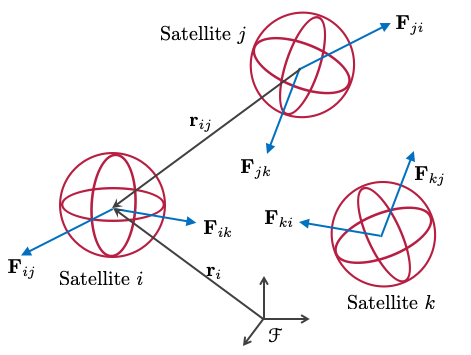}
    \caption{Each satellite is equipped with an electromagnetic actuation system consisting of three orthogonal coils. }
    \vspace{-10pt}
    \label{fig:Satellites_formation}
\end{figure}

\section{Problem Formulation}
\label{sec:Problem_Formulation}

Consider a system of $n$ satellites, where each satellite has mass $m$, as shown in Fig.~\ref{fig:Satellites_formation}.
The position $\mathbf r_i$ locates the mass center of satellite $i$ relative to the origin of an inertial frame $\CMcal{F}$ that consists of the right-handed set of mutually orthogonal unit vectors $\begin{bmatrix} \mathbf{i} &\mathbf{j} &\mathbf{k} \end{bmatrix}$. 
The velocity $\mathbf{v_i}$ and acceleration $\dot{\mathbf{v}}_i$ are the first and the second time-derivatives of $\mathbf{r}_i$ with respect to $\CMcal{F}$. 
The relative position $\mathbf{r}_{ij} \triangleq \mathbf{r}_i - \mathbf{r}_j$ locates the mass center of satellite $i$ relative to the mass center of satellite~$j$.

Each satellite has an electromagnetic actuation system (i.e., multiple electromagnetic coils), which can create a magnetic field. 
These magnetic fields interact to produce intersatellite forces, which can control the satellites' relative positions. 
More information on electromagnetic actuation for satellites is available in \cite{Kong2004,KWON2010,Porter2014,Abbasi2022}.
Each coil creates a magnetic dipole, and the resulting intersatellite force applied to satellite $i$ by satellite $j$ is given by
\begin{equation}
  \mathbf{F}_{ij} \triangleq \frac{3 \mu_0}{4 \pi |\mathbf r_{ij}|^4} \mathbf{f} (\mathbf{r}_{ij}, \mathbf{u}_i, \mathbf{u}_j), \label{F_ij}
\end{equation}
where $\mu_0$ is the vacuum permeability constant, $\mathbf{u}_i$ is the magnetic moment of satellite $i$, and 
\begin{align}
    \mathbf{f}(\mathbf{r}_{ij},&\mathbf{u}_{i},\mathbf{u}_j) \triangleq  
    \left( \mathbf{u}_j \cdot \frac{\mathbf{r}_{ij}}{|\mathbf{r}_{ij}|} \right) \mathbf{u}_i + 
 \left(\mathbf{u}_i \cdot \frac{\mathbf{r}_{ij}}{|\mathbf{r}_{ij}|} \right) \mathbf{u}_j 
 \notag 
   \\
 &+ \left[(\mathbf{u}_i \cdot \mathbf{u}_j)-5 \left(\mathbf{u}_i \cdot \frac{\mathbf{r}_{ij}}{|\mathbf{r}_{ij}|} \right) \left(\mathbf{u}_j \cdot \frac{\mathbf{r}_{ij}}{|\mathbf{r}_{ij}|} \right) \right]\frac{\mathbf{r}_{ij}}{|\mathbf{r}_{ij}|} \label{eq:f}.
\end{align}
The magnetic moment $\mathbf{u}_i$ is a function of the current supplied to the electromagnetic coils, and this is the control for satellite~$i$. 
 Note that \Cref{F_ij,eq:f} is a far-field model, which is valid for $|\mathbf{r}_{ij}|$ greater than 6--8 coil radii \cite{Schweighart2005Th}. 
This article focuses on formation maneuvers, where the the minimum allowable intersatellite distance satisfies the far-field model assumption.
More details on the model is available in \cite{Ahsun2006,Schweighart2010}.  For clarity of presentation, this article does not explicitly include electromagnetic coil dynamics; however, \Cref{section:conclusions} discusses how those effects can be address with the method of this article.

The translational dynamics of satellite $i$ are
\begin{align}
    \dot{\mathbf{v}}_i = \frac{c_0}{m} \sum_{j \in \CMcal{I} \setminus \{ i \}}\frac{1}{|\mathbf{r}_{ij}|^4}\mathbf{f}(\mathbf{r}_{ij},\mathbf{u}_{i},\mathbf{u}_j) + \frac{1}{m} \mathbf{G}(\mathbf{r}_i),
    \label{accel_i}
\end{align}
 where the gravitational force is
\begin{equation}
        \mathbf{G}(\mathbf{r}_{i}) = -\frac{\mu_{\mathrm{g}} m m_{\mathrm{e}}}{|\mathbf{r}_{i}|^3} \mathbf{r}_{i} \label{eq:gravitational_force}
\end{equation}
and $c_0 \triangleq 3 \mu_0 / (4 \pi)$, $\mu_\rmg = 6.67 \times 10^{-11}$~N-m$^2/$kg$^2$, and $m_{\mathrm{e}}>>m$ is the mass of a nearby massive celestial body (e.g., Earth).
Without loss of generality, the origin of $\mathcal{F}$ is the mass center of this celestial body.

It follows from \Cref{F_ij,eq:f} that the force applied to satellite $j$ by satellite $i$ is equal and opposite to the force applied to $i$ by $j$, that is, $\mathbf{F}_{ij} = - \mathbf{F}_{ji}$.
Hence, \eqref{accel_i} implies that $ \sum_{i \in \CMcal{I}} m \dot{\mathbf{v}}_i (t) = \sum_{i \in \CMcal{I}} \mathbf{G}(\mathbf{r}_i)$, which implies that the linear momentum of the formation is conserved in the absence of gravity. 
The electromagnetic forces $\mathbf{F}_{ij}$ can alter relative positions, but they have no effect on the position of the total mass center of the satellites. 
Thus, in practice, EMFF would most likely be used in combination with other actuation systems. 
For example, propellant thrusters on at least one satellite might be needed to address orbit maintenance in the presence of disturbances (e.g., atmospheric drag in low Earth orbit) or allow for orbit transfers, which cannot be readily accomplished with EMFF. 
Similarly, a traditional attitude control system such as reaction wheels could be used to maintain attitude in the presence of disturbances (e.g., torques introduced due to interaction with Earth's magnetic field).

For a satellite in circular orbit with radius $r_{\rmo}>0$, the gravitational force is
\begin{equation}
        \tilde{\mathbf{G}}(\mathbf{r}_{i}) = -\frac{\mu_{\mathrm{g}} m m_{\mathrm{e}}}{r_{\rmo}^3} \mathbf{r}_{i}. \label{eq:approx_gravitational_force}
\end{equation}
If all satellites have approximately the same nominal orbit, then ${\mathbf{G}}(\mathbf{r}_{i}) \approx \tilde{\mathbf{G}}(\mathbf{r}_{i})$.
The control development in \Cref{sec:Problem_Formulation,Section:Allocation,sec:Formation control with state and input constraints} uses the approximation \eqref{eq:approx_gravitational_force}; however, \eqref{eq:gravitational_force} is used for the simulation results in \Cref{sec:Formation Flying Simulation Results}. 
The method in this paper can be easily extended to the situation where $r_\rmo$ is time varying (e.g., elliptic orbits) as long as the nominal orbit $r_\rmo(t)$ is known (e.g., eccentricity, semi-major axis, and true anomaly are known). 
For clarity of presentation, we focus on the case where $r_\rmo$ is constant.


\subsection{Piecewise-Sinusoidal Controls}
\label{sub_sec:Piecewise-Sinusoidal Controls}

This section reviews the piecewise-sinusoidal magnetic-moment approach in~\cite{Abbasi2022}, which is used to address the coupling that occurs between the electromagnetic fields generated by all satellites. 
In this approach, each satellite uses a piecewise-sinusoidal magnetic moment $\mathbf{u}_i$ that is the sum of $n-1$ sinusoids with $n-1$ unique frequencies, where each unique frequency is common to only one pair of satellites. 
Thus, there are a total of $\ell \triangleq n(n-1)/2$ unique frequencies.
In this approach, the amplitudes of each common-frequency pair of sinusoids are selected to prescribe the average intersatellite force between the associated satellite pair.

For each $(i, j ) \in \mathcal{P}$, let $\omega_{ij} > 0$ be the interaction frequency, where $\omega_{ij} = \omega_{ji}$ is unique. 
Next, let $T > 0$ be a common multiple of $\{ 2\pi/\omega_{ij}: (i,j) \in \SP \}$. 
Then, for all $k \in \mathbb{N}$ and $t \in [ kT, kT + T )$, we consider the piecewise-sinusoidal control
\begin{align}
    \mathbf{u}_i(t)=\sum_{j \in \mathcal{I} \setminus \{i\} } \mathbf p_{ij,k} \sin \omega_{ij}t,
    \label{eqn:u_i}
\end{align}
where the amplitude sequences $\{ \mathbf{p}_{ij,k} \}_{k=0}^\infty$ are the control variables. 
Notice that satellite $i$ now has $n-1$ control sequences, namely, $\{ \mathbf{p}_{ij,k} \}_{k=0}^\infty$ for all $j \ne i$. 
 This article focuses on all-to-all interaction forces with $\ell$ unique frequencies.
For a formation with many satellites, all-to-all interaction may not be practical because of the large number of unique frequencies and associated practical limitations (e.g., spectral crowding). 
In this case, the method in this paper can be applied to multiple smaller formations within the overall formation. 
Since $\mathbf{F}_{ij}$ is inversely proportional to $|\mathbf{r}_{ij}|^4$, it follows that $| \mathbf{F}_{ij} |$ is negligible at large intersatellite distances.
Thus, each satellite interacts primarily with its nearby neighbors and removing the interaction frequencies with distant satellites is expected to have minimal impact. 
Analyzing this approach is outside the scope of the current article.

The next result demonstrates that the control \eqref{eqn:u_i} approximately decouples the time-averaged force between each pair of satellites. 
This result is~\cite[Prop. 1]{Abbasi2022} and is an immediate consequence of averaging the product of sinusoids over an integer number of periods.

\begin{proposition}\label{Proposition_1}
Consider $\mathbf{f}$ and $\mathbf{u}_i$ given by (\ref{eq:f}) and (\ref{eqn:u_i}). 
Let $\mathbf{r}$ be constant, and let $i, j \in \mathbb{N}$.
Then, for all $k \in \mathbb{N}$, 
\begin{equation}
        \frac{1}{T} \int^{kT+T}_{kT} \mathbf f(\mathbf{r}, \mathbf{u}_i(t),\mathbf{u}_j(t) ) \, {\rm d}t = \frac{1}{2} \mathbf f(\mathbf{r}, \mathbf{p}_{ij,k},\mathbf{p}_{ji,k}).
    \notag
\end{equation}
\end{proposition}

\subsection{Time-Averaged Dynamics}
\label{sec:time_avg_dyn}

We use \Cref{Proposition_1} to develop an approximate model for \eqref{accel_i} and \eqref{eqn:u_i}. 
First, integrating \eqref{accel_i} over the interval $[kT, kT + T ]$ yields
\begin{equation}
    \mathbf{v}_i(kT+T) =  \mathbf{v}_i(kT) + \frac{T}{m} \sum_{j \in \mathcal{I} \backslash \{ i \}} \mathbf{\bar{F}}_{ij} (k) + \frac{T}{m}  \mathbf{\bar{G}}(k),
\label{discrete_velocity}    
\end{equation}
where the average intersatellite force is
\begin{equation}
    \bar{\mathbf{F}}_{ij} (k) \triangleq \frac{1}{T} \int_{kT}^{kT+T} \frac{c_0}{| \mathbf{r}_{ij} (t) |^4} \mathbf{f}(  \mathbf{r}_{ij} (t), \mathbf{u}_{i} (t), \mathbf{u}_{j} (t)) \, {\rm d}t,
    \label{avg_F_ij}
\end{equation}

and the average gravitational force is
\begin{equation}
    \mathbf{\bar{G}}(k) \triangleq -\frac{1}{T} \int_{kT}^{kT+T} \frac{\mu_{\mathrm{g}} m m_{\mathrm{e}}}{|\mathbf{r}_{i}(t)|^3} \mathbf{r}_{i}(t) \, {\rm d}t.
\end{equation}
For all $k \in \mathbb{N}$ and $t \in [ kT, kT + T )$, define the \textit{approximate average intersatellite force}
\begin{equation}
     \tilde{\mathbf{F}}_{ij} (t) \triangleq \frac{c_0}{2 | \mathbf{r}_{ij} (t) |^4} \mathbf{f}(  \mathbf{r}_{ij} (t), \mathbf{p}_{ij,k},\mathbf{p}_{ji,k}).
     \label{avg_approx_F_ij}
\end{equation}
\Cref{Proposition_1} implies that if $\mathbf{r}_{ij} (t)$ is constant on $[ kT, kT + T )$, then $\tilde{\mathbf{F}}_{ij} (t) = \bar{\mathbf{F}}_{ij} (k)$ on $[ kT, kT + T )$. Thus, if $\mathbf{r}_{ij}$ does not change significantly over each period $T$, then \eqref{accel_i} and \eqref{eqn:u_i} is approximated by 
\begin{equation}
    \dot{\tilde{\mathbf{v}}}_i = \frac{c_0}{2 m} \sum_{j \in \CMcal{I} \setminus \{ i \}}\frac{1}{|\tilde{\mathbf{r}}_{ij} |^4}\mathbf{f}(\tilde{\mathbf{r}}_{ij},\mathbf{p}_{ij,k},\mathbf{p}_{ji,k}) - \frac{\mu_{\mathrm{g}} m_{\mathrm{e}}}{r_{\rmo}^3} \mathbf{\tilde{r}}_{i}, \label{eq:avg_model}
\end{equation}
where $\tilde{\cdot}$ is the approximate variable.

Next, we resolve \eqref{eq:avg_model} in the inertial frame $\CMcal{F}$. 
Specifically, let $r_i \triangleq [\tilde{\mathbf{r}}_i]_{\CMcal{F}}$ and $v_i \triangleq [\tilde{\mathbf{v}}_i]_{\CMcal{F}}$, and it follows from \eqref{eq:avg_model} that 
\begin{align}
    \dot{r}_{i} &= v_{i}, \label{eq:model.1}\\
    \dot{ v }_{i} & = \frac{c_0}{2m} \sum_{j \in \CMcal{I} \setminus \{ i \}}\frac{1}{ \norm{r_{ij}}^4 } f( r_{ij}, p_{ij}, p_{ji}) - \frac{\mu_{\mathrm{g}} m_{\mathrm{e}}}{r_{\rmo}^3} r_{i}, 
    \label{eq:model.2}
\end{align}
where $r_{ij} \triangleq r_{i} - r_{j}$, and
\begin{align}
    f( r_{ij}, p_{ij}, p_{ji}) &\triangleq \frac{ p^{\mathrm{T}}_{ji} r_{ij}}{ \norm{r_{ij}} } p_{ij} + \frac{p^{\mathrm{T}}_{ij}  r_{ij}}{\norm{r_{ij}}} p_{ji} +  \frac{ p^{\mathrm{T}}_{ij} p_{ji}}{ \norm{r_{ij}} } r_{ij}
 \notag 
   \\
 &\qquad 
 - 5 \frac{ p^{\mathrm{T}}_{ij} r_{ij} p^{\mathrm{T}}_{ji} r_{ij} }{ \norm{r_{ij}}^3 } r_{ij},
\label{eq:model.3}
\end{align}
and $p_{ij} \colon [0,\infty) \to \BBR^3$ are the controls, which are sampled in order to generate the amplitudes in \eqref{eqn:u_i}.
Specifically, the sinusoidal amplitudes are 
\begin{equation}
    \mathbf p_{ij,k} = \begin{bmatrix}
    \mathbf{i} &\mathbf{j} &\mathbf{k}
\end{bmatrix} p_{ij}(kT). \label{eq:p_ij_k}
\end{equation}
Note that~\cite[Sec. V]{Abbasi2022} shows that the amplitude pair $(p_{ij},p_{ji})$ can be selected such that $f( r_{ij}, p_{ij}, p_{ji})$ takes on any prescribed value. 
Thus, this paper uses \cref{eq:model.1,eq:model.2} to design a desired $f( r_{ij}, p_{ij}, p_{ji})$ that achieves formation flying while satisfying state and input constraints.
Then, we compute an amplitude pair $(p_{ij},p_{ji})$ such that $f( r_{ij}, p_{ij}, p_{ji})$ is equal to its desired value. 
The time-averaged dynamics \Cref{eq:model.1,eq:model.2,eq:model.3} are used for the control development in \Cref{sec:ctrl_obj}--\Cref{sec:Formation control with state and input constraints}. However, \eqref{F_ij}--\eqref{eq:gravitational_force} and \eqref{eqn:u_i} are used for the simulations in Section~VI.


\subsection{Control Objective}
\label{sec:ctrl_obj}

Let $\mathbf{d}_{ij}$ be the desired position of satellite $i$ relative to satellite $j$. 
Note that $\mathbf{d}_{ij}$ is potentially time varying, and note that $\mathbf{d}_{ji} = -\mathbf{d}_{ij}$.
The formation control objective is to design piecewise-sinusoidal magnetic-moment controls \eqref{eqn:u_i} such that the relative positions ${r}_{ij}$ converge to the desired relative positions $d_{ij} \triangleq [\mathbf{d}_{ij}]_{\CMcal{F}}$, and the relative velocities $v_{ij} \triangleq v_{i} - v_{j}$ converge to $\dot{d}_{ij}$.

In addition to the formation objective, the satellites must satisfy state and control-input constraints.  
First, let $\ubar{r} > 0$ denote the collision radius, that is, the minimum acceptable distance between any 2 satellites. 
 Note that $\ubar{r}$ is selected to prevent physical collision and satisfy the far-field model assumption.
Next, let $\bar{s} > 0$ denote the maximum intersatellite speed, that is, the maximum acceptable relative speed between any 2 satellites. 
Finally, let $\bar q > 0$ denote the maximum apparent power capability of each satellite. 
The apparent power of the $i$th satellite is approximately 
\begin{equation}
    q_i(t) \triangleq \frac{1}{N^2 \sigma^2} \sum_{j \in \SI\backslash \{ i \} } {Z_{ij}} \| p_{ij}(t) \|^2, \label{eq:apparent_power}
\end{equation}
where $Z_{ij} >0$ is the impedance of the coils at frequency $\omega_{ij}$, $\sigma>0$ is the coil cross-sectional area, and $N>0$ is the number of turns in the coil. 
In summary, the control objectives are:
\begin{enumerate}[leftmargin=0.8cm]
	\renewcommand{\labelenumi}{(O\arabic{enumi})}
	\renewcommand{\theenumi}{(O\arabic{enumi})}

\item\label{obj1}
$\lim_{t \to \infty} [ {r}_{ij}(t) - {d}_{ij}(t) ] =0$ and $\lim_{t \to \infty} [ v_{ij}(t) - \dot{d}_{ij}(t)]=0$.

\item\label{obj3}
For all $t \ge 0$, $\| {r}_{ij}(t) \| \ge \ubar{r}$.

\item\label{obj4}
For all $t \ge 0$, $\| {v}_{ij}(t) \| \le \bar{s}$.

\item\label{obj5}
For all $t \ge 0$, $q_i(t) \le \bar q$.
\end{enumerate}

The upper limit $\bar{s}$ on intersatellite speed is selected such that $\mathbf{r}_{ij}$ does not change significantly over the period $T$ and the time-averaged model \eqref{eq:avg_model} is valid. 
Smaller $T$ allows for larger $\bar{s}$; however, this tends to require higher frequencies $\omega_{ij}$. 
Since impedance $Z_{ij}$ scales with frequency $\omega_{ij}$, selecting smaller $T$ increases apparent power $q_{i}$, which in turn, limits control authority due to the power constraint \ref{obj5}. 
Thus, selecting $T$ involves a trade-off between power capability/control authority and the maximum allowable intersatellite speed. 

For clarity of presentation, this article does not explicitly address thermal management of the electromagnetic coils; however, the method in this article can be extended to account for thermal management by including a thermal management constraint, which is similar to the power constraint \ref{obj5} but is a function of the integral of the current amplitude squared. 
This extension is discussed in more detail in \Cref{section:conclusions}.

To account for uncertainty in $N$, $\sigma$, and $Z_{ij}$, the apparent power \Cref{eq:apparent_power} can be computed by replacing the nominal values of $N$, $\sigma$, and $Z_{ij}$ with the bounds $\ubar{N} < N$, $\ubar{\sigma}<\sigma$, and $\bar{Z}_{ij} >Z_{ij}$. 
In this case, \Cref{eq:apparent_power} computed using $\ubar{N}$, $\ubar{\sigma}$, and 
$\bar{Z}_{ij}$ is an upper bound on the apparent power.

%
\section{Control Amplitude Pair to Achieve Prescribed Intersatellite Force} 
\label{Section:Allocation}

This section provides a construction for the amplitude pair $(p_{ij},p_{ji})$ such that $f( r_{ij}, p_{ij}, p_{ji})$ has a prescribed value. 
The construction in~\cite{Abbasi2022} has multiple discontinuities, which are not conducive to address the input constraint \ref{obj5}. 
Thus, this section presents a simpler and new construction.

Consider $R: \mathbb{R}^3 \times \mathbb{R}^3 \rightarrow \mathbb{R}^{3 \times 3}$ given by
\begin{equation}
    R(r, f_*) \triangleq \begin{cases}
        \begin{bsmallmatrix}
        \frac{r}{\norm{r}} &
        \frac{  r^{\mathrm{T}} r f_* - r^{\mathrm{T}} f_* r  }{ \norm{r}\| [r]_{\times}  f_* \| } & 
        \frac{ [r]_{\times}  f_*  }{\| [r]_{\times}  f_* \|} 
    \end{bsmallmatrix}^\rmT, & [r]_{\times} f_* \neq 0, \\
         \begin{bsmallmatrix}
        \frac{r}{\norm{r}} &
        0 & 
        0
    \end{bsmallmatrix}^\rmT, & [r]_{\times} f_* = 0.
    \end{cases} 
    \label{eq:R}
\end{equation}
If $[r]_{\times} f_* \ne 0$, then $R(r,f_*)$ is the rotation from $\mathcal{F}$ to a frame whose first axis is parallel to $r$, and whose remaining axes are orthogonal to $r$ and $f_*$.
Next, consider $a_x, a_y, b_x, b_y \colon \mathbb{R}^3 \times \mathbb{R}^3 \rightarrow \mathbb{R}$ given by  
\begin{align}
    a_x(r,f_*) &\triangleq -\frac{\sgn(r^{\mathrm{T}} f_*)}{2} 
     \left( \frac{|r^{\mathrm{T}} f_*| + \Phi_1(r,f_*) }{\norm{r}} \right)^{\frac{1}{2}}, \label{eq:ax}
    \\
    a_y(r,f_*) &\triangleq \frac{1}{ \sqrt{2} }  \left( \dfrac{ -|r^{\mathrm{T}} f_*| +\Phi_2(r,f_*) }{\norm{r}} \right)^{\frac{1}{2}}, \label{eq:ay}
    \\
    b_x(r,f_*) &\triangleq  \frac{1}{2} \left( \dfrac{ |r^{\mathrm{T}} f_*| +\Phi_2(r,f_*) }{  \norm{r} } \right)^{ \frac{1}{2} }, \label{eq:bx}
    \
    \\
    b_y(r,f_*) &\triangleq - \frac{\sgn \left( r^{\mathrm{T}} f_* \right) }{ \sqrt{2} }  \bigg( \dfrac{ -|r^{\mathrm{T}} f_*|+\Phi_1(r,f_*)}{\norm{r}}  \bigg)^{ \frac{1}{2} }, \label{eq:by}
\end{align}
where $\Phi_1, \Phi_2 \colon \mathbb{R}^3 \times \mathbb{R}^3 \rightarrow \mathbb{R}$ are defined by
\begin{align}    
 \Phi_1(r,f_*) &\triangleq\sqrt{ \| [r]_{\times}  f_* \|^2 + \norm{r}^2 \norm{f_*}^2 },\label{eq:Phi1}
 \\
 \Phi_2(r,f_*) &\triangleq (2-\sgn(r^{\mathrm{T}} f_*)^2) \Phi_1(r,f_*),\label{eq:Phi2}
\end{align}
Finally, consider $c_1,c_2 \colon \mathbb{R}^3 \times \mathbb{R}^3 \rightarrow \mathbb{R}$ given by
\begin{align}
    c_1(r,f_*) &\triangleq R(r,f_*)^{\mathrm{T}} 
     a(r,f_*) , \label{eq:c1}
    \\
    c_2(r,f_*) &\triangleq R(r,f_*)^{\mathrm{T}} 
    b(r,f_*), \label{eq:c2}
\end{align}
where
\begin{equation}
    a(r,f_*) \triangleq \begin{bsmallmatrix}
        a_x(r,f_*)\\
        a_y(r,f_*)\\
        0
    \end{bsmallmatrix}, \quad b(r,f_*) \triangleq \begin{bsmallmatrix}
        b_x(r,f_*)\\
        b_y(r,f_*)\\
        0
    \end{bsmallmatrix}. \label{eq:a_b_allo_mat}
\end{equation}
The next result shows that $(c_1,c_2)$ is a construction for the amplitude pair $(p_{ij},p_{ji})$ such that $f( r_{ij}, p_{ij}, p_{ji})$ takes on a prescribed value $f_*$. 
The proof is in Appendix~\ref{appen.A}.

\begin{proposition}    \label{Proposition_2}
    For all $r \in \mathbb{R}^3  \backslash \{ 0 \}$ and $f_* \in \mathbb{R}^3$, $f(r, c_1(r,f_*), c_2(r,f_*)) = f_*$.
\end{proposition}

The prescribed value of the intersatellite force $f(r_{ij}, p_{ij}, p_{ji})$ is defined as 
\begin{equation}
    f_{ij} \triangleq \| r_{ij} \|^4 \zeta_{ij}, \label{eq:fij}
\end{equation}
where $\zeta_{ij} \colon [0,\infty) \to \BBR^3$ is the control determined from the feedback algorithm given in the next section that satisfies \ref{obj1}--\ref{obj5}. 
Then, we let 
 \begin{equation}
     p_{ij} = \begin{cases}
            c_1(r_{ij}, f_{ij}), &i<j,\\
            c_2(r_{ij}, f_{ij}), &i>j,\\
        \end{cases}
        \label{eqn:p_ij_allocation}
 \end{equation}
 and it follows from \Cref{Proposition_2} that $f(r_{ij}, p_{ij}, p_{ji}) = f_{ij}$. 
 That is, the amplitude pair $(p_{ij},p_{ji})$ given by \eqref{eqn:p_ij_allocation} achieves the prescribed intersatellite force $f_{ij}$.

\section{EMFF with State and Input Constraints}
\label{sec:Formation control with state and input constraints}

This section presents a feedback algorithm for the control $\zeta_{ij}$ to satisfy \ref{obj1}--\ref{obj5}.
Consider $x \colon [0,\infty) \to \BBR^{6n}$ and $\zeta \colon [0,\infty) \to \BBR^{3 \ell}$ defined by 
\begin{equation}
    x(t) \triangleq \begin{bmatrix}
        r_1(t)\\
        r_2(t)\\
        \vdots\\
        r_n(t) \\
        v_1(t)\\
        v_2(t)\\
        \vdots\\
        v_n(t) \\
    \end{bmatrix},  \qquad \zeta(t) \triangleq \begin{bmatrix}
    \zeta_{12}(t)\\
        \vdots\\
        \zeta_{1n}(t)\\
        \zeta_{23}(t)\\
        \vdots\\
        \zeta_{2n}(t)\\
        \vdots\\
        \zeta_{n-2, n-1}(t)\\
        \zeta_{n-2, n}(t)\\
        \zeta_{n-1, n}(t)
    \end{bmatrix}, \label{eq:state_x_control_nu}
\end{equation}
which are the state and control for the entire satellite system.
Thus, \Cref{Proposition_2}, \Cref{eq:model.1,eq:model.2,eq:model.3,eqn:p_ij_allocation,eq:fij,eq:state_x_control_nu} imply that 
\begin{equation}
    \dot{x} = A x + B \zeta,
    \label{eqn:State_space_n_satellites}
\end{equation}
where $x(0) = x_0 \in \BBR^{6n}$,
\begin{gather}
  A \triangleq \begin{bmatrix}
        0_{n \times n} &I_n\\
        - \frac{\mu_{\mathrm{g}} m_{\mathrm{e}}}{r_{\rmo}^3} I_{n} &0_{n \times n}
    \end{bmatrix} \otimes I_3, 
    \quad
B \triangleq \frac{3 \mu_0}{8 \pi m} \begin{bmatrix}
        0_{n \times \ell}\\ B_0
    \end{bmatrix} \otimes I_{3},
    \label{eq:A}\\
    B_0 \triangleq  \begin{bsmallmatrix}
        {1}^{\mathrm{T}}_{n-1} &0_{1 \times (n-2)}  &\dots &0_{1 \times 2} &0
        \\
        -I_{n-1} &\begin{bsmallmatrix} {1}^{\mathrm{T}}_{n-2} \\
        -I_{n-2} \end{bsmallmatrix} &\dots &\begin{bsmallmatrix} 
        0_{(n-4) \times 2} \\
        1^{\mathrm{T}}_{2} \\
        -I_2 
        \end{bsmallmatrix}
        &\begin{bsmallmatrix} 
        0_{n-3} \\
        1 \\
        -1 
        \end{bsmallmatrix} \end{bsmallmatrix}. \label{eq:B}
\end{gather} 

Objective \ref{obj5} imposes constraints on the magnetic moment amplitudes $p_{ij}$, which are algebraically related to $r_{ij}$ and $\zeta_{ij}$ according to \Cref{eqn:p_ij_allocation,eq:fij}. 
Thus, \ref{obj5} is a constraint involving the state $x$ and control input $\zeta$. 
We address this combined state and input constraint by applying the method in \cite{rabiee2024b,rabiee2024a}, which uses control dynamics to transform input constraints into controller-state constraints.

Let the control $\zeta$ be generated by the control dynamics
\begin{equation}
    \dot{\zeta}(t) = A_{\mathrm{c}} \zeta(t) + B_{\mathrm{c}} \mu(t),
    \label{eqn:controller_state_space}
\end{equation}
where $A_{\rmc}\in \mathbb{R}^{3 \ell \times  
3 \ell}$, $B_{\mathrm{c}} \in \mathbb{R}^{3 \ell \times 3 \ell}$ is nonsingular, $\zeta(0) =\zeta_0 \in \BBR^{3 \ell}$ is the initial condition, and $\mu \colon [0,\infty) \to \BBR^{3 \ell}$ is the \textit{surrogate control}, that is, the input to \eqref{eqn:controller_state_space}.

The cascade of \Cref{eqn:State_space_n_satellites,eq:A,eq:B,eqn:controller_state_space} is 
\begin{equation}
\dot{\tilde{x}} = \tilde{A} \tilde{x}  + \tilde B \mu, \label{eq:cascade_1}
\end{equation}
where 
\begin{equation}
    \tilde{x} \triangleq \begin{bmatrix}
        x\\
        \zeta
    \end{bmatrix}, \quad 
    \tilde A  \triangleq \begin{bmatrix}
        A &B\\
        0_{3 \ell \times 6n} &A_{\mathrm{c}} 
        \end{bmatrix}, \quad
    \tilde B \triangleq  \begin{bmatrix}
        0_{6n \times 3 \ell}\\
        B_{\mathrm{c}} 
    \end{bmatrix}. \label{eq:cascade_2}
\end{equation}
Notice that the control dynamics \eqref{eqn:controller_state_space} make $\zeta$ into a state of the cascade \Cref{eq:cascade_1,eq:cascade_2}. 
Thus, \ref{obj5} is transformed from an input constraint into a constraint on the state of the cascade.

\begin{figure*}[t]
  \includegraphics[width=1.0\textwidth,,clip=true,trim= 0.01in 0in 0in 0.in]{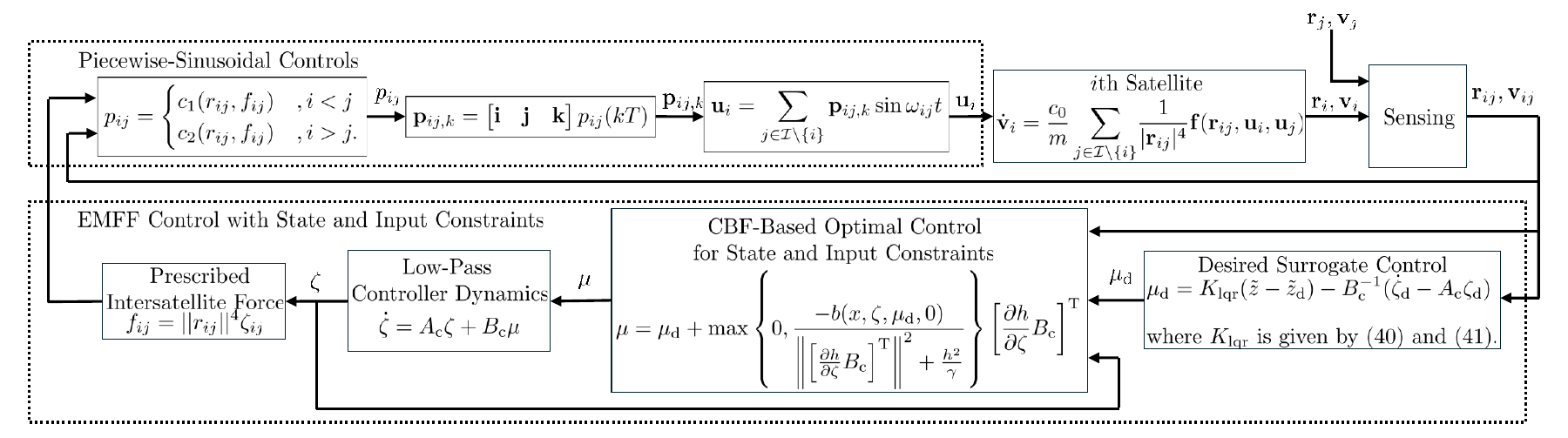}
  \centering
  \caption{EMFF control with state and input constraints using AMFF.}
  \label{fig:block_diagram_cbf_emff}
\end{figure*}

The remainder of this section presents an algorithm for $\mu$ such that the magnetic moment control $\mathbf{u}_i$ given by \Cref{eqn:u_i,eq:p_ij_k,eqn:p_ij_allocation,eq:fij,eqn:controller_state_space} achieves formation and respects the state and input constraints.
The overall control architecture is illustrated in \Cref{fig:block_diagram_cbf_emff}.

\vspace{-0.5em}

\subsection{Desired Formation Control}
\label{sec:LQR}

This section presents a desired surrogate control (i.e., a desired value for $\mu$) that achieves the formation objective \ref{obj1} but does not explicitly account for the constraints \ref{obj3}--\ref{obj5}.
To derive the desired surrogate control, we express the dynamics \Cref{eqn:State_space_n_satellites,eq:A,eq:B} in intersatellite coordinates rather than absolute coordinates. 
Specifically, define
\begin{equation}
    z \triangleq \begin{bmatrix}
        r_{12}^\rmT & \cdots & r_{1n}^\rmT & v_{12}^\rmT &  \cdots & v_{1n}^\rmT 
    \end{bmatrix}^\rmT. \label{eq:z}
\end{equation}
Next, define 
\begin{equation}
T \triangleq \begin{bmatrix}
        \begin{bsmallmatrix}
        1_{n-1} &-I_{n-1}
        \end{bsmallmatrix}
        &0_{n-1 \times n} 
        \\
        0_{n-1 \times n} & \begin{bsmallmatrix}
        1_{n-1} &-I_{n-1}
        \end{bsmallmatrix} 
    \end{bmatrix} \otimes I_3, 
    \label{eq:T}
\end{equation}
and note that $z = T x$, which implies that 
\begin{equation}
\dot{z} = T A x + G \zeta, \label{eq:reduced}
\end{equation}
where
\begin{equation*}
    G \triangleq T B = \begin{bmatrix}
        0_{n-1) \times \ell}\\
        G_0
    \end{bmatrix} \otimes I_3,
    \quad
G_0 \triangleq \frac{3 \mu_0}{8 \pi m} \begin{bmatrix}
        1_{n-1} & -I_{n-1 }
    \end{bmatrix} B_0.
\end{equation*}
Next, it follows from \Cref{eq:A,eq:z,eq:T,eq:state_x_control_nu} that
\begin{equation}
     T A  x= F z, \label{eq:tilde_x_z_transform}
\end{equation}
where
\begin{equation*}
    F \triangleq \begin{bmatrix}
        0_{n-1 \times n-1} &I_{n-1} \\
       - \frac{\mu_{\mathrm{g}} m_{\mathrm{e}}}{r_{\rmo}^3} I_{n-1} &0_{n-1 \times n-1} 
    \end{bmatrix} \otimes I_3. 
\end{equation*}
Thus, substituting \eqref{eq:tilde_x_z_transform} into \eqref{eq:reduced} yields the reduced-order relative dynamics 
\begin{equation}
    \dot{z} = F z + G \zeta. 
    \label{eq:reduced_state_z}
\end{equation}
The cascade of \Cref{eq:reduced_state_z} with \Cref{eqn:controller_state_space} is 
\begin{equation}
\dot{\tilde{z}} = \tilde{F} \tilde{z}  + \tilde G \mu, \label{eq:LQR_cascade.1}
\end{equation}
where 
\begin{equation}
    \tilde{z} \triangleq \begin{bmatrix}
        z\\
        \zeta
    \end{bmatrix}, \quad 
    \tilde F  \triangleq \begin{bmatrix}
        F & G\\
        0_{3 \ell \times 6(n-1)} &A_{\mathrm{c}} 
        \end{bmatrix}, \quad
    \tilde G \triangleq  \begin{bmatrix}
        0_{6(n-1) \times 3 \ell}\\
        B_{\mathrm{c}} 
    \end{bmatrix}. \label{eq:LQR_cascade.2}
\end{equation}

The desired trajectory for the state $\tilde z$ is 
\begin{equation*}
   \tilde z_\rmd \triangleq \begin{bmatrix}
        d^\rmT &
        \dot d^\rmT&
        \zeta_\rmd^\rmT
    \end{bmatrix}^\rmT
\end{equation*}
where 
\begin{gather*}
    d \triangleq  \begin{bmatrix}
        d_{12}^\rmT &
        \cdots&
        d_{1n}^\rmT
    \end{bmatrix}^\rmT, \\ 
    \zeta_\rmd \triangleq (G_0^{\rmT} (G_0 G_0^{\rmT})^{-1} \otimes I_3) \left(\ddot d + \frac{\mu_{\mathrm{g}} m_{\mathrm{e}}}{r_{\rmo}^3} d \right),
\end{gather*}
and $G_0 G_0^{\rmT} = n \left(\frac{3 \mu_0}{8 \pi m} \right)^2 \left( I_{n-1} + 1_{n-1 \times n-1} \right)$ is nonsingular. 
 Note that desired intersatellite force $\zeta_{d}$ is a function of the nominal orbit.

Next we consider the cost 
\begin{align*}
  \bar{J}(\hat{\mu}(\cdot))  &\triangleq \int_{0}^{ \infty } \left( \left [ \tilde z(\tau) - \tilde z_{\mathrm{d}}(\tau) \right ]^\rmT W_{z} \left [  \tilde z(\tau) - \tilde z_{\mathrm{d}}(\tau) \right ] \nn \right. \\
    & \qquad+ \left. \hat \mu(\tau)^\rmT W_\mu \hat \mu(\tau) \right) \, \mathrm{d} \tau,
\end{align*}
where $W_z \in \BBR^{(6(n-1) + 3 \ell) \times (6(n-1) + 3 \ell)}$ is positive semidefinite and $W_\mu \in \BBR^{3 \ell \times 3 \ell}$ is positive definite.
 A convenient choice for the weights is
\begin{equation}
    W_z = \mathrm{diag}(w_{r}I_{3n-3}, w_{v}I_{3n-3} ,w_{\zeta} I_{3 \ell},\quad W_{\mu} = w_{\mu} I_{3 \ell}, \label{LQR_weights}
\end{equation}
where $w_r,w_v, w_{\zeta}, w_{\mu} >0$. 
In this case, large $w_r$ and $w_v$ prioritizes fast formation maneuvers, whereas large $w_\zeta$ tends to reduce the magnitude of the intersatellite forces and large $w_{\mu}$ tends to reduce the magnitude of the derivative of the intersatellite forces.

Finally, we define the \textit{desired surrogate control}
\begin{equation}
    \mu_{\rmd}(x,\zeta) \triangleq K_{\rm{lqr}}(\tilde z- \tilde z_{\rmd}) - B^{-1}_{\rmc}  ( \dot{\zeta}_{\rmd} -A_{\rmc} \zeta_\rmd ), \label{eq:mu_d}
\end{equation}
where the LQR control gain is given by
\begin{equation}
   K_{\mathrm{lqr}} \triangleq -W^{-1}_{\mu} \tilde G^{\rm{T}} P, \label{eq:lqr_gain}
\end{equation}
and $P$ is the positive-semidefinite solution to the Riccati equation
\begin{equation}
  \tilde F^{\rmT} P + P \tilde F + W_{z} - P \tilde G W_{\mu }^{-1} \tilde G^{\rmT} P = 0.\label{eq:P}
\end{equation}
The next result follows immediately from  \cref{eq:LQR_cascade.1,eq:LQR_cascade.2,eq:mu_d,eq:lqr_gain}.

\begin{proposition} \label{prop:desired}
Consider \cref{eq:LQR_cascade.1,eq:LQR_cascade.2}, where $\mu=\mu_\rmd$.
Then, the error $\tilde e \triangleq \tilde z- \tilde z_{\rmd}$ satisfies
\begin{equation*}
    \dot {\tilde e} = (\tilde{F}+\tilde{G} K_{\rm{lqr}}) \tilde e,
\end{equation*}
where $\tilde{F}+\tilde{G} K_{\rm{lqr}}$ is asymptotically stable. 
\end{proposition}

\Cref{prop:desired} implies that $\tilde z$ converges exponentially to $\tilde z_{\rmd}$. 
Thus, the desired surrogate control $\mu_\rmd$ achieves \ref{obj1}; however, this control does not address \ref{obj3}--\ref{obj5}. 
The remainder of this section presents a method for constructing the control $\mu$ that is as close as possible to $\mu_\rmd$, while satisfying the state and input constraints \ref{obj3}--\ref{obj5}.

\subsection{Soft-Minimum Relaxed CBF for State and Input Constraints}

To address \ref{obj3}, consider the continuously differentiable candidate CBF
\begin{equation}
    R_{ij}(x) \triangleq \frac{1}{2} \left ( \|r_{ij}\|^2 - \ubar{r}^2 \right ), \label{eq:Rij}
\end{equation}
and note \ref{obj3} is satisfied if and only if for all $t \ge 0$, $R_{ij}(x(t)) \ge 0$.

To address \ref{obj4}, consider the continuously differentiable candidate CBF
\begin{equation}
    V_{ij}(x) \triangleq \frac{1}{2} \left ( \bar{s}^2 - \|v_{ij}\|^2 \right ), \label{eq:Vij}
\end{equation}
and note \ref{obj4} is satisfied if and only if for all $t \ge 0$, $V_{ij}(x(t)) \ge 0$.

Objective \ref{obj5} is a constraint related to $\| p_{ij} \|^2$, which is not continuous in $(r_{ij},\zeta_{ij})$.
Thus, we consider a continuously differentiable function $\psi: \mathbb{R}^{3} \times \mathbb{R}^{3} \rightarrow \mathbb{R}$ that upper bounds $\| p_{ij} \|^2$. 
Specifically, define
\begin{align}
    \psi(r_{ij}, \zeta_{ij}) &\triangleq  -\frac{\norm{r_{ij}}^3 r_{ij}^{\mathrm{T}} \zeta_{ij}}{4}  \tanh \left( \frac{\norm{r_{ij}}^3 r_{ij}^{\mathrm{T}} \zeta_{ij} }{\epsilon_1 } \right)
    \notag 
    \\
    &\qquad + \sqrt{ \norm{ r_{ij} }^6 \Phi_1(r_{ij},\zeta_{ij})^2  + \epsilon_2},
    \label{eq:psi}
\end{align}
where $0<\epsilon_1<<1$ and $0<\epsilon_2<<1$.
The next result relates $\| p_{ij} \|^2$ and $\psi(r_{ij}, \zeta_{ij})$.
The proof is in Appendix~\ref{appen.A}.

\begin{proposition}    \label{Prop:amplitude_bound}
Let $r \in \mathbb{R}^{3} \backslash \{0 \}$ and $f_* \in \mathbb{R}^3$. 
Then, 
\begin{equation*}
\psi \left ( r, \frac{f_*}{\| r \|^4} \right ) > \| c_1(r, f_*)\|^2 = (2-\sgn(r^{\mathrm{T}} f_*)^2)\norm{c_2(r,f_*)}^2.
\end{equation*}

\end{proposition}

\Cref{Prop:amplitude_bound} shows that $\psi(r_{ij}, \zeta_{ij})$ is a continuously differentiable upper bound on $\| p_{ij} \|^2$ and $\| p_{ji} \|^2$. 
The parameters $\epsilon_1$ and $\epsilon_2$ are introduced to make this upper bound smooth. 
Specifically, $\norm{r_{ij}}^3 r_{ij}^{\mathrm{T}} \zeta_{ij}  \tanh ( \norm{r_{ij}}^3 r_{ij}^{\mathrm{T}} \zeta_{ij}/ \epsilon_1 )$ is a smooth approximation of $\norm{r_{ij}}^3 \lvert r_{ij}^{\mathrm{T}} \zeta_{ij} \rvert$ and $(\norm{ r_{ij} }^6 \Phi_1(r_{ij},\zeta_{ij})^2  + \epsilon_2)^{1/2}$ is a smooth approximation of $( \norm{ r_{ij} }^6 \Phi_1(r_{ij},\zeta_{ij})^2)^{1/2}$. 
If $\epsilon_1$ and $ \epsilon_2$ are large, then $\psi(r_{ij}, \zeta_{ij})$ is a conservative approximation of $\| p_{ij}\|^2$. 
However, if $\epsilon_1$ and $ \epsilon_2$ are small, then $\|\psi^{\prime}(r_{ij},\zeta_{ij})\|^2$ is large at points where $\| p_{ij}\|^2$ is not differentiable. 
Thus, selecting $\epsilon_1$ and $\epsilon_2$ is a trade off between conservativeness of $\psi(r_{ij},\zeta_{ij})$ and the size of $\|\psi^{\prime}(r_{ij},\zeta_{ij})\|$.
For the examples in \Cref{sec:Formation Flying Simulation Results}, we determined that 
$\epsilon_1$ and $\epsilon_2$ on the order of $10^{-3}$ are good values.

To address \ref{obj5}, we consider the continuously differentiable candidate CBF
\begin{equation}
    Q_i(x,\zeta) \triangleq \bar q - \frac{1}{N^2 \sigma^2}\sum_{j \in \SI\backslash \{ i \} } Z_{ij} \psi(r_{ij}, \zeta_{ij}).\label{eq:Qi} 
\end{equation}
\Cref{Prop:amplitude_bound} combined with \Cref{eqn:p_ij_allocation,eq:fij} implies that if for all $t \ge 0$, $Q_i(x(t),\zeta(t)) \ge 0$, then \ref{obj5} is satisfied. 
Since the control dynamics \eqref{eqn:controller_state_space} make $\zeta$ a state of the cascade \Cref{eq:cascade_1,eq:cascade_2}, it follows that the constraint $Q_i(x(t),\zeta(t)) \ge 0$, which is sufficient to satisfy \ref{obj5}, is a constraint on the state of the cascade.

Next, we define 
\begin{align*}
    \SSS_\rms &\triangleq \{ (x,\zeta) \in \BBR^{6n} \times \BBR^{3\ell} \colon \mbox{for all } i \in \SI, \ Q_{i}(x,\zeta) \ge 0, \nn\\
&\qquad \mbox{and for all } (i,j) \in \SP, V_{ij}(x) \ge 0 \mbox{ and } R_{ij}(x) \ge 0\}.
\end{align*}
and it follows that if for all $t \ge 0$, $(x(t),\zeta(t)) \in \SSS_\rms$, then \ref{obj3}--\ref{obj5} are satisfied.

We note that $R_{ij}$, $V_{ij}$, and $Q_i$ have relative degree 3, 2, and one with respect to \Cref{eq:cascade_1,eq:cascade_2} with input $\mu$. 
Since $R_{ij}$ and $V_{ij}$ have relative degree 3 and 2, we use a higher-order approach (e.g., \cite{tan2021}) to construct higher-order candidate CBFs from $R_{ij}$ and $V_{ij}$.
Let $\alpha_{0},\alpha_1 \colon \mathbb{R} \to \mathbb{R}$ be extended class-$\mathcal{K}$ functions that are $3$-times and $2$-times continuously differentiable. 
Consider $R_{ij,1} \colon \mathbb{R}^{6n} \to \mathbb{R}$ defined by
\begin{align}
R_{ij,1}(x) & \triangleq  R_{ij}^\prime(x) \left [ A x +B \zeta \right ] + \alpha_{0}(R_{ij}(x))\nn\\
&= r_{ij}^\rmT v_{ij} + \alpha_{0}(R_{ij}(x)), \label{eq:Rij_1}
\end{align}
and consider $R_{ij,2} \colon \colon \mathbb{R}^{6n} \times \BBR^{3 \ell} \to \mathbb{R}$ defined by
\begin{align}
R_{ij,2}(x,\zeta) & \triangleq R_{ij,1}^\prime(x) \left [ A x + B\zeta \right ] + \alpha_{1}(R_{ij,1}(x)).
\label{eq:Rij_2}
\end{align}
Similarly, let $\alpha_{v} \colon \mathbb{R} \to \mathbb{R}$ be a $2$-times continuously differentiable extended class-$\mathcal{K}$ function, and consider $V_{ij,1} \colon \mathbb{R}^{6n} \times \BBR^{3 \ell} \to \mathbb{R}$ defined by
\begin{align}
V_{ij,1}(x,\zeta) & \triangleq V_{ij}^\prime(x) \left [ Ax + B\zeta \right ] + \alpha_{v}(V_{ij}(x))\nn\\
&=V_{ij}^\prime(x) B\zeta + \alpha_{v}(V_{ij}(x)).
\label{eq:Vij_1}
\end{align}

The intersection of the zero-superlevel sets of $Q_{i}$,  $V_{ij,1}$, and $R_{ij,2}$ is given by  
\begin{align*}
    \bar \SH &\triangleq \{ (x,\zeta) \in \BBR^{6n} \times \BBR^{3\ell} \colon \mbox{for all } i \in \SI, \ Q_{i}(x,\zeta) \ge 0, \nn\\
&\quad \mbox{and for all } (i,j) \in \SP, \ V_{ij,1}(x) \ge 0 \mbox{ and } R_{ij,2}(x) \ge 0\}.
\end{align*}
Similarly, the intersection of the zero-superlevel sets of $Q_{i}$,  $V_{ij}$, $V_{ij,1}$, $R_{ij}$, $R_{ij,1}$, and $R_{ij,2}$ is given by  
\begin{equation*}
        \bar \SSS \triangleq \{ (x,\zeta) \in \bar \SH \cap \SSS_\rms \colon \mbox{for all }  (i,j) \in \SP, R_{ij,1}(x) \ge 0\}.
\end{equation*}
Proposition~1 in \cite{tan2021} implies that if $(x_0,\zeta_0) \in \bar \SSS$ and for all $t\ge 0$, $(x(t),\zeta(t)) \in \bar \SH$, then for all $t\ge 0$, $(x(t),\zeta(t)) \in \bar \SSS \subset \SSS_\rms$.
In this case, \ref{obj3}--\ref{obj5} are satisfied. 
Thus, we consider a candidate CBF whose zero-superlevel set is a subset of $\bar \SH$.
Let $\rho>0$, and consider $h \colon \BBR^{6n} \times \BBR^{3 \ell} \rightarrow \mathbb{R}$ defined by
\begin{align}
h(x,\zeta) &\triangleq \mathrm{softmin}_{\rho} \Big ( R_{12,2}(x,\zeta),\ldots,R_{1n,2}(x,\zeta),\nn\\
&\qquad R_{23,2}(x,\zeta),\ldots, R_{2n,2}(x,\zeta),\ldots, R_{n-1n,2}(x,\zeta),\nn\\
&\qquad V_{12,1}(x,\zeta),\ldots,V_{1n,1}(x,\zeta), V_{23,1}(x,\zeta),\nn\\
&\qquad \ldots, V_{2n,1}(x,\zeta), \ldots, V_{n-1n,1}(x,\zeta),\nn\\
&\qquad Q_1(x,\zeta),\ldots,Q_n(x,\zeta) \Big ), 
\label{eq:h}
\end{align}
where 
\begin{equation}\label{eq:softmin}
\mbox{softmin}_\rho (z_1,\ldots,z_N) \triangleq -\frac{1}{\rho}\log\sum_{i=1}^Ne^{-\rho z_i},
\end{equation}
is the log-sum-exponential \textit{soft minimum}.

The soft minimum \eqref{eq:softmin} provides a lower bound on the minimum (e.g., \cite{rabiee2024c,rabiee2024b}).
The worst case conservativeness of the soft-minimum approximation is $(\log N)/\rho$ \cite{safari2024b}. 
Thus, $\rho$ can be selected to limit conservativeness based on $N$ and the order of magnitude of the arguments of the soft minimum. 
If $\rho$ is small, then the soft minimum is a conservative approximation of the minimum. 
However, if $\rho$ is large, then $\|h'(x,\zeta)\|$ is large at points where the minimum is not differentiable. 
Thus, selecting $\rho$ is a trade-off between the conservativeness of $h$ and the size of $\|h'(x, \zeta)\|$. 
The examples in \Cref{sec:Formation Flying Simulation Results} use $\rho=20$, which results in negligible conservativeness while maintaining sufficient smoothness.

The zero-superlevel set of $h$ is 
\begin{equation}
\mathcal{H} \triangleq \{ (x,\zeta) \in \mathrm{R}^{ 6n } \times \BBR^{3 \ell} \colon h(x,\zeta) \geq 0 \},
\end{equation}
which is a subset of $\bar \SH$.
In fact, \cite[Prop. 2]{safari2024b} shows that as $\rho \to \infty$, $\SH \to \bar \SH$.
Thus, for sufficiently large $\rho>0$, $\SH$ approximates $\bar \SH$.

Define
\begin{equation*}
    \mathcal{B} \triangleq \{ (x,\zeta) \in \mathrm{bd} \ \mathcal{H} :  h^{\prime}(x,\zeta) \tilde A \tilde x \leq 0 \},
\end{equation*}
which is the subset of $\mathrm{bd} \, \SH$, where the uncontrolled (i.e., $\mu=0$) flow of \Cref{eqn:State_space_n_satellites,eq:A,eq:B,eqn:controller_state_space} leaves $\SH$.
The next result is from \cite[Prop. 3]{rabiee2024b} and provides conditions such that $h$ is a relaxed CBF, that is, $h$ satisfies the CBF condition on~$\SB$.

\begin{proposition}\label{prop:Relexed_CBF}
Assume that for all $(x,\zeta) \in \SB$, $\frac{\partial h(x,\zeta)}{\partial \zeta} \ne 0$. 
Then, for all $(x,\zeta) \in \SB$,
\begin{equation*}
    \sup_{\hat \mu\in \BBR^{3 \ell}} h^{\prime} (x,\zeta) \left [ \tilde A \tilde x + \tilde B \hat \mu \right ] \ge 0.
\end{equation*}
\end{proposition}

\Cref{prop:Relexed_CBF} and Nagumo's theorem imply that $\SH$ is control forward invariant if for all $(x,\zeta) \in \SB$, $\frac{\partial h(x,\zeta)}{\partial \zeta} \ne 0$.
The next result follows from \cite[Prop. 3]{rabiee2024b} and shows that forward invariance of $\SH$ implies forward invariance of 
\begin{equation*}
        \SSS \triangleq \SH \cap \bar \SSS,
\end{equation*}
which is a subset of $\SSS_\rms$. 

\begin{proposition}
Consider \Cref{eqn:State_space_n_satellites,eq:A,eq:B,eqn:controller_state_space}, where $(x_0,\zeta_0) \in \SSS$. 
Assume there exists $\bar t \in (0,\infty]$ such that for all $t \in [0,\bar t)$, $(x(t),\zeta(t)) \in \SH$.
Then, for all $t \in [0,\bar t)$, $(x(t),\zeta(t)) \in \SSS$.
\end{proposition}

\subsection{Optimal Control Subject to State and Input Constraints}

We use the relaxed CBF $h$ to construct a constraint that guarantees \ref{obj3}--\ref{obj5}.
Then, we present a surrogate control that is as close as possible to the desired surrogate control $\mu_\rmd$ while satisfying the constraint that guarantees that \ref{obj3}--\ref{obj5}.

Consider the constraint function $b \colon \BBR^{6n} \times \BBR^{3 \ell} \times \BBR^{3 \ell} \times \BBR \to \BBR$ given by 
\begin{align}
    b(x,\zeta,\hat \mu, \hat \eta) &\triangleq h^{\prime} (x,\zeta) \left [ \tilde A \tilde x + \tilde B \hat \mu \right ] + \alpha ( h( x,\zeta )) + \hat{\eta} h(x,\zeta) \nn\\
    &= \frac{\partial h(x,\zeta)}{\partial x} \left [ A x + B \zeta \right ] + \frac{\partial h(x,\zeta)}{\partial \zeta} \left [ A_\rmc \zeta + B_\rmc \hat \mu \right ] \nn\\
    &\qquad + \alpha ( h( x,\zeta )) + \hat{\eta} h(x,\zeta), \label{eq:constraint_b}
\end{align}
where $\hat \mu$ is the control variable, $\hat \eta$ is a slack variable, and $\alpha \colon \BBR \to \BBR$ is locally Lipschitz and nondecreasing such that $\alpha(0)=0$. 
Next, let $\gamma >0$ and consider the cost function 
\begin{equation*}
    \SJ(x,\zeta,\hat \mu,\hat \eta) = \frac{1}{2} \| \hat \mu - \mu_\rmd(x,\zeta) \|^2 + \frac{1}{2} \gamma \hat{\eta}^2.
\end{equation*}
The objective is to find $(\hat \mu,\hat \eta)$ that minimizes the cost $\SJ$ subject to the relaxed CBF constraint $b(x,\zeta,\hat \mu,\hat \eta) \ge 0$.

For each $(x,\zeta) \in \BBR^{6n} \times \BBR^{3 \ell}$, the minimizer of $\SJ$ subject to  $b(x,\zeta,\hat \mu,\hat \eta) \ge 0$ can be derived from the first-order necessary conditions for optimality (e.g., see \cite{rabiee2024b,safari2024b,ames2016,wieland2007,cortez2022}).
This yields the control $\mu_* \colon \BBR^{6n} \times \BBR^{3 \ell} \to \BBR^{3 \ell}$ defined by
\begin{equation}    
\mu_*(x,\zeta) = \mu_{\mathrm{d}} (x,\zeta) + \lambda(x,\zeta) \left [ \frac{\partial h(x,\zeta)}{\partial \zeta} B_\rmc \right ]^{\mathrm{T}}, \label{eq:mu*}
\end{equation}
where 
\begin{equation}
\lambda(x,\zeta) \triangleq \max \left \{ 0,\dfrac{- b(x,\zeta,\mu_\rmd(x,\zeta),0)}{ \left  \| \left [ \frac{\partial h(x,\zeta)}{\partial \zeta} B_\rmc \right ]^{\mathrm{T}} \right \|^2 + \frac{h(x,\zeta)^2}{\gamma} }\right \}, \label{eq:lambda}
\end{equation}
and the slack variable $\eta_* \colon \BBR^{6n} \times \BBR^{3 \ell} \to \BBR$ given by
\begin{equation*}
    \eta_*(x,\zeta) \triangleq \gamma^{-1} h(x,\zeta) \lambda(x,\zeta).
\end{equation*}

The next result shows that $(\mu_*(x,\zeta),\eta_*(x,\zeta))$ is the unique global minimizer of $\SJ(x,\zeta,\hat \mu,\hat \eta)$ subject to $b(x,\zeta,\hat \mu,\hat \eta) \ge 0$.
The proof is similar to that of \cite[Theorem~1]{safari2024b}.

\begin{theorem} 
Assume that for all $(x,\zeta) \in \SB$, $\frac{\partial h(x,\zeta)}{\partial \zeta} \ne 0$. 
Let $(x,\zeta) \in \BBR^{6n} \times \BBR^{3 \ell}$, and let $(\hat \mu,\hat \eta) \in \BBR^{3 \ell} \times \BBR$ be such that $b(x,\zeta,\hat \mu, \hat \eta) \ge 0$ and $( \hat \mu, \hat \eta) \ne (\mu_*(x,\zeta), \eta_*(x,\zeta))$. 
Then,
\begin{equation*}
    \SJ(x,\zeta,\hat \mu,\hat \eta) >      \SJ(x,\zeta,\mu_*(x,\zeta),\eta_*(x,\zeta)).
\end{equation*} \label{Theorem_1}
\end{theorem}

The following theorem is the main result on constraint satisfaction. 
It demonstrates that the control makes $\SSS$ forward invariant. 
This result follows from \cite[Corollary 3]{rabiee2024b}

\begin{theorem} \label{thm:main_result_fwd_invariance}
Consider \Cref{eqn:State_space_n_satellites,eq:A,eq:B}, where the control $\zeta$ is given by \Cref{eqn:controller_state_space} with $\mu = \mu_*$, which is given by \cref{eq:mu*,eq:lambda}. 
Assume that for all $(x,\zeta) \in \SB$, $\frac{\partial h(x,\zeta)}{\partial \zeta} \ne 0$, and assume that $h^\prime$ is locally Lipschitz. 
Then, for all $(x_0,\zeta_0) \in \SSS$, the following statements hold: 
\begin{enumerate}
[leftmargin=0.8cm]
	\renewcommand{\labelenumi}{\roman{enumi})}
	\renewcommand{\theenumi}{\roman{enumi})}
    
    \item There exists a maximum value $t_{\rm m} \in (0 ,\infty]$ such that \Cref{eqn:State_space_n_satellites,eq:A,eq:B,eqn:controller_state_space} with $\mu = \mu_*$ has a unique solution on $[0, t_{\rm m} )$.
    
    \item For all $t \in [0, t_{\rm m} )$, $(x(t),\zeta(t)) \in \SSS \subset \SSS_\rms$.

\end{enumerate}
\end{theorem}

Theorems 1 and 2 demonstrate that the control \Cref{eqn:controller_state_space,eq:mu*,eq:lambda} with $\mu = \mu_*$ satisfies \ref{obj3}--\ref{obj5}, and yields $\mu$ that is as close as possible to $\mu_\rmd$, which is designed to satisfy~\ref{obj1}.
\Cref{fig:block_diagram_cbf_emff} illustrates the implementation of the control \Cref{eqn:controller_state_space,eq:mu_d,eq:h,eq:mu*,eq:lambda} with $\mu = \mu_*$.

Theorem~\ref{thm:main_result_fwd_invariance} demonstrates that the control Theorem~\ref{eq:mu*} makes $\SSS$ forward invariant. 
Although  Theorem~\ref{thm:main_result_fwd_invariance} does not provide formal guarantees for the case where the initial condition is outside of this forward invariant set, we note that the control does attempt to force the state $(x,\zeta)$ back to $\SSS$ in this case. 
To illustrate this property, assume that at time $t$, the state $(x,\zeta)$ is such that $h(x,\zeta)<0$, which implies that that state is outside $\SSS$. 
At time $t$, the constraint function is
\begin{equation*}
    b(x,\zeta,\mu_*, \eta_*) = \dot{h}(x, \zeta ) + \alpha ( h(x, \zeta) ) + \eta_* h(x, \zeta),
\end{equation*}
where $\dot{h}$ is a derivative of $h$ along the closed-loop trajectories. 
If the weight on slack is large (i.e., $\gamma >> 1$), then $\eta_* \approx 0$ (assuming slack is needed to ensure feasibility). 
In this case, 
\begin{equation*}
    b(x,\zeta,\mu_*, \eta_*) \approx \dot{h}(x, \zeta ) + \alpha (h(x, \zeta ) ),
\end{equation*}
and the constraint $b(x,\zeta,\mu_*, \eta_*) \ge 0$ tends to cause $h$ to increase, specifically, $\dot{h}(x, \zeta) \geq \alpha ( h(x, \zeta ) ) > 0$, which drives the state to $\SH$.

The control \Cref{eq:mu*} is developed under the assumption that $m$ and $\mu_0$ are known; however, the control can be modified to address uncertainty in $m$ and $\mu_0$. 
Specifically, it suffices to have knowledge of upper bounds $\bar{m} \geq m$ and $\bar{\mu}_0 \geq \mu_0$ and lower bounds $\ubar{m} \leq m$ and $\ubar{\mu}_0 \leq \mu_0$. 
In this case, note that 
\begin{equation*}
    \frac{\partial h(x,\zeta)}{\partial x} B \zeta \geq \Gamma(x, \zeta), 
\end{equation*}
where
\begin{gather*}
    \Gamma(x, \zeta) \triangleq \begin{dcases}
        \dfrac{\ubar{\mu}_0}{\bar{m}} g(x,\zeta), & g(x,\zeta) \geq 0,
    \\
    \dfrac{\bar{\mu}_0}{\ubar{m}} g(x,\zeta) , & g(x,\zeta) < 0.\\
    \end{dcases}\\
        g(x, \zeta)  \triangleq \frac{3}{8 \pi} \frac{\partial h(x,\zeta)}{\partial x} \begin{bmatrix}
        0_{3n \times 3\ell}\\
        B_0 \otimes I_3
    \end{bmatrix} \zeta. 
\end{gather*}
Hence, $\frac{\partial h(x,\zeta)}{\partial x} B \zeta$ in $\Cref{eq:constraint_b}$ can be replaced by the bound $\Gamma(x, \zeta)$, which is sufficient to ensure constraint satisfaction with the true values $m$ and $\mu_0$. 
Specifically, Theorem~\ref{thm:main_result_fwd_invariance} holds if $\frac{\partial h(x,\zeta)}{\partial x} B \zeta$ in $\Cref{eq:constraint_b}$ is replaced by the bound $\Gamma(x, \zeta)$. 

This article used an infinite-horizon LQR solution to design the desired surrogate control $\mu_\rmd$ given by \eqref{eq:mu_d}.
This approach is well suited to the time-averaged model \eqref{eq:avg_model} developed in this article. 
However, we note that the optimal control \Cref{eq:mu*} guarantees state and input constraint satisfaction \ref{obj3}--\ref{obj5} independent of the design of desired surrogate control $\mu_\rmd$.
Thus, alternative designs that achieve the formation objective \ref{obj1} can be considered for the desired surrogate control.
For example, $\mu_\rmd$ could be designed using finite-horizon model-predictive control (MPC) with slack constraints to encourage constraint satisfaction.
In this case, the CBF-based optimal control \Cref{eq:mu*} ensures constraint satisfaction even if the desired surrogate control calculated from MPC does not.



\section{EMFF Simulations}
\label{sec:Formation Flying Simulation Results}

We consider the satellite system \Cref{F_ij,eq:f,accel_i}, where the mass is $m=15$~kg and the vacuum permeability is $\mu_0 = 4 \pi \times 10^{-7}$ H/m.
Each electromagnetic coil has $N=400$ turns of American Wire Gauge 12 wire, and the cross-sectional area is $\sigma =0.25^2 \pi$~m$^2$. These parameters are based on experimental testbeds at the University of Kentucky. Based on these parameters, the computed resistance of each coil is $3.2735$~Ohms and the inductance is $0.2$~Henries.
The maximum apparent power capacity of each satellite is $\bar{q}=10^4$~W.

For $(i,j)\in \SP$ such that $j>i$, we use the interaction frequencies $\omega_{ij} = 20\pi \nu_{ij}$~rad/s, where 
$\nu_{ij} \triangleq ((i-1)(2n-i)/2) + j-i$.
We let $T=0.1$~s, which is the least common period of the interaction frequencies.
The control dynamics \cref{eqn:controller_state_space} are implemented with $A_\rmc = -0.1 I_{3\ell}$, $B_\rmc =  I_{3\ell}$, and $\zeta_0 = 0$.
For all examples, we use $\alpha_0(h)=0.25h$, $\alpha_1(h)=\alpha_v(h)=h$, and $\alpha(h)=0.03h$.
For $(i,j)\in \SP$ such that $j>i$, we define
\begin{equation*}
    \mu_{ij}(x,\zeta) \triangleq \begin{bmatrix} 0_{3 \times 3(\nu_{ij}-1)} & I_3 & 0_{3 \times 3(\ell-\nu_{ij})} \end{bmatrix} \mu(x,\zeta),
\end{equation*}
which is the surrogate control used to generate $\zeta_{ij}$ from the control dynamics \cref{eqn:controller_state_space}. 
We implement the control \Cref{eq:h,eq:mu*,eq:lambda} with $\rho = 20$ and $\mu = \mu_*$.

\subsection{EMFF in Deep Space}
\label{deep_space_examples}

First, we consider satellites in deep space, where gravity is negligible (i.e., $\mathbf{G}=0$).
The collision radius is $\ubar{r}=2$~m, and the upper bound on intersatellite speed is $\bar{s}=0.025$~m/s. We implement the control \Cref{eq:mu*,eq:lambda} with $\gamma=10^{40}$, which effectively disables the slack $\hat \eta$.

\begin{figure}[ht!]
        \centering        \includegraphics[width=0.50\textwidth,clip=true,trim= 0.2in 0in 0.5in 0.2in]{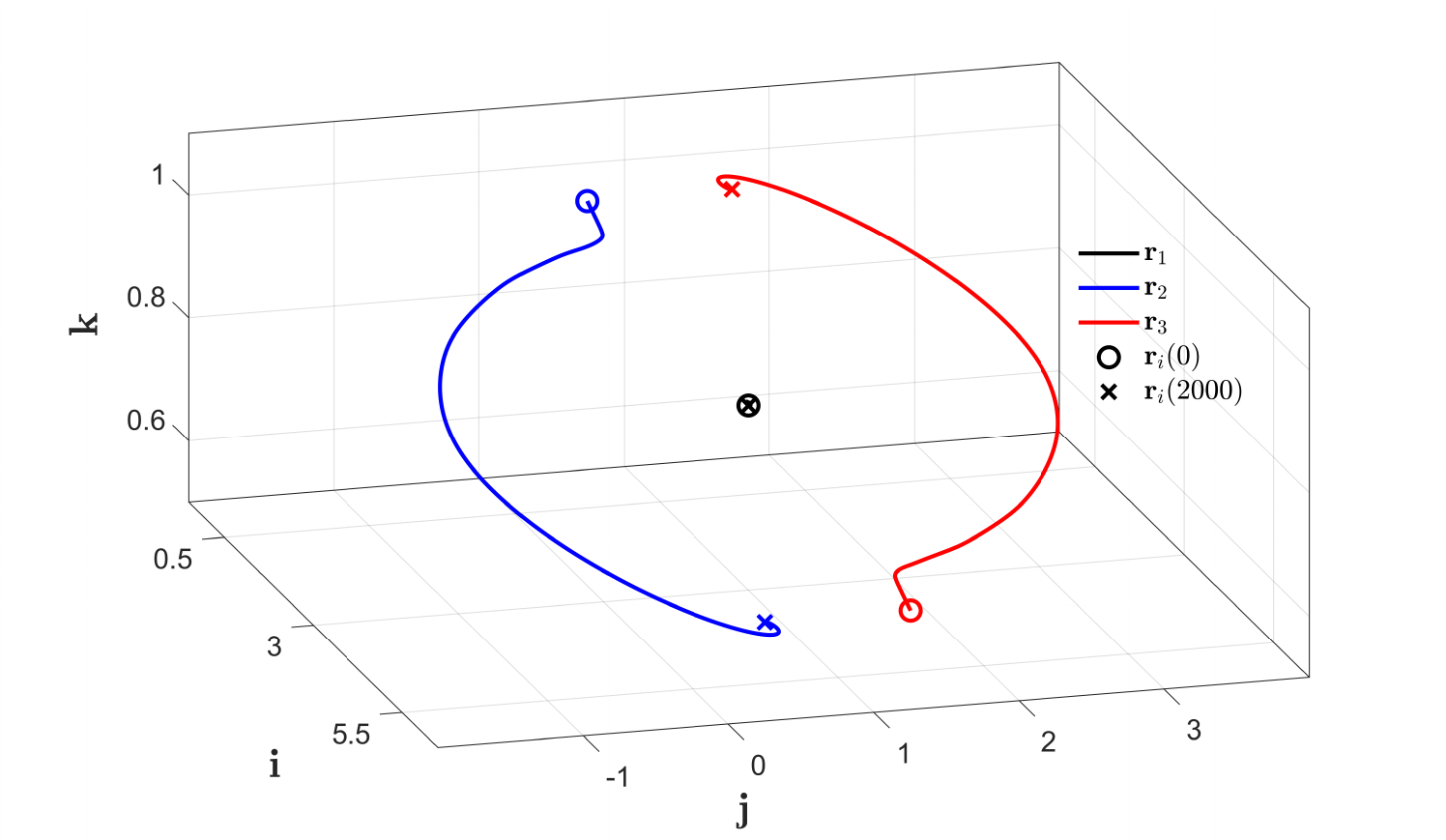}
        \centering
        \caption{Satellite trajectories $\mathbf{r}_i$ achieve the desired formation and avoid collisions.}
        \label{fig:3d_collision_avoidance}
    \end{figure}

\begin{figure}[ht!]
        \centering        
    \includegraphics[width=0.50\textwidth,clip=true,trim= 0.2in 0.03in 0.5in 0.2in]{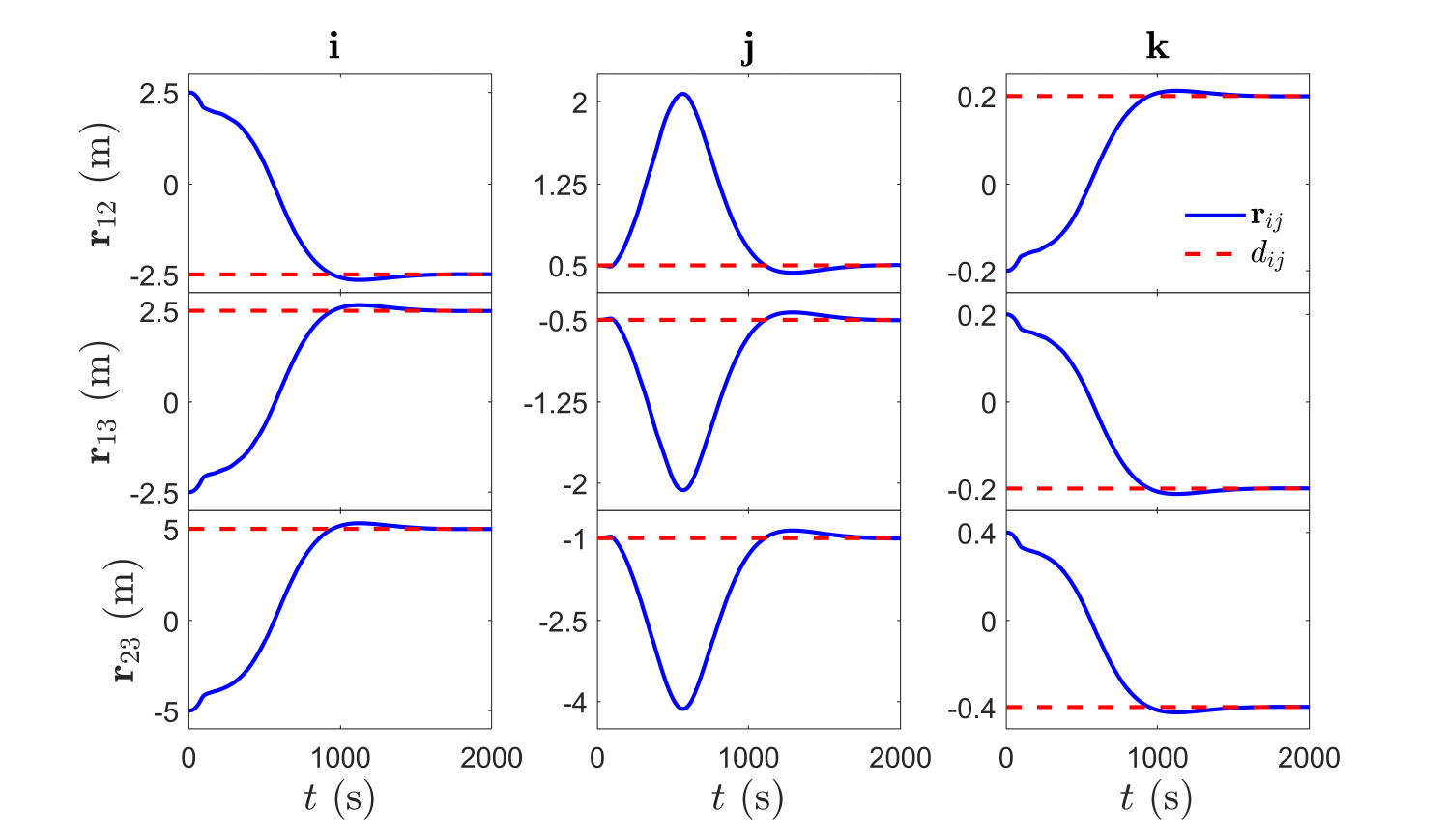}
        \caption{ Relative positions $\mathbf{r}_{ij}$ converge to desired formation $d_{ij}$.}
        \label{fig:position_power_plus_collision_avoidance}
    \end{figure}

\begin{figure}[ht!]
        \centering        \includegraphics[width=0.50\textwidth,clip=true,trim= 0.2in 0.03in 0.5in 0.2in]{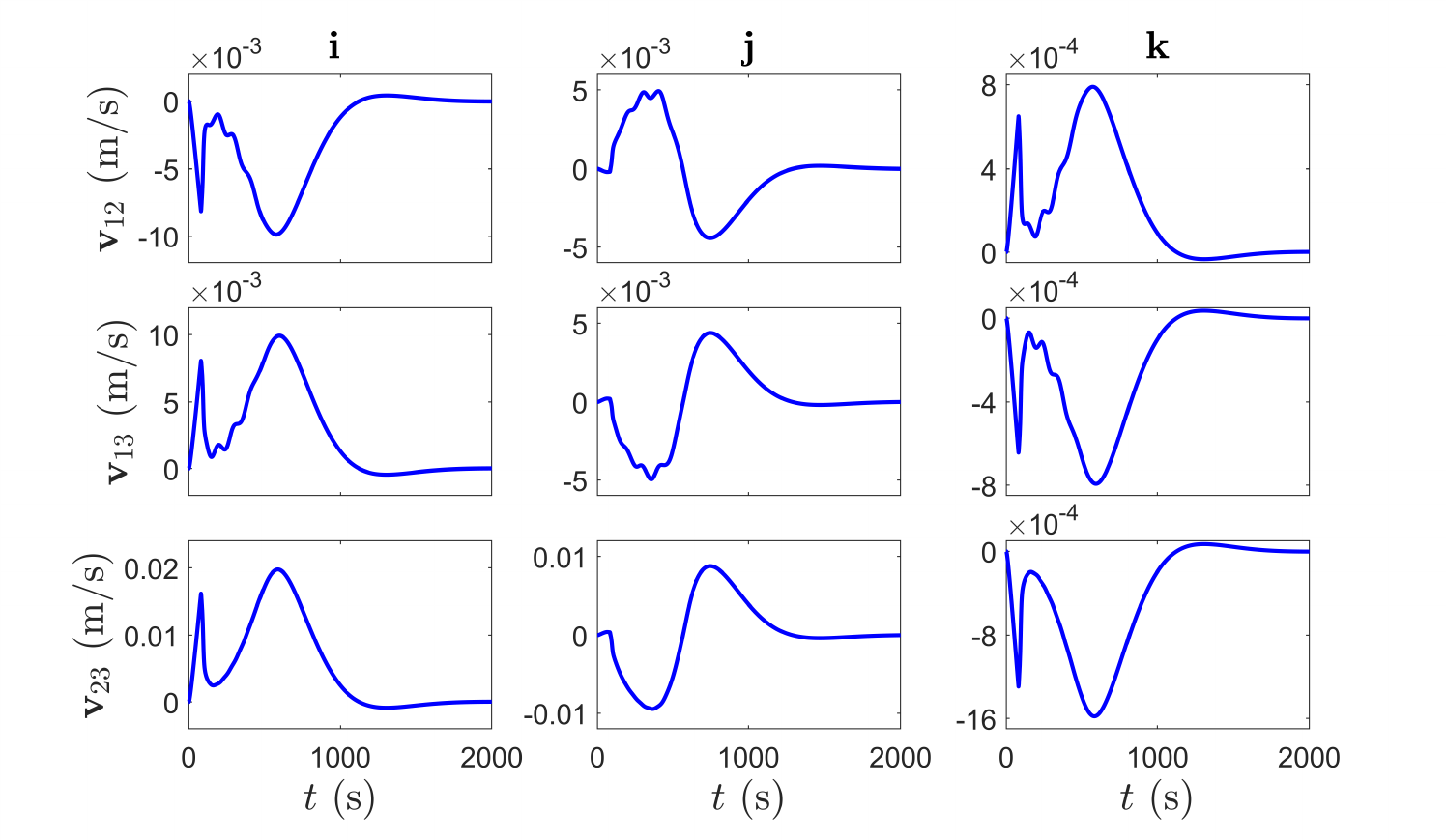}
        \caption{Relative velocities $\mathbf{v}_{ij}$ satisfy the intersatellite speed limit throughout the maneuver.}
       \label{fig:velocity_power_plus_collision_avoidance}
    \end{figure}

\begin{figure}[ht]
        \centering      \includegraphics[width=0.50\textwidth,clip=true,trim= 0.2in 0.2in 0.5in 0.6in]{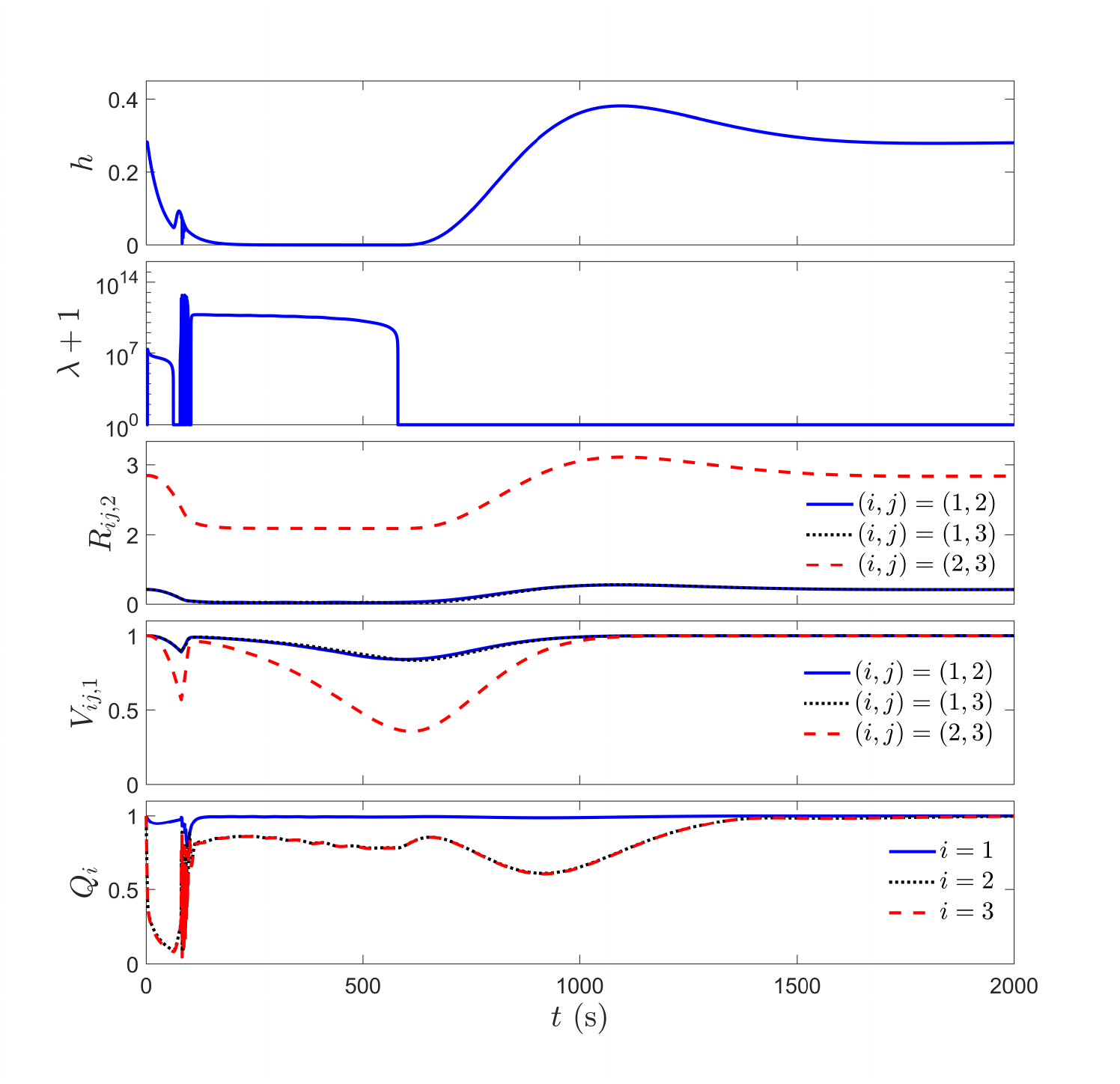}
        \centering
        \caption{Soft minimum $h$ is positive for all time, which implies that its arguments $R_{ij,2}$, $V_{ij,1}$, and $Q_i$ are positive for all time (as shown). 
        Note that $\lambda$ is positive at time instants when the control $\mu_*$ is modified from $\mu_\rmd$ in order to satisfy power limits and collision constraints.}
        \label{fig:barrier_functions_power_plus_collision_avoidance}
    \end{figure}

\begin{figure}[hbt!]
        \centering      \includegraphics[width=0.50\textwidth,clip=true,trim= 0.2in 0.1in 0.5in 0.2in]{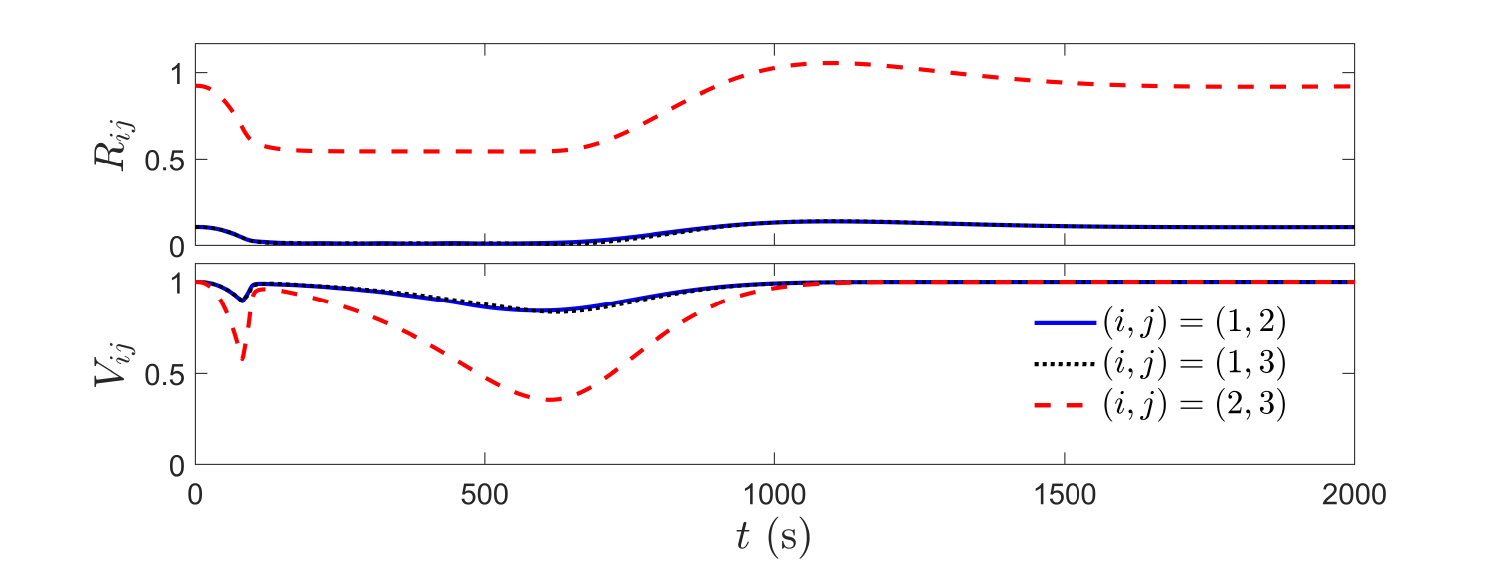}
        \centering
        \caption{CBFs $R_{ij}$ and $V_{ij}$ are positive for all time, which demonstrates that \labelcref{obj3,obj4} are satisfied.}
        \label{fig:low_barrier_functions_power_plus_collision_avoidance}
    \end{figure}

\begin{figure}[hbt!]
        \centering       \includegraphics[width=0.50\textwidth,clip=true,trim= 0.2in 0.03in 0.5in 0in]{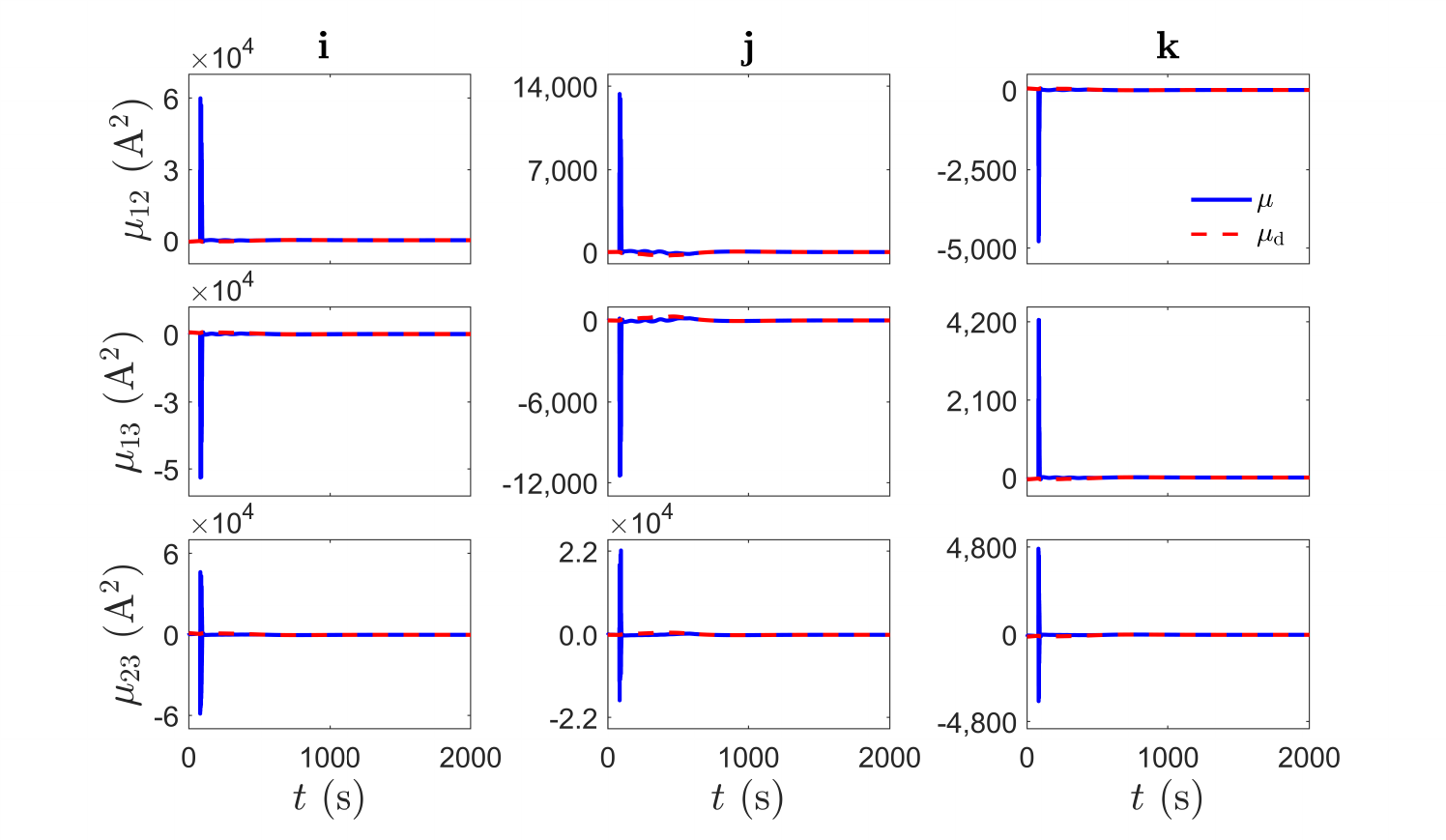}
        \caption{Surrogate control $\mu$ deviates from $\mu_{\rmd}$ when $\lambda$ is positive to satisfy power and collision constraints.}
          \label{fig:mu_power_plus_collision_avoidance}
    \end{figure}

\begin{figure}[hbt!]
        \centering       \includegraphics[width=0.50\textwidth,clip=true,trim= 0.2in 0.03in 0.5in 0in]{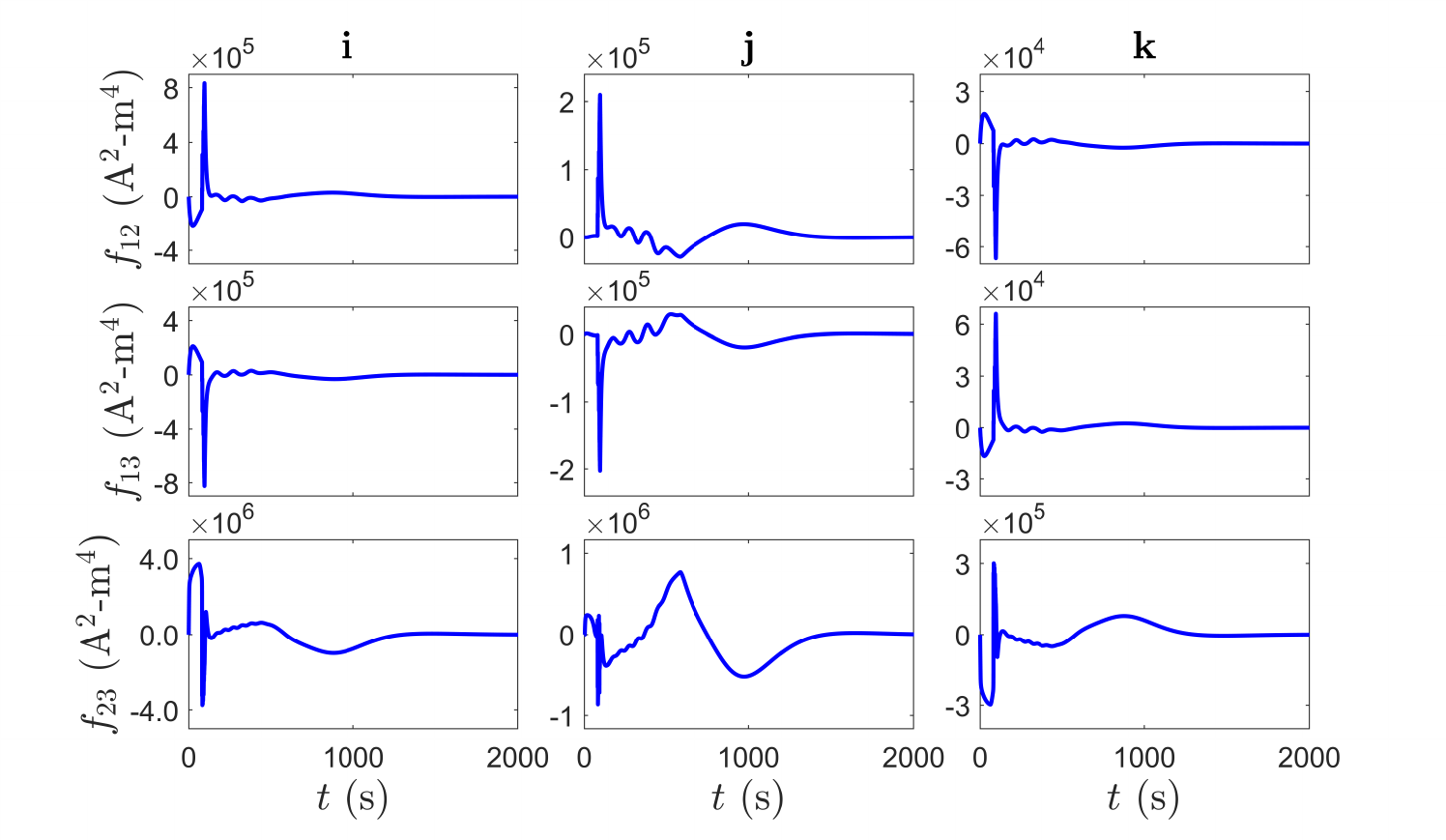}
        \caption{ Prescribed intersatellite force $f_{ij}$ computed using $\norm{r_{ij}}$ and $\zeta_{ij}$.}
          \label{fig:force_power_plus_collision_avoidance}
    \end{figure}

\begin{figure}[hbt!]
        \centering     \includegraphics[width=0.50\textwidth,clip=true,trim= 0.2in 0.03in 0.5in 0in]{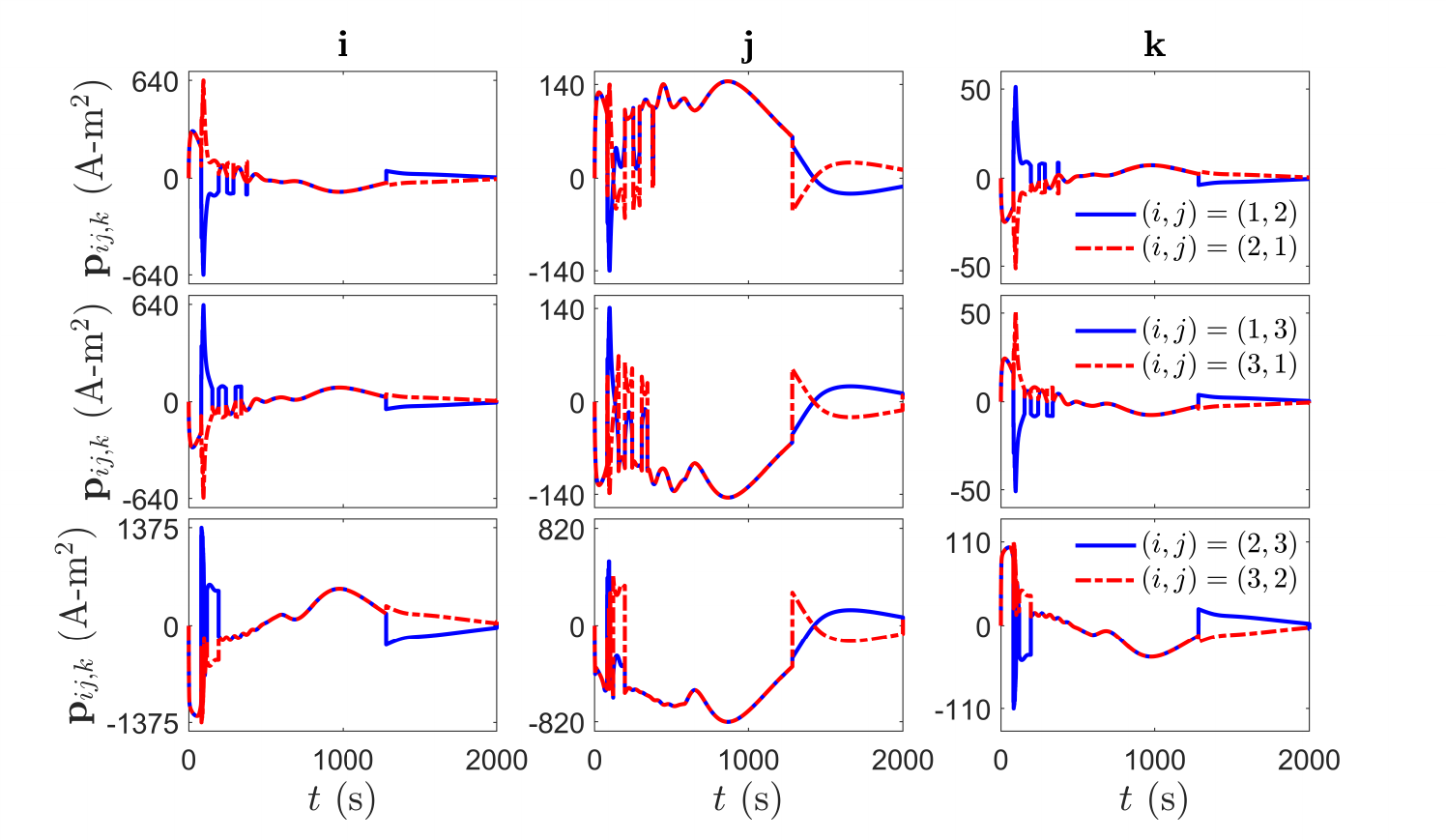}
        \caption{Amplitude $\textbf{p}_{ij,k}$ of magnetic moment control.}
        \label{fig:amplitudes_power_plus_collision_avoidance}
    \end{figure}

\begin{example}\label{3_sat_example}

This example illustrates algorithm effectiveness for power limitation and collision avoidance. 
We select initial conditions and a desired formation such that $\mu_\rmd$ would exceed the power limitation and result in collisions without the modification \Cref{eq:mu*,eq:lambda} to address these constraints. 
Specifically, let $n=3$,  
$r_{1}(0) = [ 3 \quad 1 \quad 0.8  ]^{\rm{T}}$~m, 
$r_{2}(0) = [0.5  \quad 0.5 \quad 1  ]^{\rm{T}}$~m, 
$r_3(0) = [5.5 \quad 1.5 \quad 0.6]^{\rm{T}}$~m, 
$v_1(0) = v_2(0) =v_3(0) = 0$~m/s,
$d_{12} = [-2.5 \quad 0.5 \quad 0.2 ]^{\rm{T}}$~m, 
$d_{13} = [2.5 \quad -0.5 \quad -0.2 ]^{\rm{T}}$~m, and 
$d_{23} = [5 \quad -1 \quad 0.4]^{\rm{T}}$~m.
The weights for the desired surrogate control $\mu_\rmd$ are \eqref{LQR_weights}, where $w_r = 10^6$, $w_v=1$, $w_{\zeta}=0.01$, and $w_{\mu} =20$.

\Cref{fig:3d_collision_avoidance,fig:position_power_plus_collision_avoidance,fig:velocity_power_plus_collision_avoidance} show the satellites avoid collision and achieve the desired formation.
\Cref{fig:barrier_functions_power_plus_collision_avoidance} shows that $h$ and its arguments (i.e., $R_{ij,2}$, $V_{ij,1}$, $Q_i$) are positive for all time. 
Hence, $R_{ij}$, $V_{ij}$ (\Cref{fig:low_barrier_functions_power_plus_collision_avoidance}), and $Q_i$ are positive for all time, and \ref{obj3}--\ref{obj5} are satisfied. 
\Cref{fig:barrier_functions_power_plus_collision_avoidance} also shows that $\lambda$ is positive for $t\in[3,62.6] \cup [78.3,97]\cup [103.2,580.5]$, which indicates that $\mu$ deviates from $\mu_\rmd$ during these time intervals as shown in \Cref{fig:mu_power_plus_collision_avoidance}. \Cref{fig:barrier_functions_power_plus_collision_avoidance} also shows that for $t\in[3,6.7]$, the soft minimum $h$ is dominated by $R_{12,2}$ and $R_{13,2}$, indicating that $\mu$ deviates from $\mu_\rmd$ to satisfy collision constraints. For $t\in(6.7,62.6]$, the soft minimum $h$ is dominated by $Q_2$ and $Q_3$, indicating that $\mu$ deviates from $\mu_\rmd$ to satisfy power constraints. For $t\in[78.3,97]$, the soft minimum $h$ is dominated by $R_{12,2}$, $R_{13,2}$, $Q_2$, and $Q_3$, indicating that $\mu$ deviates from $\mu_\rmd$ to satisfy both power and collision constraints. 
Finally, $h$ is dominated by $R_{12,2}$ and $R_{13,2}$ for $t\in[103.2,580.5]$, indicating that $\mu$ deviates from $\mu_\rmd$ to satisfy the collision constraints.
In fact, the position trajectories under $\mu_\rmd$ are straight lines from the initial to final positions, and such trajectories would cause collisions. \Cref{fig:force_power_plus_collision_avoidance} shows the prescribed intersatellite forces $f_{ij}$, and \Cref{fig:amplitudes_power_plus_collision_avoidance} shows the amplitude controls $\mathbf{p}_{ij,k}$ computed from $f_{ij}$ using
\Cref{eq:p_ij_k,eqn:p_ij_allocation,eq:fij}. 
\exampletriangle
\end{example}

\begin{figure}[ht!]
        \centering        \includegraphics[width=0.50\textwidth,clip=true,trim= 0.2in 0in 0.5in 0.2in]{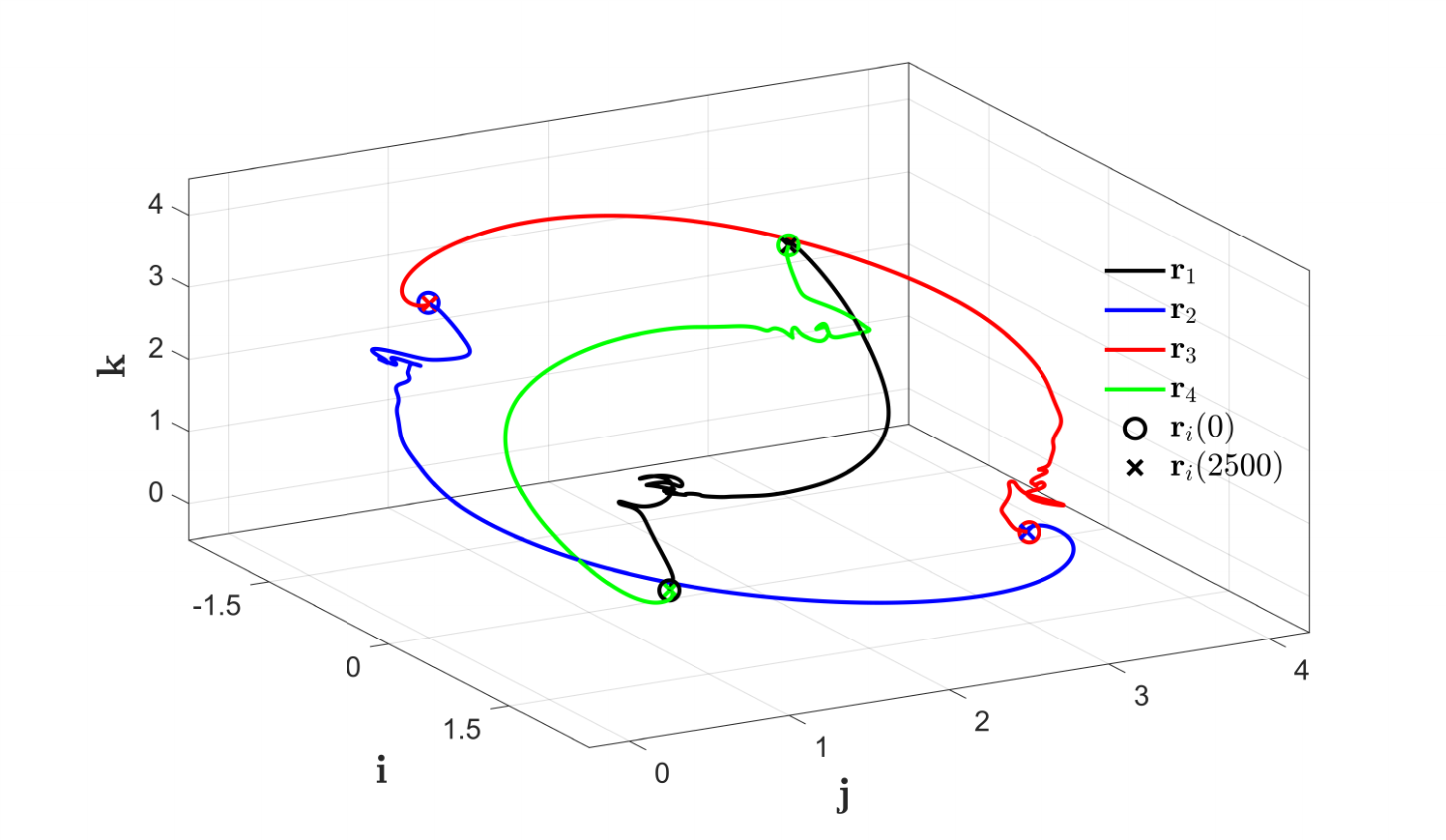}
        \centering
        \caption{Satellite trajectories $\mathbf{r}_i$ show that diagonally opposite satellites swap positions.}
        \label{fig:3d_4_sat}
    \end{figure}

\begin{figure}[ht]
        \centering        
        \includegraphics[width=0.50\textwidth,clip=true,trim= 0.2in 0.5in 0.5in 0.5in]{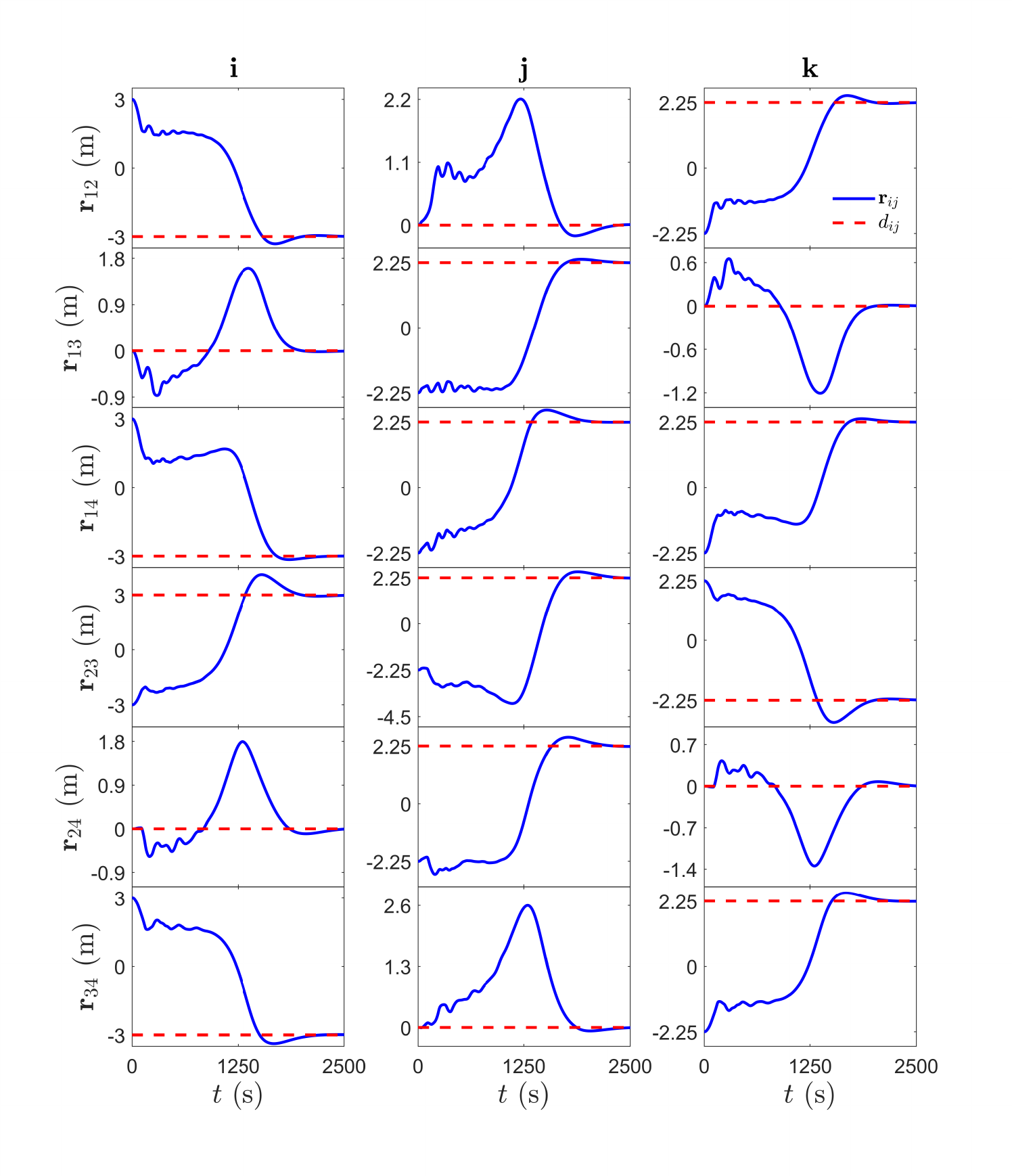}
        \caption{ Relative positions $\mathbf{r}_{ij}$ converge to desired formation $d_{ij}$.}
        \label{fig:position_4_sat}
    \end{figure}

\begin{figure}[ht]
        \centering        \includegraphics[width=0.50\textwidth,clip=true,trim= 0.2in 0.5in 0.5in 0.5in]{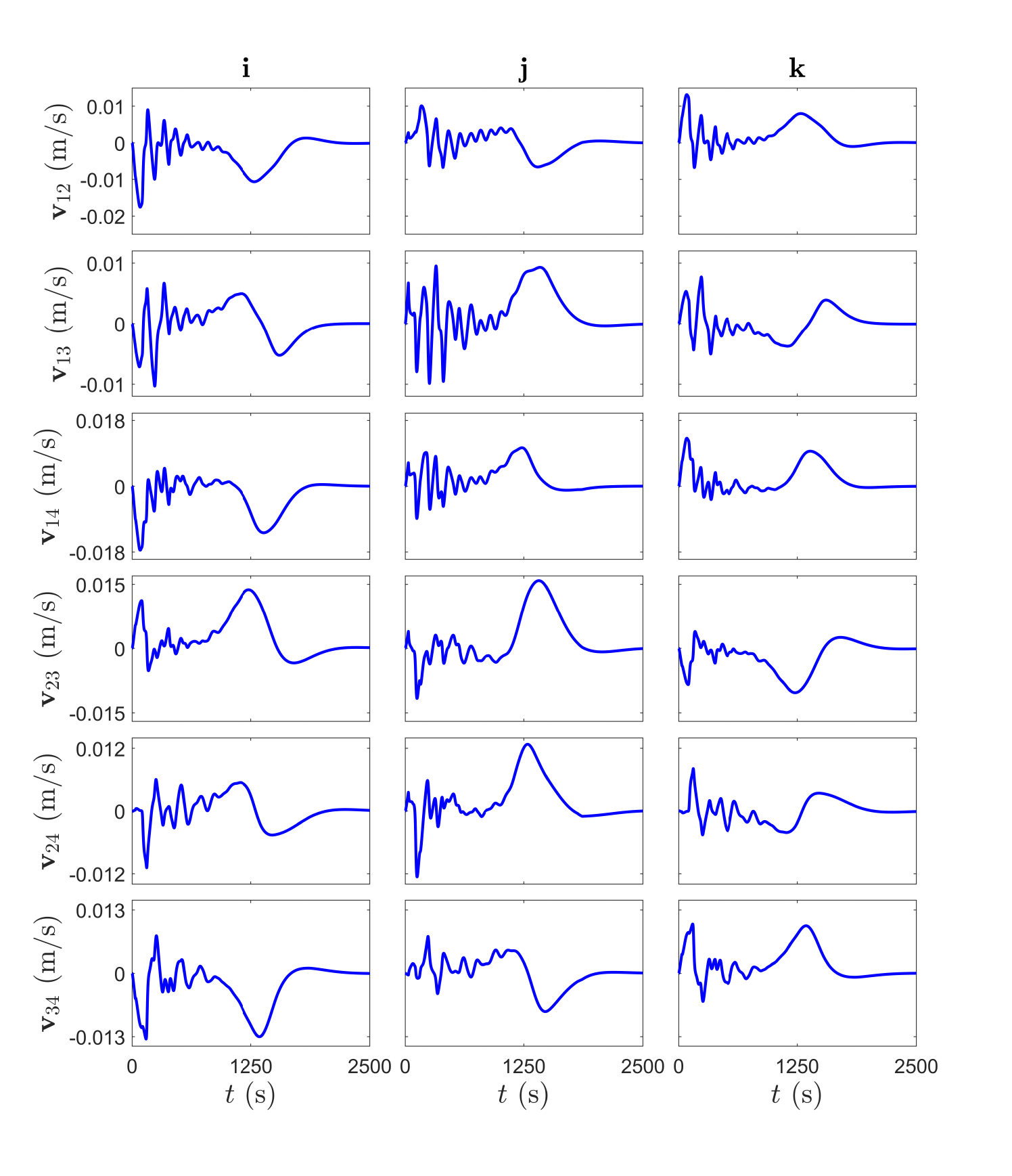}
        \caption{Relative velocities $\mathbf{v}_{ij}$ satisfy the intersatellite speed limit throughout the maneuver.}
       \label{fig:velocity_4_sat}
    \end{figure}

\begin{figure}[hbt!]
        \centering      \includegraphics[width=0.50\textwidth,clip=true,trim= 0.2in 0.3in 0.5in 0.5in]{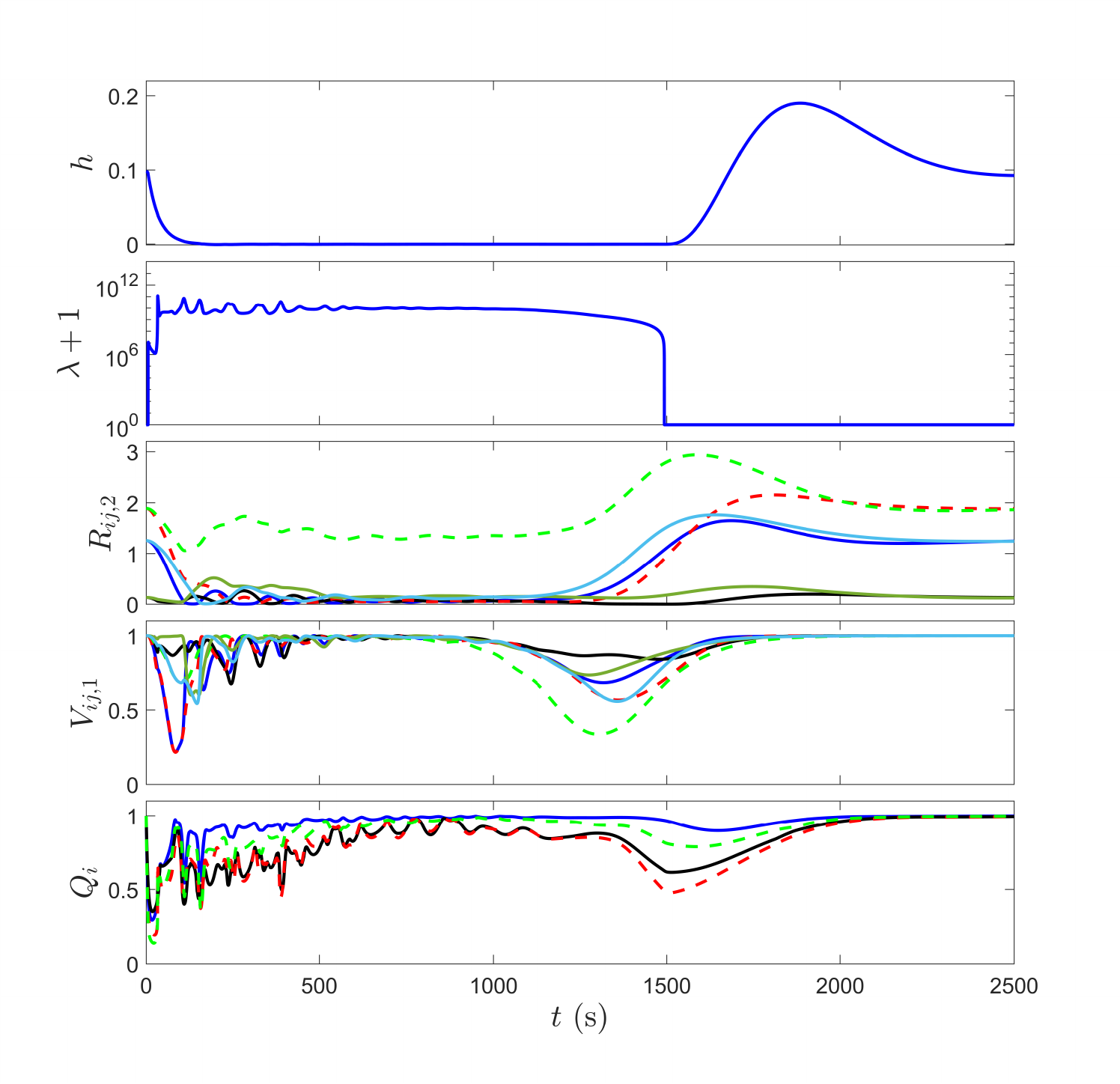}
        \centering
        \caption{Soft minimum $h$ is positive for all time, which implies that its arguments $R_{ij,2}$, $V_{ij,1}$, and $Q_i$ are positive for all time (as shown). 
        Note that $\lambda$ is positive at time instants when the control $\mu_*$ is modified from $\mu_\rmd$ in order to avoid collision.} 
        \label{fig:barrier_functions_4_sat}
    \end{figure}

\begin{figure}[hbt!]
        \centering      \includegraphics[width=0.50\textwidth,clip=true,trim= 0.2in 0.5in 0.5in 0.2in]{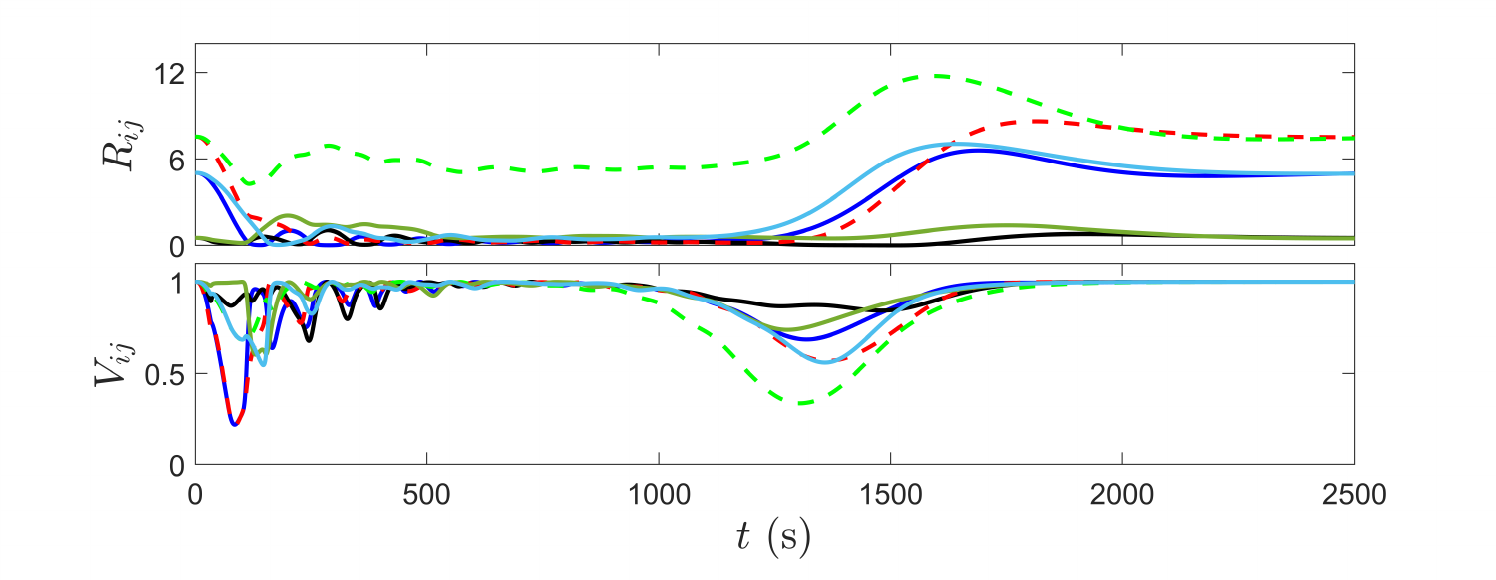}
        \centering
        \caption{CBFs $R_{ij}$ and $V_{ij}$ are positive for all time, which demonstrates that \labelcref{obj3,obj4} are satisfied.}
        \label{fig:constraints_rij_vij_4_sat}
    \end{figure}

\begin{figure}[hbt!]
        \centering       \includegraphics[width=0.50\textwidth,clip=true,trim= 0.2in 0.5in 0.5in 0.5in]{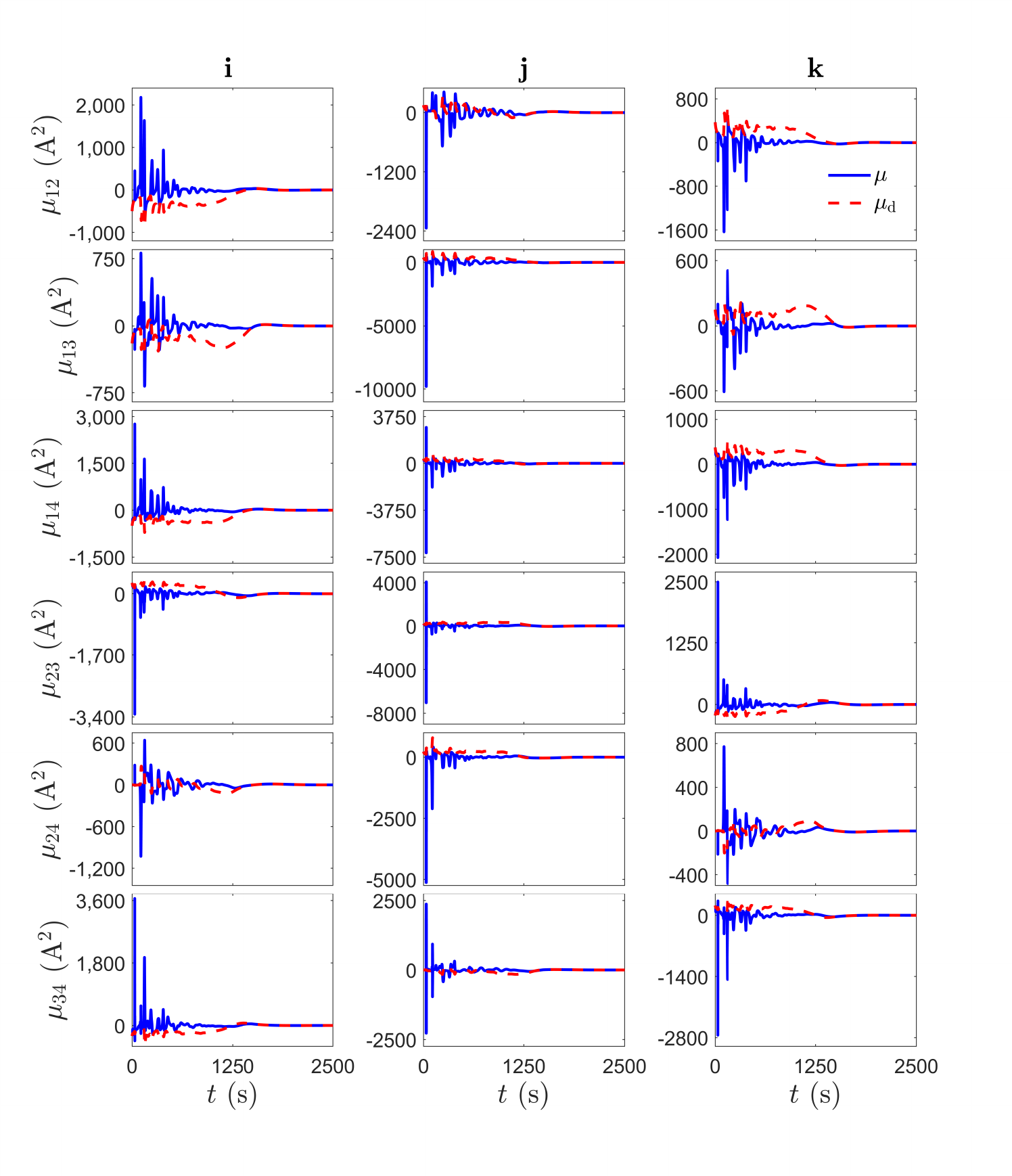}
        \caption{Surrogate control $\mu$ deviates from $\mu_{\rmd}$ when $\lambda$ is positive to avoid collision.}
          \label{fig:mu_4_sat}
    \end{figure}

\begin{figure}[hbt!]
        \centering       \includegraphics[width=0.50\textwidth,clip=true,trim= 0.2in 0.5in 0.5in 0.3in]{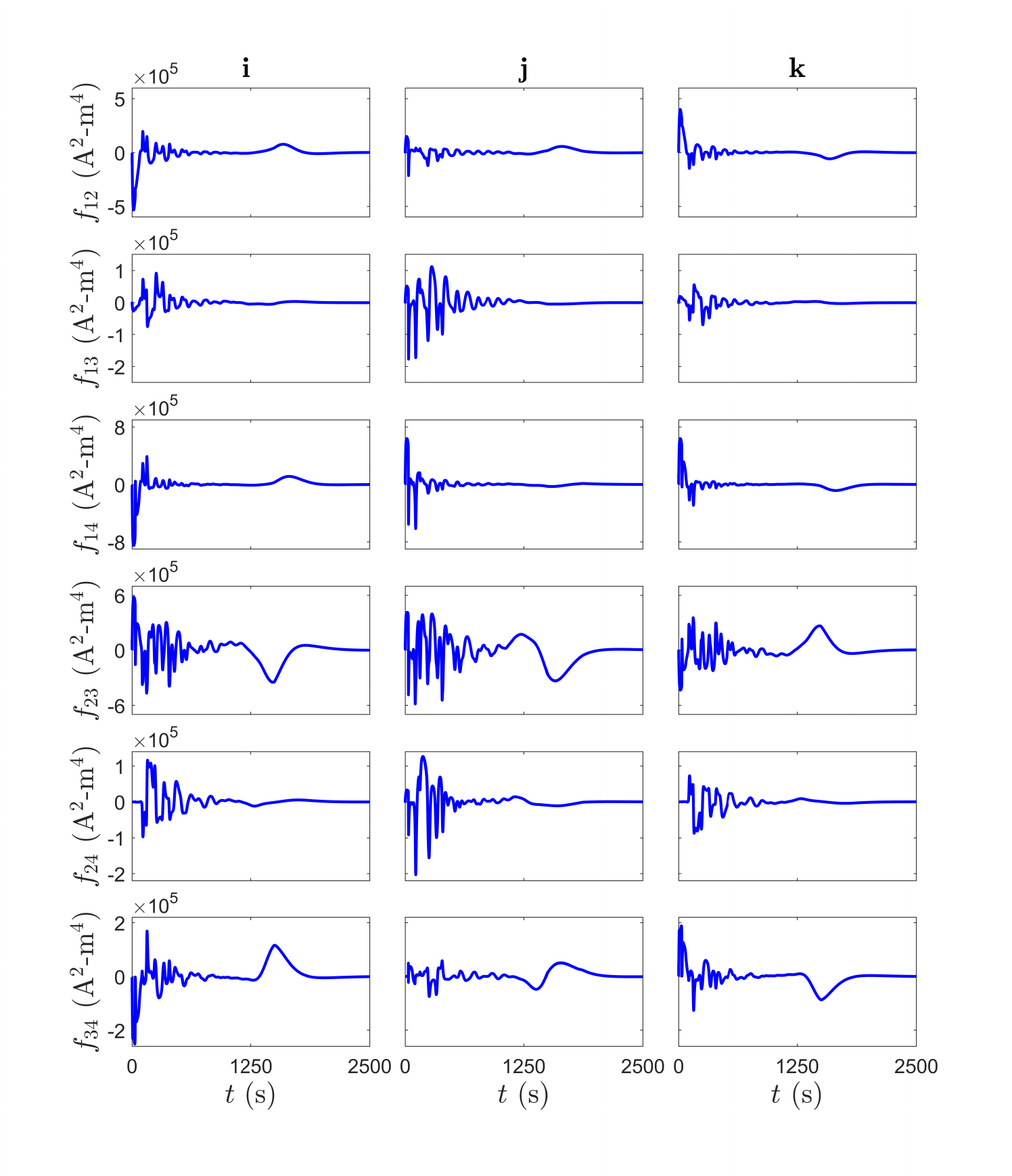}
        \caption{ Prescribed intersatellite force $f_{ij}$ computed using $\norm{r_{ij}}$ and $\zeta_{ij}$.}
          \label{fig:force_4_sat}
    \end{figure}

\begin{figure}[hbt!]
        \centering     \includegraphics[width=0.50\textwidth,clip=true,trim= 0.2in 0.5in 0.5in 0.3in]{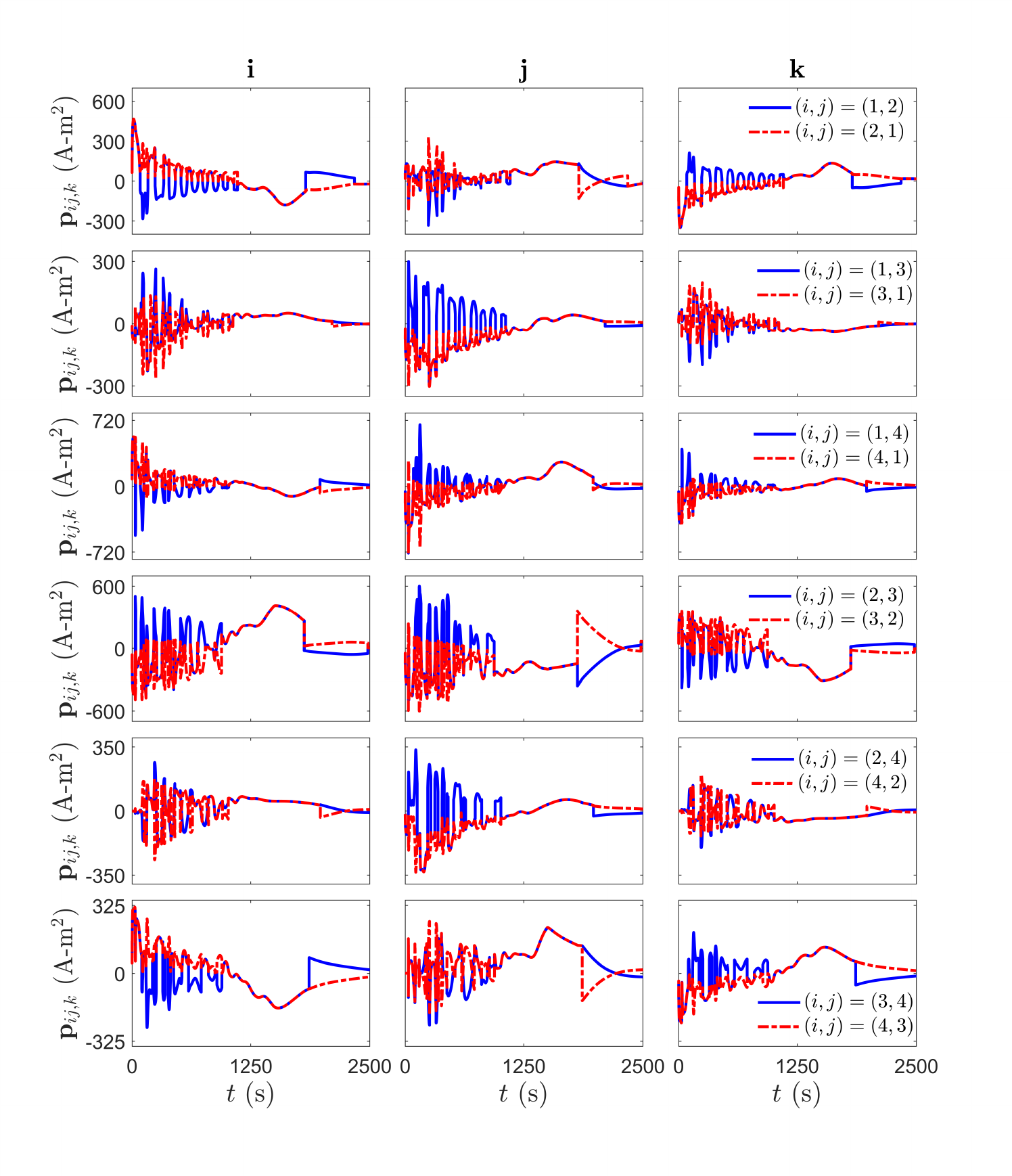}
        \caption{ Amplitude controls $\mathbf{p}_{ij,k}$ of magnetic moment control.}
        \label{fig:amplitudes_4_sat}
    \end{figure}

\begin{example}\label{4_sat_example}
This example illustrates algorithm effectiveness for handling multiple no-collision constraints. 
We select initial conditions and a desired formation such that $\mu_d$ would result in collisions. 
Specifically, let $n=4$, $r_{1}(0) = \begin{bmatrix}
            1.5 &0.75 &0.75
        \end{bmatrix}^{\rm{T}}$~m, $r_{2}(0) = \begin{bmatrix}
            -1.5  &0.75 &3  
        \end{bmatrix}^{\rm{T}}$~m, $r_3(0) = \begin{bmatrix}
            1.5  &3 &0.75
        \end{bmatrix}^{\rm{T}}$~m, $r_4(0) = \begin{bmatrix}
            -1.5  &3 &3
        \end{bmatrix}^{\rm{T}}$~m, $v_1(0) = v_2(0) =v_3(0) =v_4(0) = 0$~m/s, $d_{12} = \begin{bmatrix}
            -3 &0 &2.25
        \end{bmatrix}^{\rm{T}}$~m, $d_{13} = \begin{bmatrix}
            0 &2.25 &0   
        \end{bmatrix}^{\rm{T}}$~m, and $d_{14} = \begin{bmatrix}
            -3 &2.25 &2.25
        \end{bmatrix}^{\rm{T}}$~m. Note that the satellites are initially in a quadrilateral formation, and the desired formation involves swapping the position of the satellites on diagonally opposite vertices. The weights for the desired surrogate control $\mu_\rmd$ are \eqref{LQR_weights}, where $w_r = 10^6$, $w_v=10$, $w_{\zeta}=1$, and $w_{\mu} =10^2$.

\Cref{fig:3d_4_sat,fig:position_4_sat,fig:velocity_4_sat} show the satellites achieve the desired formation. 
\Cref{fig:barrier_functions_4_sat} shows that $h$ and its arguments (i.e., $R_{ij,2}$, $V_{ij,1}$, and $Q_i$) are positive for all time. 
Hence, $R_{ij}$ and $V_{ij}$ (\Cref{fig:constraints_rij_vij_4_sat}) are positive for all time, and \ref{obj3}--\ref{obj5} are satisfied. 
\Cref{fig:barrier_functions_4_sat} also shows that $\lambda$ is positive for $t\in [5.5,1493.2]$, which indicates that $\mu$ deviates from $\mu_\rmd$ during these time intervals as shown in \Cref{fig:mu_4_sat}. 
\Cref{fig:barrier_functions_4_sat} also shows that for $t \in [5.5,1493.2]$, $h$ is dominated by various collision avoidance CBFs $R_{ij,2}$ indicating that $\mu$ deviates from $\mu_\rmd$ to avoid collision. \Cref{fig:force_4_sat} shows $f_{ij}$, and \Cref{fig:amplitudes_4_sat} shows $\mathbf{p}_{ij,k}$.
\exampletriangle
\end{example}


\subsection{Formation Flying in Low Earth Orbit}
\label{low_earth_orbit_examples}

We consider satellites in low Earth orbit (LEO), where the initial orbits are at an altitude of 500~km from the surface of the Earth. 
The gravitational force acting on satellite $i$ is \Cref{eq:gravitational_force}, where $m_{\mathrm{e}} = 5.9 \times 10^{24}$ kg is the mass of Earth.

All the satellites are initially in the same orbital plane. 
Without loss of generality, $\CMcal{F}$ is selected such that the $\mathbf{i}$-$\mathbf{j}$ plane is the initial orbital plane, that is, $\mathbf{r}_i(0)$ and $\mathbf{v}_i(0)$ are in the $\mathbf{i}$-$\mathbf{j}$  plane. 
We consider a desired formation, where $\mathbf{d}_{ij}$ and its time derivatives are in the $\mathbf{i}$-$\mathbf{j}$ plane.

Next, we define 
\begin{equation*}
    \mathbf{r}_{\mathrm{com}} \triangleq \frac{1}{n} \sum_{i=1}^{n} \mathbf{r}_i,
\end{equation*}    
which is the position of the mass center of the formation. 
Consider the frame $\CMcal{F}_{\mathrm{com}}$ with unit vectors $ \begin{bmatrix}
    \mathbf{i}_{ \mathrm{com}} &\mathbf{j}_{\mathrm{com}} &\mathbf{k}_{\mathrm{com}}
\end{bmatrix}$, where $\mathbf{i}_{\mathrm{com}}= \mathbf{r}_{\mathrm{com}}/|\mathbf{r}_{\mathrm{com}}|$, $\mathbf{j}_{\mathrm{com}}$ is in the $\mathbf{i}$-$\mathbf{j}$ plane, and $\mathbf{k}_{\mathrm{com}}$ is such that $\mathbf{k}_{\mathrm{com}} \cdot \mathbf{k}>0$.
Let $R_{\mathrm{com}}$ be the rotation from $\CMcal{F}_{\mathrm{com}}$ to $\CMcal{F}$, and it follows that
\begin{equation*}
    d_{ij} = R_{\mathrm{com}} [\mathbf{d}_{ij}]_{\CMcal{F}_{\mathrm{com}}}, \label{eq:desired_relative_position_leo}   
\end{equation*}
If $\mathbf{r}_i$ is in the initial orbital plane for all time, then $\mathbf{k}_{\mathrm{com}}= \mathbf{k}$ and $R_{\mathrm{com}}$ is a principal rotation about $\mathbf{k}$.

 The collision radius is $\ubar{r}=2$~m, and the upper bound on intersatellite speed is $\bar{s}=0.025$~m/s. 
 We implement the control \Cref{eq:mu*,eq:lambda} with $\gamma=10^{5}$.

\begin{figure}[ht]
        \centering        \includegraphics[width=0.50\textwidth,clip=true,trim= 0.2in 0.0in 0.5in 0.2in]{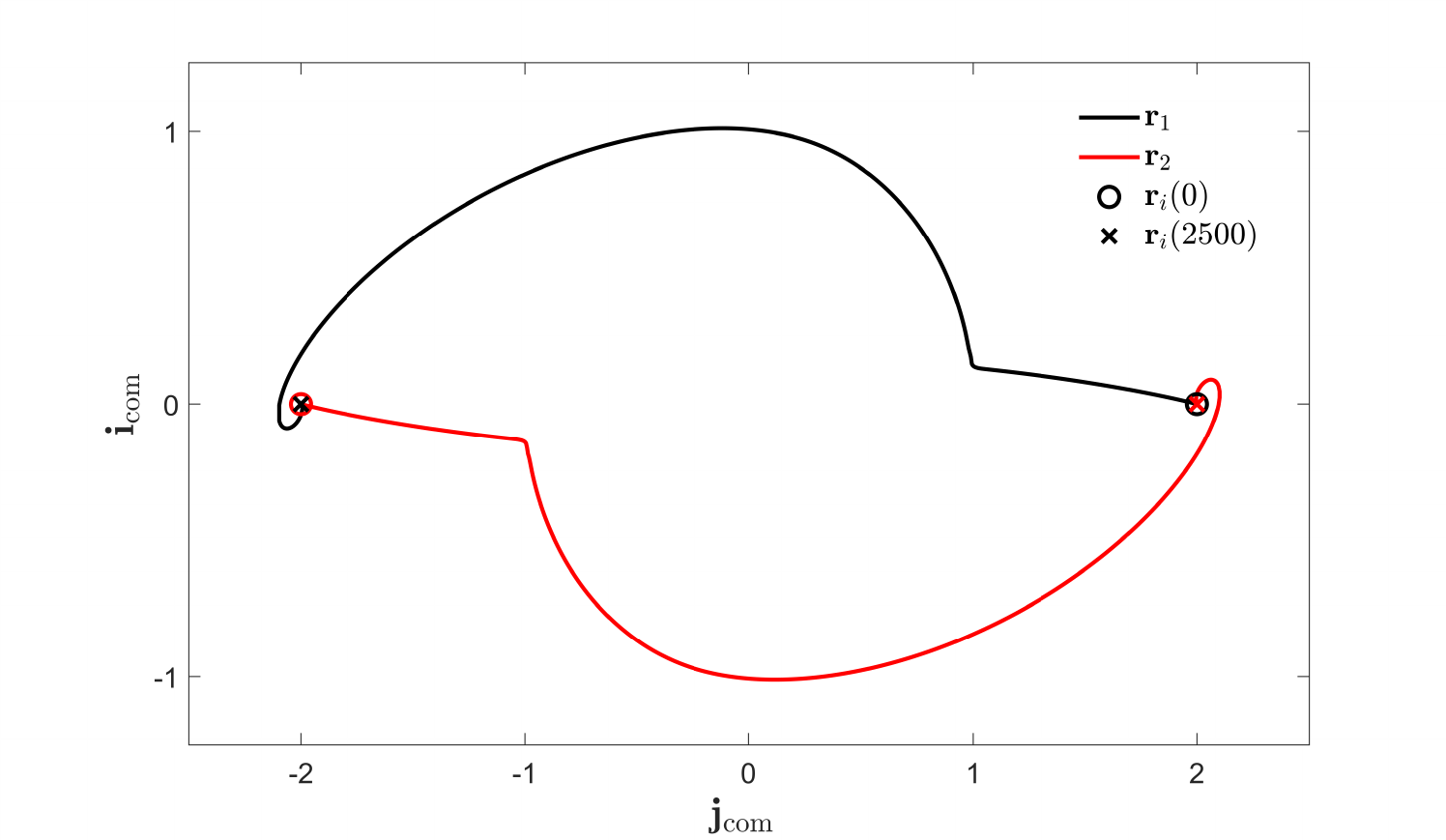}
        \centering
        \caption{Satellite trajectories $\mathbf{r}_i$ achieve the desired formation and avoid collision.}
        \label{fig:3d_2_sat}
    \end{figure}

\begin{figure}[ht]
        \centering           \includegraphics[width=0.50\textwidth,clip=true,trim= 0.2in 0.1in 0.5in 0.2in]{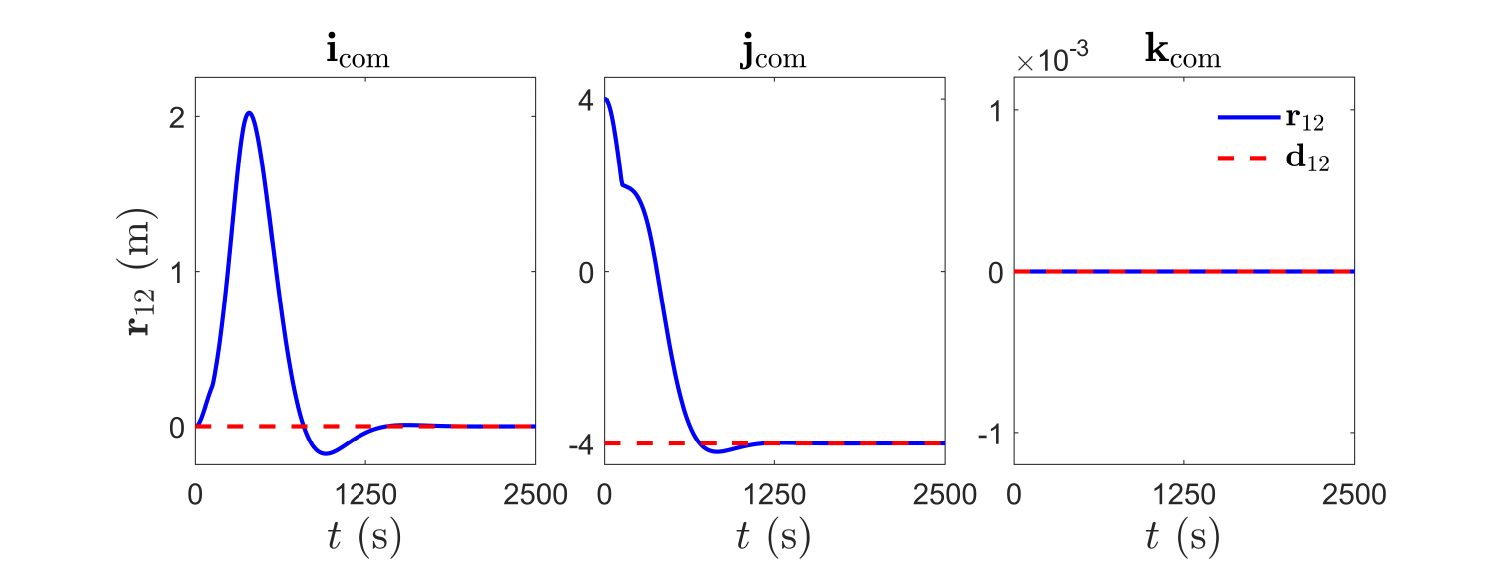}
        \caption{ Relative position $\mathbf{r}_{12}$ converges to $\mathbf{d}_{12}$.}
        \label{fig:position_2_sat}
    \end{figure}

\begin{figure}[ht]
        \centering        \includegraphics[width=0.50\textwidth,clip=true,trim= 0.2in 0.1in 0.5in 0.2in]{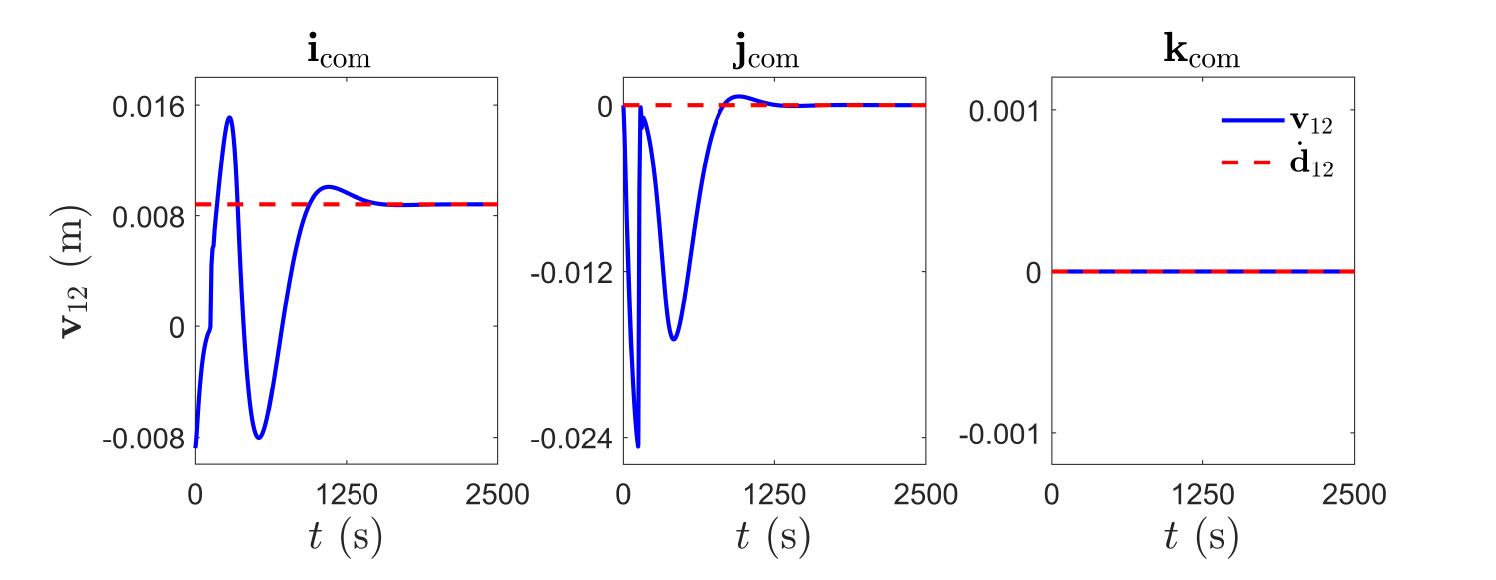}
        \caption{ Relative velocity $\mathbf{v}_{12}$ converges to $\dot{\mathbf{d}}_{12}$.}
       \label{fig:velocity_2_sat}
    \end{figure}

\begin{figure}[hbt!]
        \centering      \includegraphics[width=0.50\textwidth,clip=true,trim= 0.2in 0.1in 0.5in 0.2in]{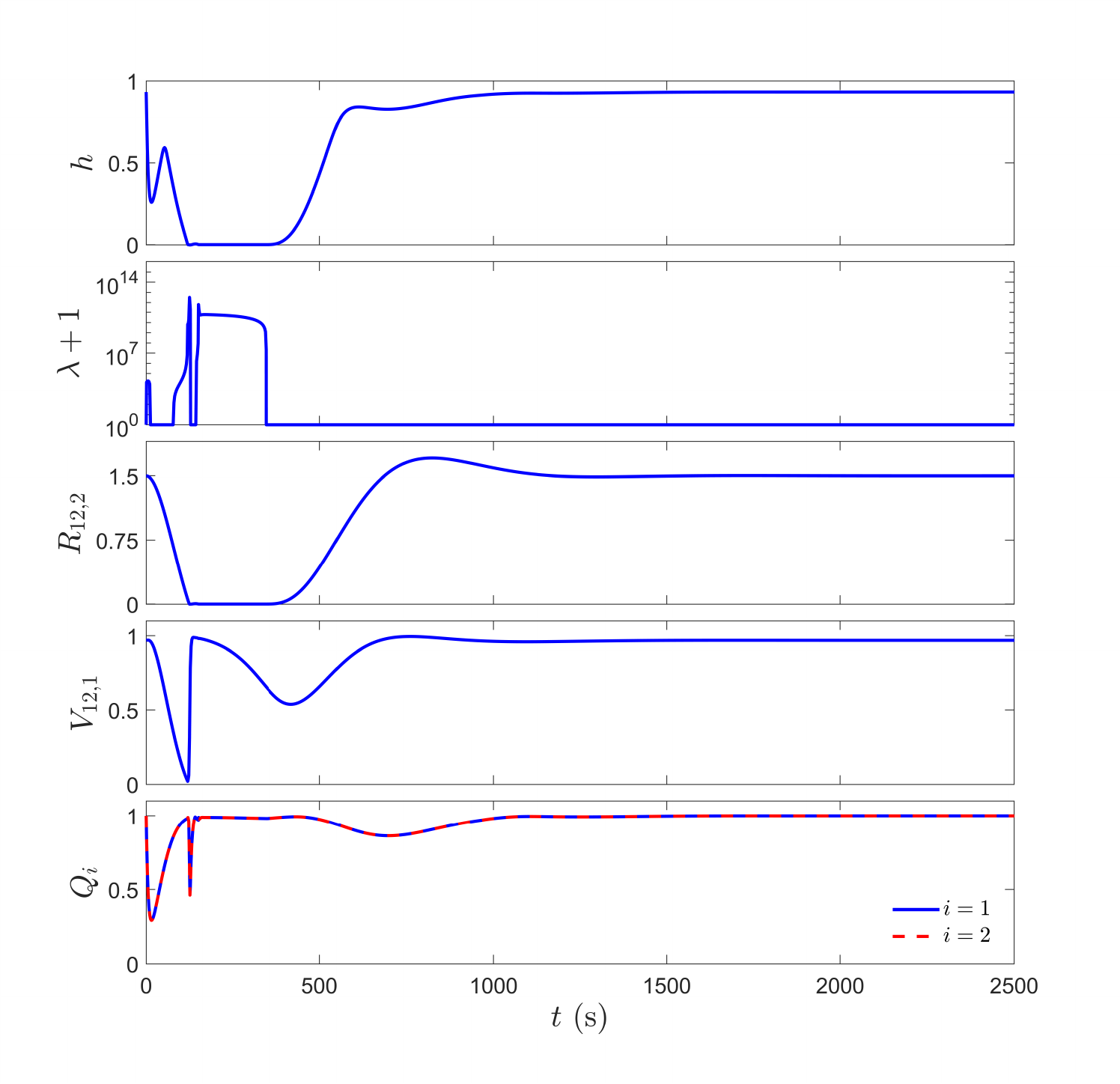}
        \centering
        \caption{Soft minimum $h$ is positive for all time, which implies that its arguments $R_{ij,2}$, $V_{ij,1}$, and $Q_i$ are positive for all time (as shown). 
        Note that $\lambda$ is positive at time instants when the control $\mu_*$ is modified from $\mu_\rmd$ in order to satisfy the state and input constraints.}        \label{fig:soft_min_barrier_functions_2_sat}
    \end{figure}

\begin{figure}[hbt!]
        \centering      \includegraphics[width=0.50\textwidth,clip=true,trim= 0.2in 0.1in 0.5in 0.2in]{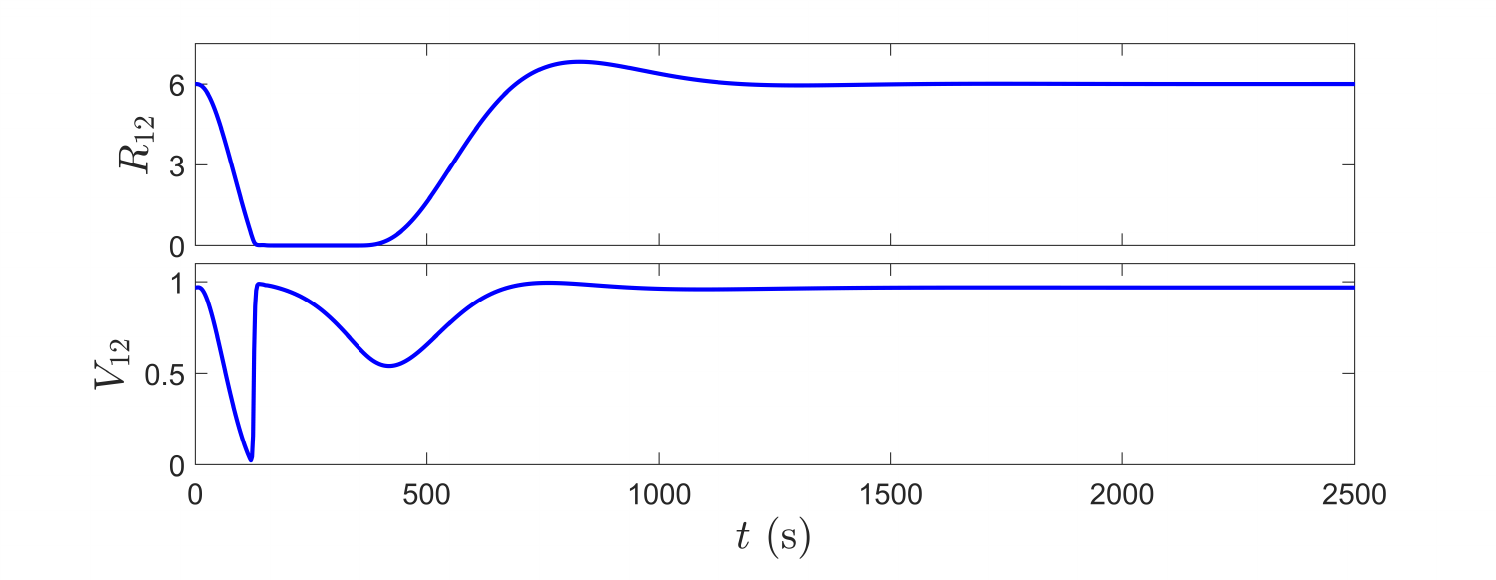}
        \centering
        \caption{CBFs $R_{ij}$ and $V_{ij}$ are positive for all time, which demonstrates that \labelcref{obj3,obj4} are satisfied.}    \label{fig:barrier_functions_2_sat}
    \end{figure}

\begin{figure}[hbt!]
        \centering       \includegraphics[width=0.50\textwidth,clip=true,trim= 0.2in 0.1in 0.5in 0.2in]{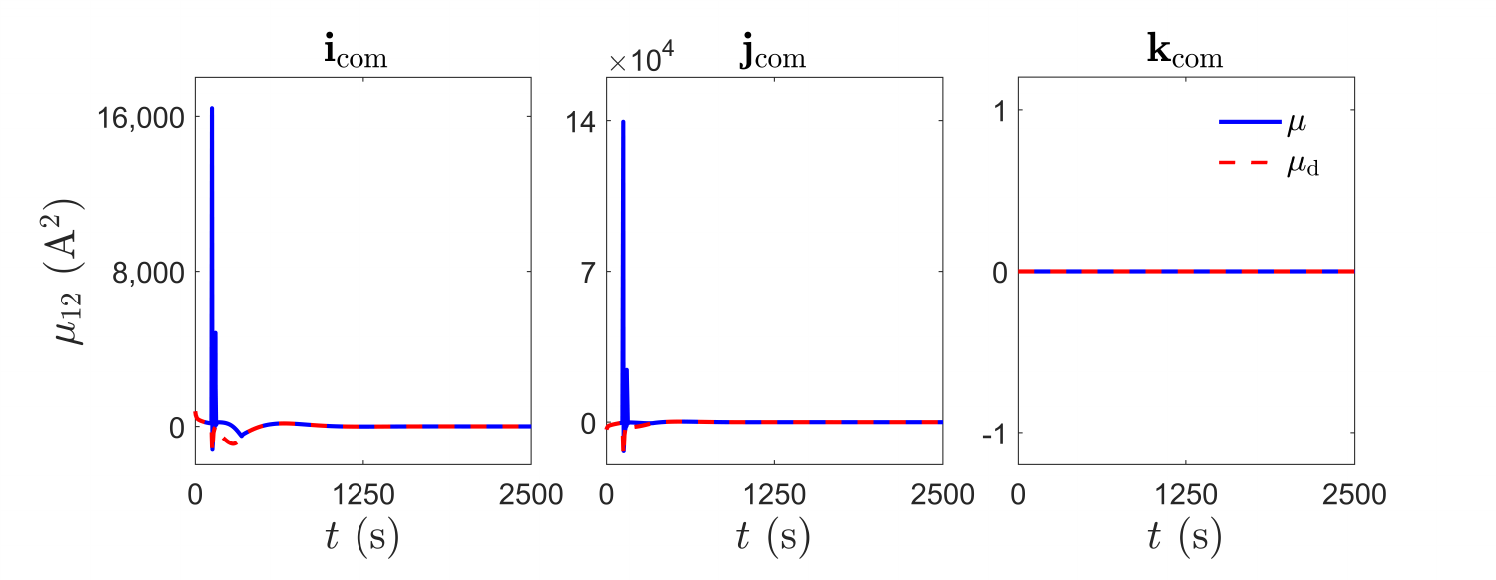}
        \caption{Surrogate control $\mu$ deviates from $\mu_{\rmd}$ to satisfy state and input constraints.}
          \label{fig:mu_2_sat}
    \end{figure}

\begin{figure}[hbt!]
        \centering       \includegraphics[width=0.50\textwidth,clip=true,trim= 0.2in 0.1in 0.5in 0.2in]{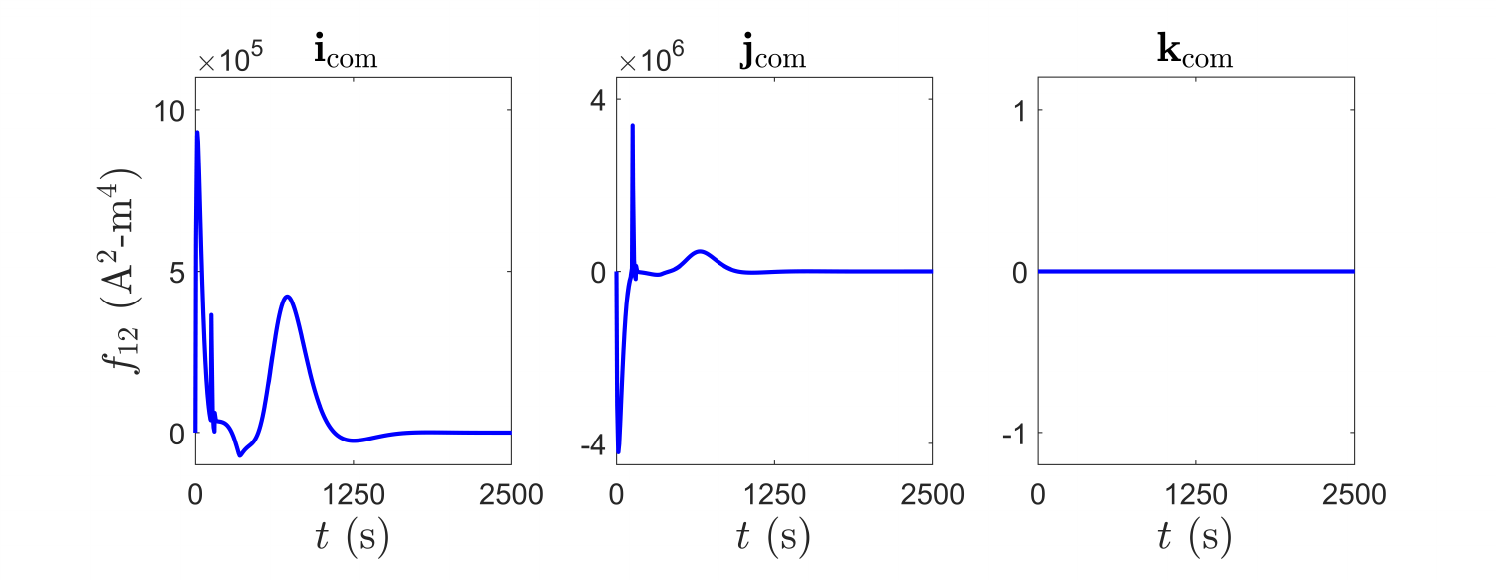}
        \caption{ Prescribed intersatellite force $f_{12}$ computed using $\norm{r_{12}}$ and $\zeta_{12}$.}
          \label{fig:force_2_sat}
    \end{figure}

\begin{figure}[hbt!]
        \centering     \includegraphics[width=0.50\textwidth,clip=true,trim= 0.2in 0.1in 0.5in 0.2in]{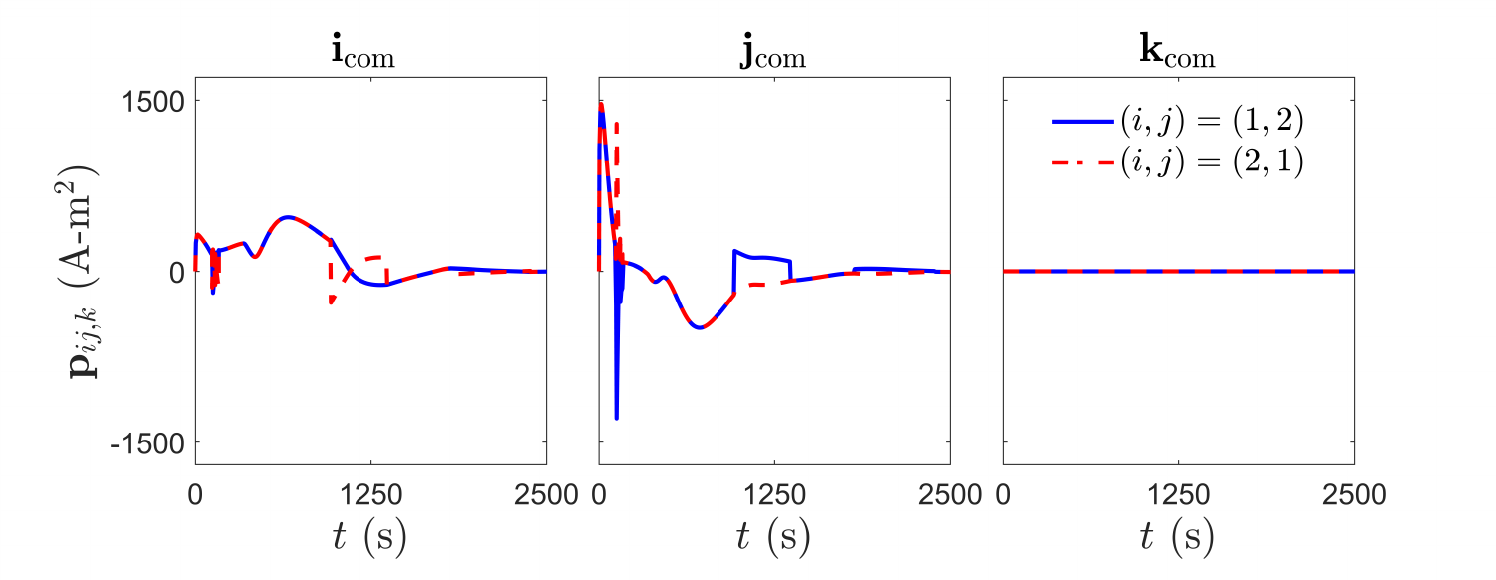}
        \caption{ Amplitude $\mathbf{p}_{ij,k}$ of magnetic moment control.}
        \label{fig:amplitudes_2_sat}
    \end{figure}

\begin{example} \label{2_sat_swap_leo_example}
This example illustrates a single maneuver in LEO. 
We select initial conditions and a desired formation such that $\mu_\rmd$ would exceed the power limit, the speed limit, and result in a collision without the modification \Cref{eq:mu*,eq:lambda} to address these constraints.
Specifically, let 
$n=2$, 
$r_{1}(0) = [6878\times 10^3 \ 2 \ 0]^{\rm{T}}$~m, 
$r_{2}(0) = [6878 \times 10^3 \ -2 \ 0 ]^{\rm{T}}$~m, 
$v_1(0) = \begin{bmatrix} -0.0022 &7579.9 &0 \end{bmatrix}^{\rm{T}}$~m/s, 
$v_2(0) = \begin{bmatrix} 0.0022 &7579.9 &0 \end{bmatrix}^{\rm{T}}$~m/s,
and 
$[ \mathbf{d}_{12} ]_{\CMcal{F}_{\mathrm{com}}} = \begin{bmatrix}
            0 &-4 &0
        \end{bmatrix}^{\rm{T}}$~m. 
The weights for the desired surrogate control $\mu_\rmd$ are \eqref{LQR_weights}, where $w_r = 10^7$, $w_v=10$, $w_{\zeta}=1$, and $w_{\mu} =50$.

\Cref{fig:3d_2_sat,fig:position_2_sat,fig:velocity_2_sat} show that the satellites avoid collision and swap positions to achieve the desired formation. 
\Cref{fig:soft_min_barrier_functions_2_sat,fig:barrier_functions_2_sat} show that \ref{obj3}--\ref{obj5} are satisfied. 
\Cref{fig:soft_min_barrier_functions_2_sat} also shows that $\lambda$ is positive for $t \in [0.1,12.4] \cup [78.1, 127.9] \cup [141.2,345.3]$, and \Cref{fig:mu_2_sat} shows that $\mu$ deviates from $\mu_\rmd$ during these intervals.
\Cref{fig:soft_min_barrier_functions_2_sat} also shows that $h$ is dominated by $Q_{1}$ and $Q_2$ for $t \in [0.1,12.4]$, $V_{12,1}$ for $t \in [78.1, 121.1]$, and by $R_{12,2}$ for $t \in (121.1,127.9] \cup [141.2,345.3]$.
Thus, this maneuver required modifications to satisfy power limits, speed limits, and collision avoidance.
\Cref{fig:force_2_sat,fig:amplitudes_2_sat} show $f_{12}$ and $\mathbf{p}_{12,k}$.
\exampletriangle
\end{example}

        

\begin{figure}[ht]
        \centering        \includegraphics[width=0.50\textwidth,clip=true,trim= 0.2in 0.0in 0.5in 0.2in]{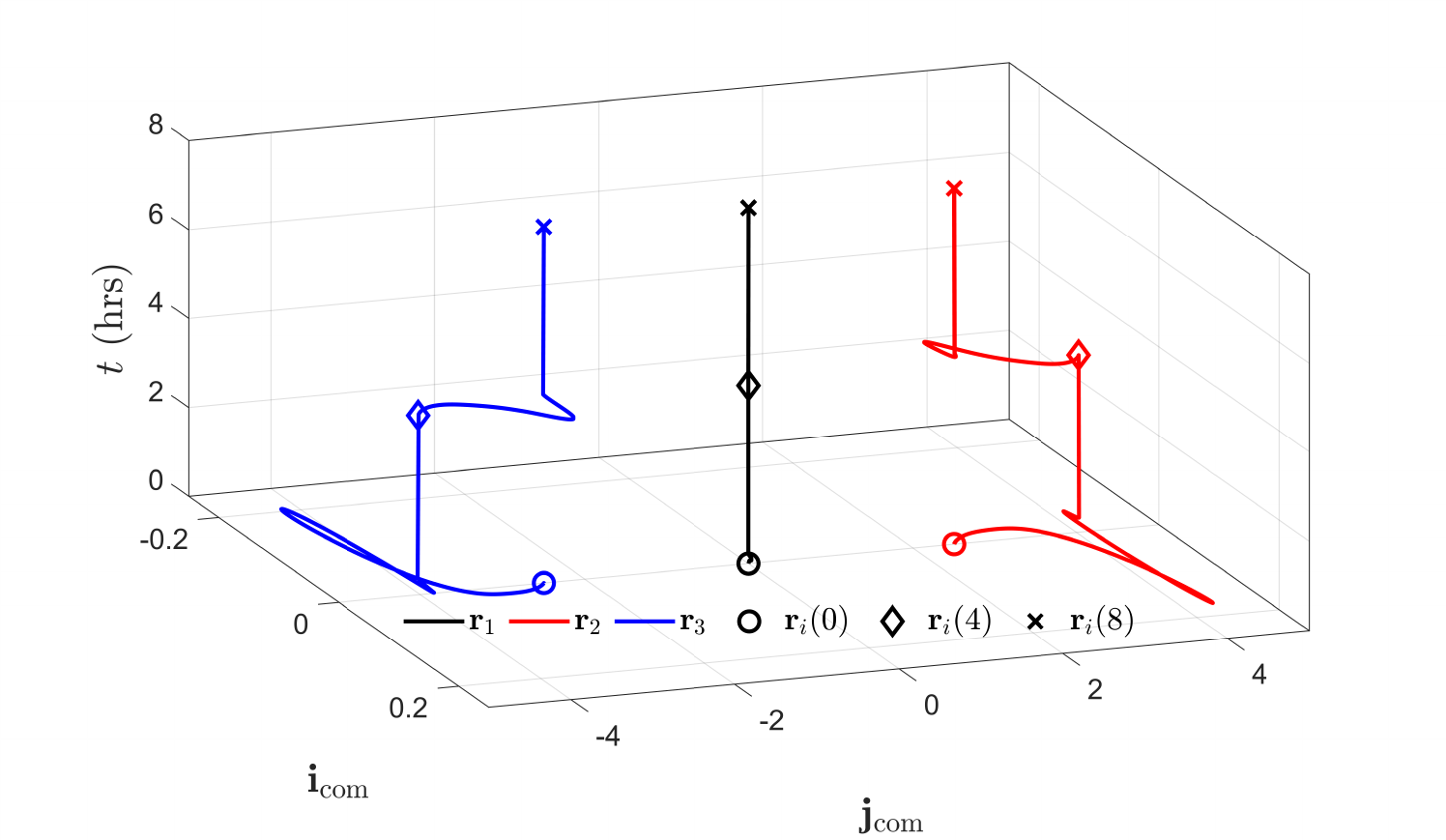}
        \centering
        \caption{Satellite trajectories $\mathbf{r}_i$ over a period of 8 hours (2 maneuvers).}
        \label{fig:3d_3_sat_trajectory_leo}
    \end{figure}

\begin{figure}[ht]
        \centering           \includegraphics[width=0.50\textwidth,clip=true,trim= 0.2in 0.1in 0.5in 0.2in]{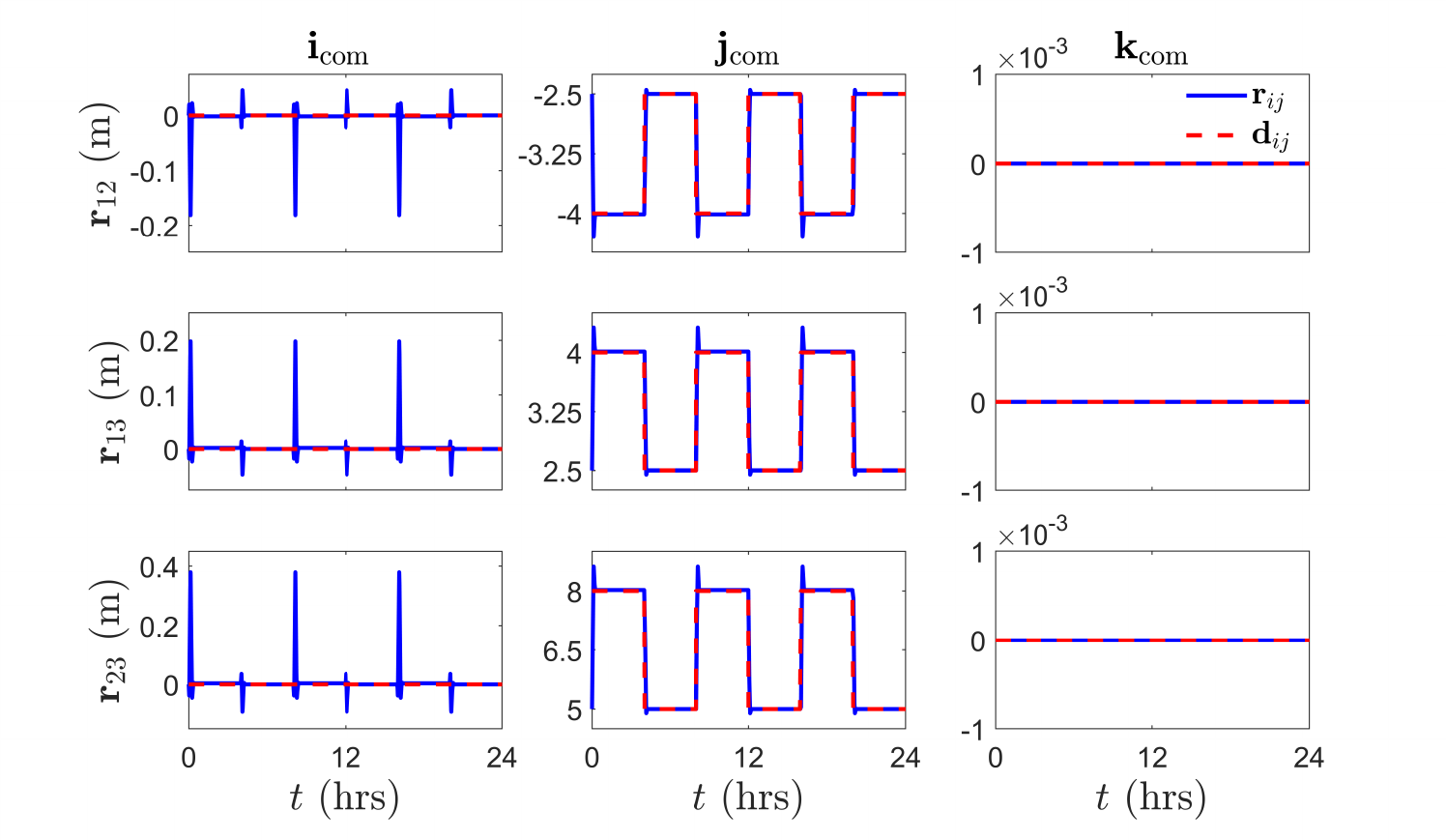}
        \caption{ Relative positions $\mathbf{r}_{ij}$ converge to desired formation $\mathbf{d}_{ij}$.}
        \label{fig:relative_position_3_sat_leo}
    \end{figure}

\begin{figure}[ht]
        \centering        \includegraphics[width=0.50\textwidth,clip=true,trim= 0.2in 0.1in 0.5in 0.2in]{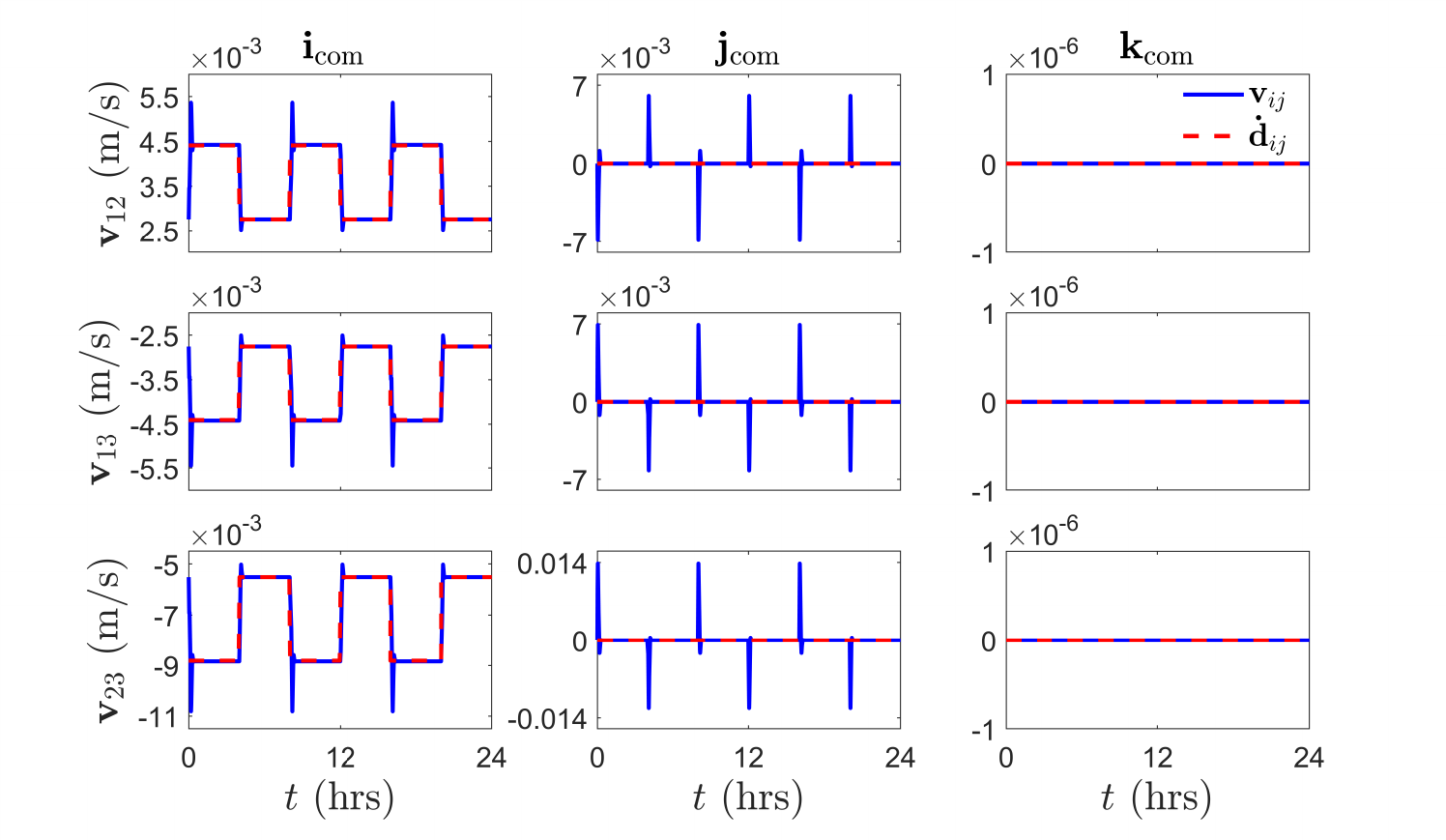}
        \caption{ Relative velocities $\mathbf{v}_{ij}$ converge to $\dot{\mathbf{d}}_{ij}$.}
       \label{fig:relative_velocity_3_sat_leo}
    \end{figure}

\begin{figure}[hbt!]
        \centering      \includegraphics[width=0.50\textwidth,clip=true,trim= 0.2in 0.3in 0.5in 0.5in]{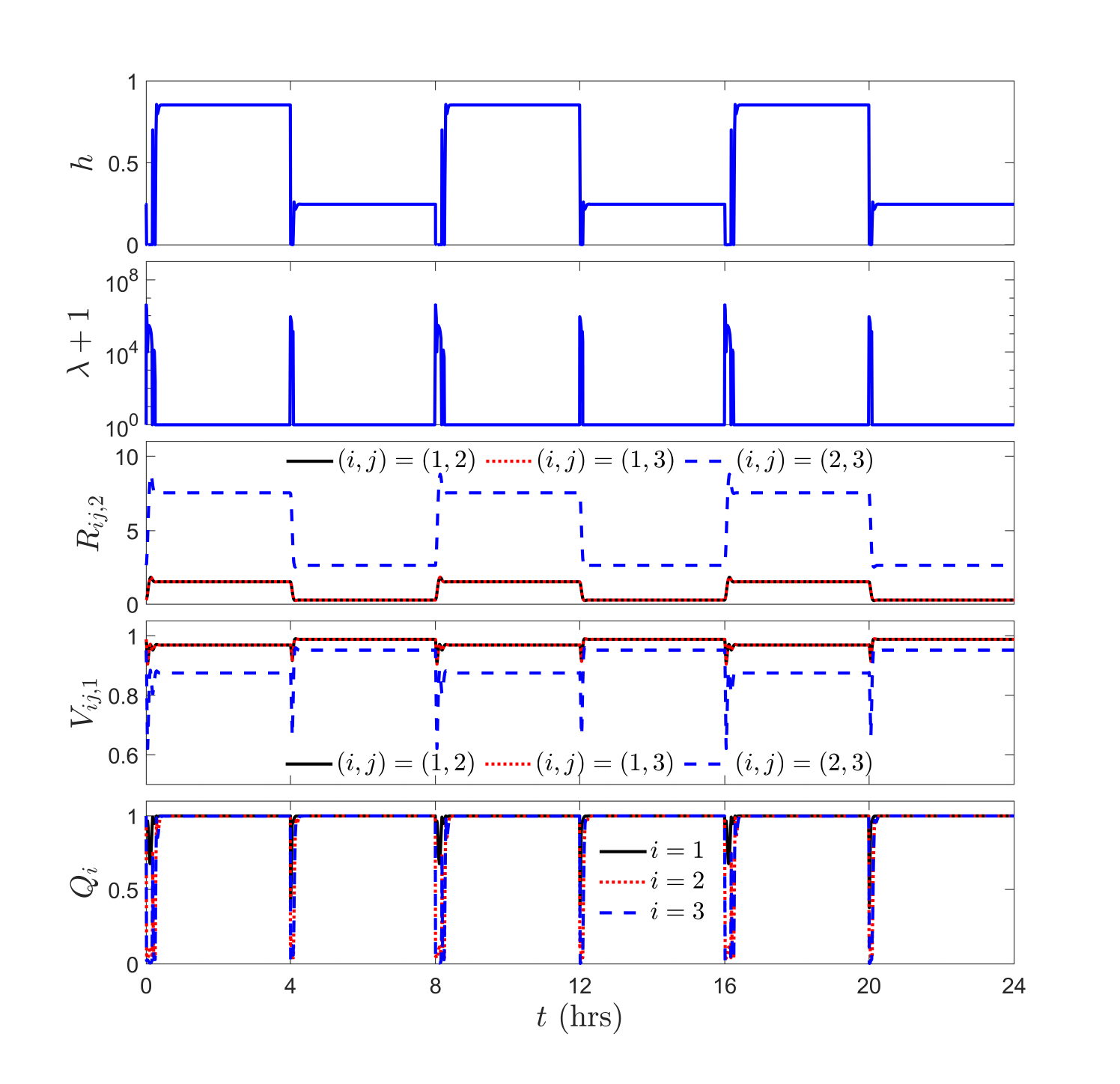}
        \centering
        \caption{Soft minimum $h$ is positive for all time, which implies that its arguments $R_{ij,2}$, $V_{ij,1}$, and $Q_i$ are positive for all time (as shown). 
        Note that $\lambda$ is positive at time instants when the control $\mu_*$ is modified from $\mu_\rmd$ in order to satisfy the power limits.}       \label{fig:barrier_functions_3_sat_leo}
    \end{figure}

\begin{figure}[hbt!]
        \centering      \includegraphics[width=0.50\textwidth,clip=true,trim= 0.2in 0.5in 0.5in 0.2in]{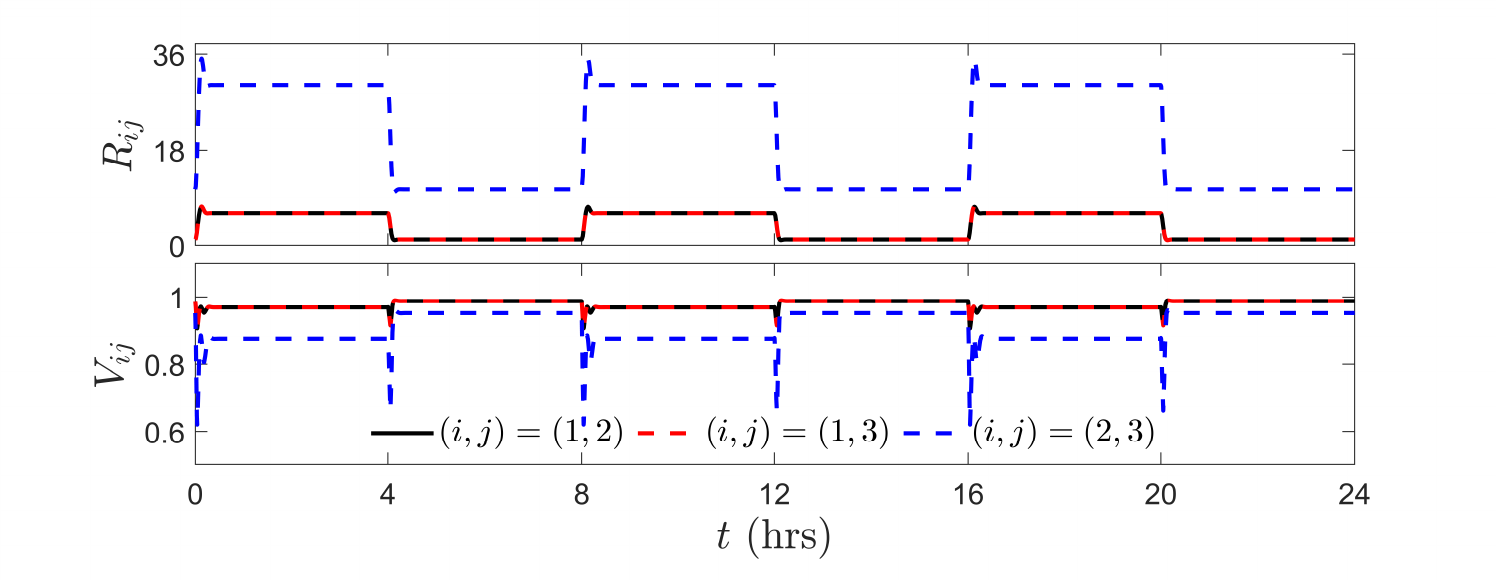}
        \centering
        \caption{CBFs $R_{ij}$ and $V_{ij}$ are positive for all time, which demonstrates that \labelcref{obj3,obj4} are satisfied.}        \label{fig:pos_vel_const_3_sat_leo}
    \end{figure}

\begin{figure}[hbt!]
        \centering       \includegraphics[width=0.50\textwidth,clip=true,trim= 0.2in 0.1in 0.5in 0.2in]{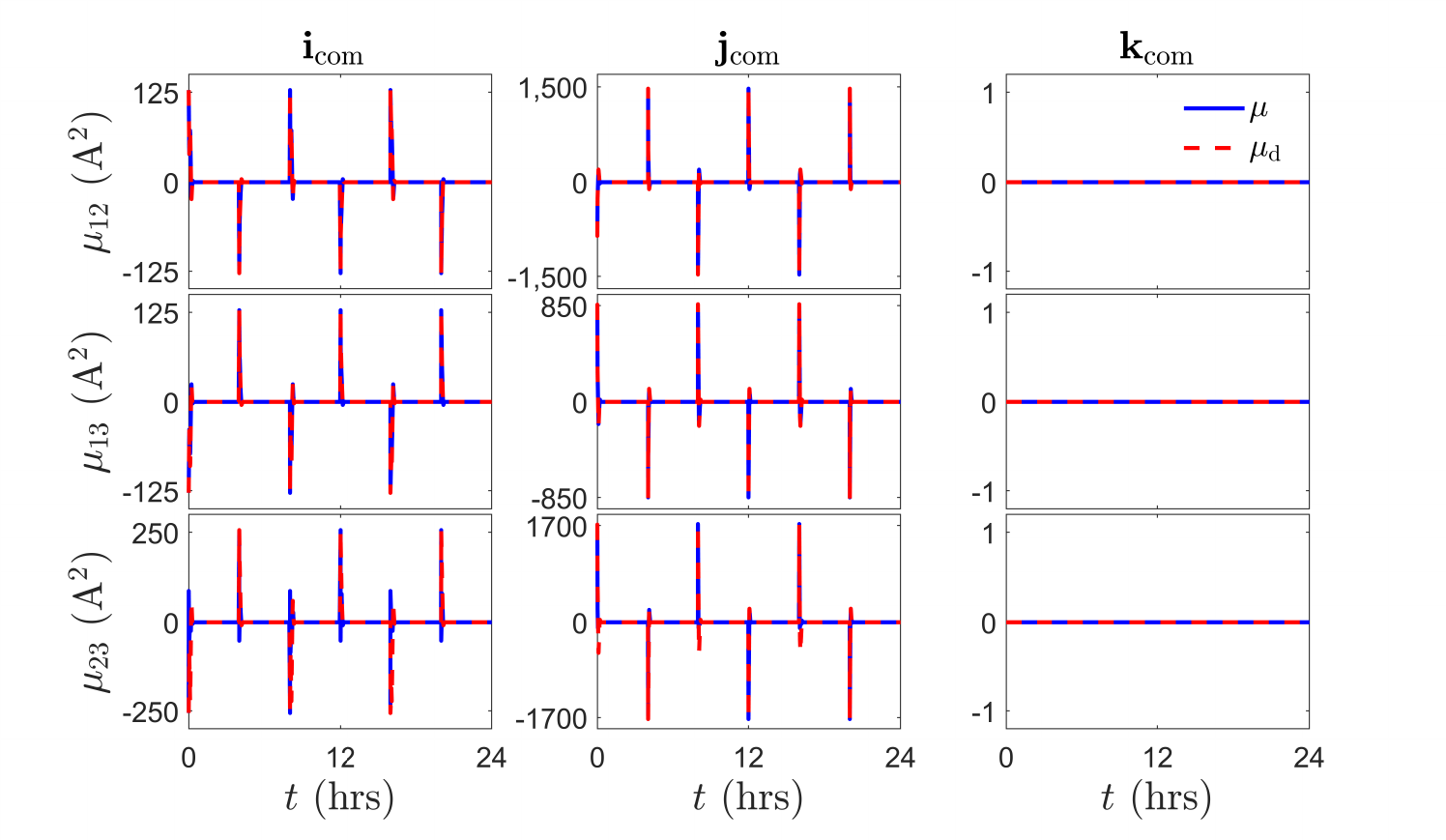}
        \caption{ Surrogate control $\mu$ deviates from $\mu_{\rmd}$ when $\lambda$ is positive to satisfy power limits.}
          \label{fig:mu_3_sat_leo}
    \end{figure}

\begin{figure}[hbt!]
        \centering       \includegraphics[width=0.50\textwidth,clip=true,trim= 0.2in 0.1in 0.5in 0.2in]{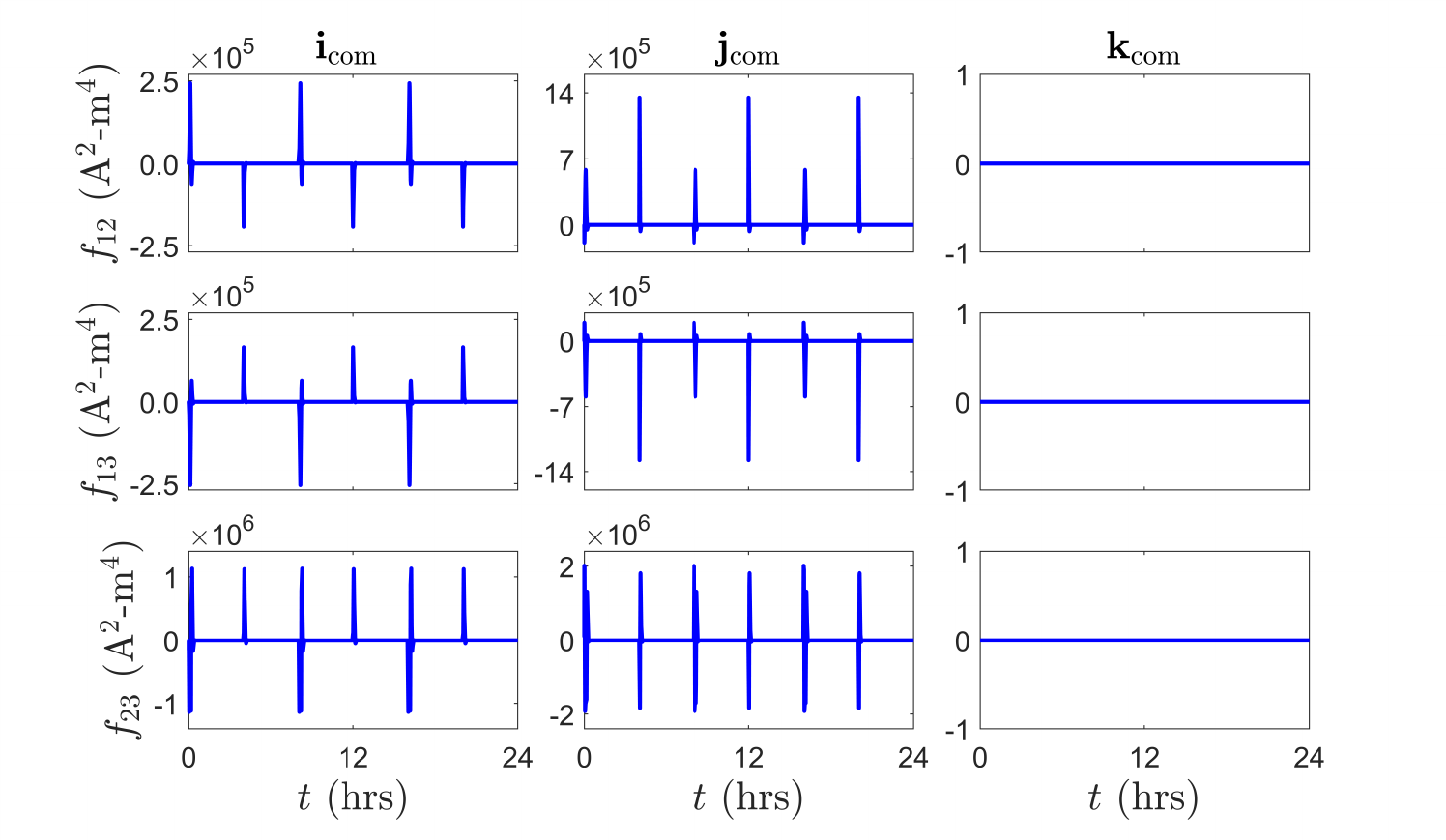}
        \caption{ Prescribed intersatellite forces $f_{ij}$ computed using $\norm{r_{ij}}$ and $\zeta_{ij}$.}
          \label{fig:force_3_sat_leo}
    \end{figure}

\begin{figure}[hbt!]
        \centering     \includegraphics[width=0.50\textwidth,clip=true,trim= 0.2in 0.1in 0.5in 0.2in]{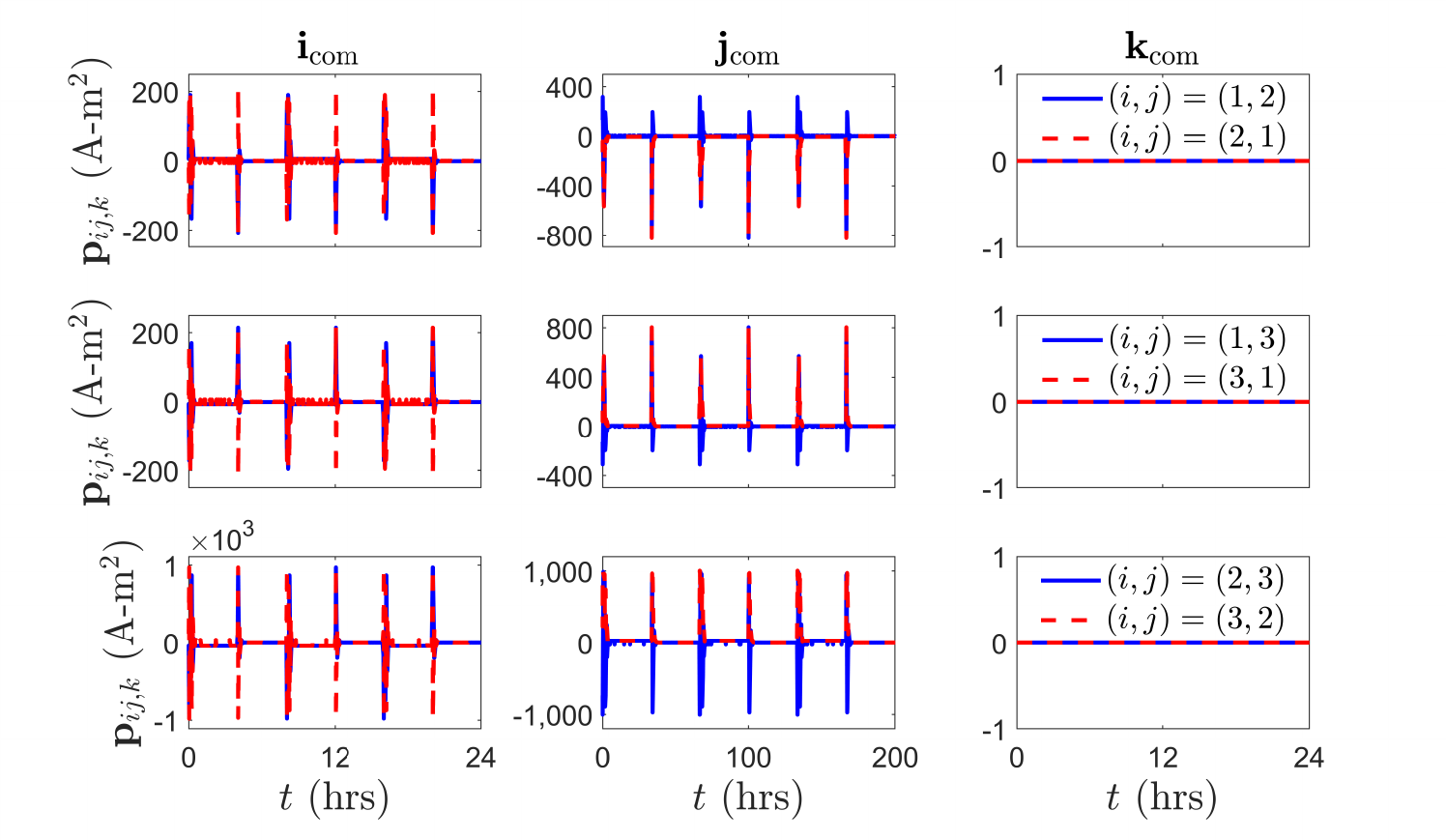}
        \caption{ Amplitude $\mathbf{p}_{ij,k}$ of magnetic moment control.}
        \label{fig:amplitudes_3_sat_leo}
    \end{figure}

\begin{example}  \label{3_sat_leo_example}

This example illustrates periodic maneuvers in LEO for a period of 24 hours. 
Let $n=3$, $r_{1}(0) = [
            6878\times 10^3 \ 0 \ 0
        ]^{\rm{T}}$~m, $r_{2}(0) = [
            6878 \times 10^3 \ 2.5 \ 0  
        ]^{\rm{T}}$~m, $r_{3}(0) = [
            6878 \times 10^3 \ -2.5 \ 0  
        ]^{\rm{T}}$~m, $v_1(0) = \begin{bmatrix}
            0 &7579.9 &0
        \end{bmatrix}^{\rm{T}}$~m/s, $v_2(0) = \begin{bmatrix}
            -0.0027 &7579.9 &0
        \end{bmatrix}^{\rm{T}}$~m/s, and $v_3(0) = \begin{bmatrix}
            0.0027 &7579.9 &0
        \end{bmatrix}^{\rm{T}}$~m/s.
Every 4 hours, the desired positions $[ \mathbf{d}_{12} ]_{\CMcal{F}_{\mathrm{com}}}$ and $[ \mathbf{d}_{13} ]_{\CMcal{F}_{\mathrm{com}}}$ switch between 
$\begin{bmatrix}
            1.2 \times 10^{-6} &-4 &0
        \end{bmatrix}^{\rm{T}}$~m and $\begin{bmatrix}
            1.2 \times 10^{-6} &4 &0
        \end{bmatrix}^{\rm{T}}$~m (repulsion), and $\begin{bmatrix}
            4.5 \times 10^{-7} &-2.5 &0
        \end{bmatrix}^{\rm{T}}$~m and $\begin{bmatrix}
            4.5 \times 10^{-7} &2.5 &0
        \end{bmatrix}^{\rm{T}}$~m (attraction).
Thus, there are 6 maneuvers over 24 hours. 
The weights for the desired surrogate control $\mu_\rmd$ are \eqref{LQR_weights}, where $w_r = 10^8$, $w_v=10^2$, $w_{\zeta}=1$, and $w_{\mu} =10^2$.

\Cref{fig:3d_3_sat_trajectory_leo} shows the trajectories of the satellites over the first 8 hours, where we observe the 2 formation maneuvers.
\Cref{fig:relative_position_3_sat_leo,fig:relative_velocity_3_sat_leo} show that the satellites  achieve formation. \Cref{fig:barrier_functions_3_sat_leo,fig:pos_vel_const_3_sat_leo}) show that \ref{obj3}--\ref{obj5} are satisfied. 
\Cref{fig:barrier_functions_3_sat_leo} also shows that $\lambda$ is positive immediately after the desired formation changes, and during these times, $h$ is dominated by $Q_1$ and $Q_3$, indicating that $\mu$ deviates from $\mu_\rmd$ (\Cref{fig:mu_3_sat_leo}) to satisfy power constraints. \Cref{fig:force_3_sat_leo,fig:amplitudes_3_sat_leo} show $f_{ij}$ and $\mathbf{p}_{ij,k}$.
\exampletriangle
\end{example}



\section{Concluding Remarks}\label{section:conclusions}

This article introduced a feedback control algorithm for EMFF with constraints on intersatellite distance, intersatellite speed, and apparent power. 
Theorem~\ref{thm:main_result_fwd_invariance} provides sufficient conditions such that all constraints are satisfied, and  Theorem~\ref{Theorem_1} demonstrated that that the algorithm generates a control that is as close as possible to the desired formation control while satisfying constraints.

The soft-minimum relaxed CBF approach in this article can be used to incorporate additional constraints. 
For example, this article did not directly address thermal management of the electromagnetic coils, which is important because high currents can lead to heating damage. 
However, thermal management can be addressed by incorporating additional constraints related to heat into the soft-minimum relaxed CBF. 
To demonstrate this approach, let $\bar \theta >0$ denote the maximum acceptable temperature of the coils from nominal temperature. 
The temperature change of the coils on the $i$th satellite is approximately 
\begin{equation*}
\theta_i = \frac{1}{C_\rmh} \int_0^t q_{\rmh, i}(\tau) \rmd \tau
\end{equation*}
where $C_\rmh$ is thermal capacitance, and the heat flow rate is
\begin{equation*}
    q_{\rmh, i} \triangleq \frac{1}{N^2 \sigma^2} \sum_{j \in \mathcal{I} \backslash i} R_\rmh(\theta_i) \norm{p_{ij}}^2 - \chi, 
\end{equation*}
where $R_\rmh$ is the temperature-dependent coil resistance and $\chi$ is the heat dissipation rate. 
Similar to the technique used for power, consider 
\begin{equation*}
\bar \theta_i \triangleq \frac{1}{C_\rmh} \int_0^t \left [ \frac{1}{N^2 \sigma^2} \sum_{j \in \mathcal{I} \backslash i} \bar R_\rmh \psi(r_{ij}(\tau),\zeta_{ij}(\tau)) - \chi \right ] \rmd \tau,
\end{equation*}
where $\bar R_\rmh$ is a known upper bound on $R_\rmh$, and it follows that $\bar \theta_i$ is a 2-times continuously differentiable upper bound on $\theta_i$. 
To address the heat constraint, we consider the 2-times continuously differentiable candidate CBF $\Theta_i \triangleq \bar \theta - \bar \theta_i$, which is relative degree 2. 
Let $\alpha_\rmh \colon \mathbb{R} \to \mathbb{R}$ be a $2$-times continuously differentiable extended class-$\mathcal{K}$ function, and consider the relative-degree-one CBF
\begin{align*}
    \Theta_{i,1}(x,\zeta) &\triangleq \dot \Theta_i + \alpha_\rmh(\Theta_i) \\
    &= \frac{1}{C_\rmh N^2 \sigma^2} \sum_{j \in \mathcal{I} \backslash i} \bar R_\rmh \psi(r_{ij},\zeta_{ij}) - \chi + \alpha_\rmh(\Theta_i). 
\end{align*}
To enforce heat constraints, $\Theta_{1,1}(x,\zeta),\cdots,\Theta_{n,1}(x,\zeta)$ are included as additional arguments in the soft-minimum relaxed CBF \eqref{eq:h}.

Another important practical consideration not addressed in this article is electromagnetic coil dynamics. 
Specifically, this article assumed ideal coil behavior; however, an inner-loop feedback architecture could be adopted in combination with the methods in this paper to directly address nonlinear coil dynamics. 
For example, consider the nonlinear coil dynamics 
\begin{align}
    \dot{\bar{x}}_i &= \bar{f}(\bar{x}_i) + \bar{g}(\bar{x}_i) \nu_i, \label{eq:coil_dyn_1}\\
    u_i &= \bar{h}(\bar{x}_i) \label{eq:coil_dyn_2},
\end{align}
where $\bar{x}_i(t) \in \BBR^{\bar n_i}$ is the state, $u_i$ is the magnetic moment generate by the coil resolved in $\SF$, and $\nu_i$ is the commanded magnetic moment. 
For simplicity, assume \Cref{eq:coil_dyn_1,eq:coil_dyn_2} is relative degree one; however, note that the approach described here can be extended to higher relative degree. 
Next, the desired trajectory for $u_i$ is 
\begin{equation}
    u_{\rmd,i} (t) \triangleq \sum_{j \in \mathcal{I} \backslash \{i\} } p_{ij}(t) \sin \omega_{ij} t,
\end{equation}
where the amplitudes $p_{ij}$ are generated from the control method presented in this article. 
One option to generate the command $\nu_i$ is to use a feedback linearizing control 
\begin{equation*}
    \nu_i = - L_{\bar{g}} \bar{h}(\bar x_i)^{-1} \left[ L_{\bar{f}} \bar{h}(\bar x_i) - \dot{u}_{\rmd,i} + \beta (u_i - u_{\rmd ,i}) \right],
\end{equation*}
where $\beta > 0$. 
This feedback control forces the command-following error $u_{i}-u_{\rmd,i}$ 
to converge to zero exponentially. 
The feedback linearizing control is just one feedback option for addressing the coil dynamics. 
The feedback linearizing approach has drawbacks. 
For example, it is model based and relies on feel state measurements. 
If there is significant model uncertainty and/or the state cannot be accurately estimated, then robust feedback linearizing control methods such as \cite{Hoagg2013} could be used.
Other approaches may also be appropriate based on coil dynamics; however, a detailed study is beyond the scope of this article.

%
\appendices

\section{Proofs of \Cref{Proposition_2,Prop:amplitude_bound}}
\label{appen.A}

\begin{proof}[\indent Proof of \Cref{Proposition_2}]
It follows from \eqref{eq:R} that $R r = \left[ \norm{r} \quad 0  \quad 0 \right]^{\mathrm{T}}$. 
Thus, substituting \Cref{eq:c1,eq:c2,eq:a_b_allo_mat} into \eqref{eq:model.3} yields
\begin{equation}
    f( r, c_1, c_2) 
 = a_x R^{\rm T} b + b_x R^{\rm T} a +  \frac{a^{\rm T} R R^{\rm T} b - 5 a_x b_x}{\norm{r} } r,
 \label{eq:f(r,a,b)}
\end{equation}
where the arguments $r$ and $f_*$ are omitted.
We consider 2 cases: (i) $[r]_{\times} f_* \neq 0$ and (ii) $[r]_{\times} f_* = 0$.

First, consider $[r]_{\times} f_* \neq 0$, and substituting \Cref{eq:R,eq:a_b_allo_mat} into \eqref{eq:f(r,a,b)} yields 
\begin{align}
    f( r, c_1, c_2) &= \left( -2 a_x b_x + a_y b_y - \frac{(a_x b_y + a_y b_x) r^{\mathrm{T}} f_*}{\| [r]_{\times}  f_* \|} \right) \frac{r}{\norm{r}} \notag
    \\
    &\qquad  + \frac{(a_x b_y + a_y b_x) \norm{r}}{\| [r]_{\times}  f_* \|} f_*.
    \label{eq:f(ax_bx_ay_by)}
\end{align}
We consider 2 subcases: $r^{\mathrm{T}} f_*=0$ and $r^{\mathrm{T}}f_* \neq 0$. 
For $r^{\mathrm{T}} f_*=0$, it follows from \Cref{eq:ax,eq:ay,eq:bx,eq:by} that $a_x = b_y=0$ and $a_y b_x = \| [r]_{\times}  f_* \|/\norm{r}$.
Thus, substituting into \eqref{eq:f(ax_bx_ay_by)} yields $f( r_{ij}, c_1, c_2) =  f_*$.
For $r^{\mathrm{T}}f_* \neq 0$, it follows from \Cref{eq:Phi1,eq:Phi2} that $\Phi_2=\Phi_1$, and using \Cref{eq:ax,eq:ay,eq:bx,eq:by} yields
\begin{gather*}
    a_x b_x = - \frac{\sgn(r^{\mathrm{T}} f_*)(|r^{\mathrm{T}} f_*| +\Phi_1)}{4 \norm{r}}, 
    \\ 
    a_x b_y = a_y b_x = \frac{\| [r]_{\times}  f_* \|}{2 \norm{r}},
    \\
     a_y b_y = -\frac{\sgn(r^{\mathrm{T}} f_*) (-|r^{\mathrm{T}} f_*| +\Phi_1)}{2 \norm{r}},
\end{gather*}
Thus, substituting into \eqref{eq:f(ax_bx_ay_by)} yields $f( r_{ij}, c_1, c_2) =  f_*$, which confirms the result for $[r]_{\times} f_* \neq 0$.

Next, we consider $[r]_{\times} f_* = 0$, and substituting \Cref{eq:ax,eq:bx,eq:R,eq:a_b_allo_mat} into \eqref{eq:f(r,a,b)} yields
\begin{align}
    f( r, c_1, c_2) 
    &= \frac{\sgn(r^{\mathrm{T}} f_*)(|r^{\mathrm{T}} f_*| +\Phi_1)^{\frac{1}{2}}(|r^{\mathrm{T}} f_*| +\Phi_2)^{\frac{1}{2}}}{2 \norm{r}^2} r. \label{eq:f(r,c1,c2)_r_x_c_neq_0}
\end{align}
We consider 2 subcases: $r^{\mathrm{T}} f_*=0$ and $r^{\mathrm{T}}f_* \neq 0$. 
For $r^{\mathrm{T}} f_*=0$, it follows from \eqref{eq:f(r,c1,c2)_r_x_c_neq_0} that $f( r, c_1, c_2) =0$. 
Since $r^{\mathrm{T}} f_*=0$ and $[r]_{\times} f_* = 0$, it follows that $f_*=0$, which implies that $f( r, c_1, c_2) =f_*$. 
For $r^{\mathrm{T}}f_* \neq 0$, it follows from \Cref{eq:Phi1,eq:Phi2} that $\Phi_2=\Phi_1$. 
Since $r^{\mathrm{T}}f_* \neq 0$ and $[r]_{\times} f_* = 0$, it follows that $r = (\sgn(r^{\mathrm{T}} f_*) \|r\|/ \|f_*\| )f_*$ and $|r^{\mathrm{T}} f_*| +\Phi_1 = 2\| r\| \| f_* \|$.
Thus, substituting into \eqref{eq:f(r,c1,c2)_r_x_c_neq_0} yields $f( r_{ij}, c_1, c_2) =  f_*$, which confirms the result for $[r]_{\times} f_* = 0$.   
\end{proof}

\begin{proof}[\indent Proof of \Cref{Prop:amplitude_bound}]
It follows from \Cref{eq:psi,eq:Phi1} that
\begin{align}
    \psi(r,f_*/\norm{r}^4) &= - \frac{r^{\mathrm{T}} f_*}{4 \norm{r}}   \tanh \left( \frac{ r^{\mathrm{T}} f_* }{\epsilon_1 \norm{r} } \right) \nn
    \\
    & \qquad + \sqrt{ \Phi_1(r,f_*)^2/\norm{r}^2  + \epsilon_2 } \nn\\
    &\ge - \frac{| r^{\mathrm{T}} f_* |}{4 \norm{r}}  + \frac{\Phi_1(r,f_*)}{\norm{r}}.
    \label{eq:psi(r,f/r4)}
\end{align}
We consider 2 cases: (i) $r^{\mathrm{T}} f_* \neq 0$ and (ii) $r^{\mathrm{T}} f_* = 0$.

First, consider $r^{\mathrm{T}} f_* \neq 0$, and
\Cref{eq:ax,eq:ay,eq:bx,eq:by,eq:Phi1,eq:Phi2,eq:c1,eq:c2,eq:a_b_allo_mat} imply that 
\begin{equation*}
    \norm{c_1(r,f_*)}^2 =\norm{c_2(r,f_*)}^2 = -  \frac{|r^{\mathrm{T}} f_*|}{4\norm{r}}    
    + \frac{3 \Phi_1(r,f_*) }{4 \norm{r}},
\end{equation*}
which combined with \eqref{eq:psi(r,f/r4)} confirms the result for $r^{\mathrm{T}} f_* \neq 0$.

Next, consider $r^{\rmT}f_*=0$, and 
\Cref{eq:ax,eq:ay,eq:bx,eq:by,eq:Phi1,eq:Phi2,eq:c1,eq:c2,eq:a_b_allo_mat} imply that 
\begin{equation*}
    \norm{c_1(r,f_*)}^2 = 2 \norm{c_2(r,f_*)}^2 = \frac{\Phi_1(r,f_*) }{\norm{r}},
\end{equation*}
which combined with \eqref{eq:psi(r,f/r4)} confirms the result for $r^{\mathrm{T}} f_* = 0$. 
\end{proof}

\bibliographystyle{ieeetr} 

 \bibliography{EMFF_CBF}

\end{document}